\documentclass[a4paper,11pt]{article}
\usepackage{jinstpub} 
\usepackage{lineno}
\usepackage{circuitikz}

\usepackage{amssymb}
\usepackage{amsmath}
\usepackage{siunitx}
\usepackage{csquotes}
\usepackage{booktabs}
\usepackage{verbatim}
\usepackage{graphicx}
\usepackage{xcolor}
\usepackage{mathastext}
\definecolor{darkorange}{RGB}{230,159,0}
\definecolor{reddishpurple}{RGB}{204,121,167}
\definecolor{bluishgreen}{RGB}{0,158,115}

\title{A Wide Bandwidth Trans-impedance Amplifier for Picosecond-Scale SiPM Characterization in a Wide Temperature Range}

\author[a,b]{P. Carniti,}
\author[a,b]{C. Gotti,}
\author[a,b]{G. Pessina}
\author[a,b,1]{and D. Trotta\note{Corresponding author.}}
\affiliation[a]{INFN Milano-Bicocca,\\Piazza della Scienza 3, Milano, Italy}
\affiliation[b]{University of Milano-Bicocca,\\Piazza della Scienza 3, Milano, Italy}

\emailAdd{d.trotta3@campus.unimib.it}

\abstract{
Future high-energy physics experiments using SiPMs as photosensitive elements may require operation at low temperatures (down to 80 K) to measure single photons with high time resolution in a highly radioactive environment. 
This calls for a complete characterization of these sensors over a wide temperature range to find the best compromise between detector performance and cooling requirements.
This paper presents the design of a transimpedance amplifier featuring high gain ($\sim 7500\ V/A$), very high speed ($<500\ ps$ rise time) and low input noise ($\lesssim 20\ pA/\sqrt{Hz}$), able to faithfully reproduce all the features of SiPM signals without affecting its noise or jitter performance. 
These features make the amplifier suitable for precise measurements of the time-of-arrival of single-photon signals, as well as gain and recovery time.
This article provides a detailed and thorough analysis of the circuit.
The network was simulated and measured in two configurations that differ in their open-loop gain and dominant pole frequencies. 
After selecting the best configuration for our purposes, the amplifier was characterized in detail at ambient temperature and at 80 K. 
Finally, we evaluated the amplifier using a SiPM operated at low over-voltage. 
While SiPMs are typically characterized at high over-voltage to enhance gain and minimize timing jitter, testing at low over-voltage allowed us to assess the amplifier's performance under more challenging and realistic conditions for single-photon timing.

}

\arxivnumber{}

\begin{document}

\maketitle
\flushbottom

\section*{List of Symbols and Abbreviations}

\begin{description}
    \item[\boldmath{$A_{AMP}$}] Amplifier open-loop gain
    \item[\boldmath{$A_{AMP_0}$}] Amplifier open-loop gain at low frequency
    \item[\textbf{CFOA}] Current Feedback Operational Amplifier
    \item[\boldmath{$C_o$}] internal CFOA capacitor
    \item[\textbf{DCR}] Dark Count Rate
    \item[\textbf{ENC}] Equivalent Noise Charge
    \item[\textbf{GBW}] Gain Bandwidth Product
    \item[\boldmath{$G_{CFOA}$}] CFOA closed-loop gain
    \item[\boldmath{$G_{CL}$}] Amplifier closed-loop gain
    \item[\boldmath{$G_{\infty}$}] Amplifier closed-loop gain when the loop gain is considered infinite
    \item[\boldmath{$G_{SiPM}$}] SiPM electron gain
    \item[\textbf{HBT}] Heterojunction Bipolar Transistor
    \item[\textbf{OA}] Operational Amplifier
    \item[\textbf{ODP}] OA Dominant Pole configuration
    \item[\textbf{PCB}] Printed circuit board
    \item[\textbf{SNR}] Signal-to-Noise Ratio
    \item[\boldmath{$T_{AMP}$}] Amplifier loop gain
    \item[\boldmath{$T_{AMP_0}$}] Amplifier loop gain at low frequency
    \item[\textbf{TIA}] Transimpedance Amplifier
    \item[\textbf{TDP}] Transistor Dominant Pole configuration
    \item[\boldmath{$\tau_{1_{CL}}$}] First closed-loop gain time constant
    \item[\boldmath{$\tau_{2_{CL}}$}] Second closed-loop gain time constant
    \item[\boldmath{$\tau_{1_{LG}}$}] Amplifier first loop gain time constant
    \item[\boldmath{$\tau_{2_{LG}}$}] Amplifier second loop gain time constant
    \item[\boldmath{$\tau_{1_{ODP}}$}] Amplifier first closed-loop gain time constant, ODP configuration
    \item[\boldmath{$\tau_{2_{ODP}}$}] Amplifier second closed-loop gain time constant, ODP configuration
    \item[\boldmath{$\tau_{1_{TDP}}$}] Amplifier first closed-loop gain time constant, TDP configuration
    \item[\boldmath{$\tau_{2_{TDP}}$}] Amplifier second closed-loop gain time constant, TDP configuration
    \item[\boldmath{$\tau_{cable}$}] Cable bandwidth time constant
    \item[\boldmath{$\tau_{gen}$}] Signal generator time constant
    \item[\boldmath{$\tau_{osc}$}] Oscilloscope time constant
    \item[\boldmath{$\tau_{HF}$}] CFOA second closed-loop gain time constant
    \item[\boldmath{$\tau_s$}] SiPM recharge time constant
    \item[\textbf{SiPM}] Silicon PhotoMultiplier
    \item[\boldmath{$\omega_{T1}$}] Amplifier angular frequency when the loop gain magnitude is equal to one
\end{description}

\section{Introduction}
\label{sec:intro}

Recent developments in physics show a growing interest in solid-state single-photon detectors. 
They are used in calorimeters, trackers, ring-imaging Cherenkov detectors, time-of-flight systems and others.  

The most common are Silicon PhotoMultipliers (SiPMs). These sensors are sensitive to single photons, are rather compact, immune to magnetic fields and need relatively low voltage bias, all characteristics that make them attractive when evaluating the requirements of the detector and the extreme conditions in which these sensors would be used. 

However, the major downside of using these sensors in high-energy physics experiments is that radiation can hinder their single-photon sensitivity \cite{IrradiatedSiPMs}. 
As with many solid-state sensors, they are sensitive to displacement damage caused by hadron fluence, conventionally rescaled to the effects of equivalent 1-MeV neutrons.
The primary effect in SiPMs is an increase in the rate of spurious single-photon signals, or dark count rate (DCR), which is often mitigated by lowering the temperature. 
At high fluences in the range $10^{12}\sim 10^{13}\ \mathrm{cm^{-2}}$ (1-MeV-equivalent), the cooling requirements can extend down to cryogenic temperatures \cite{IrradiatedSiPMs_nostro}. 

For this reason, to assess the application of SiPMs in all scenarios, these devices need to be characterized at different temperatures to find the best compromise between operating temperature and DCR. 
For best results with timing measurements, the readout electronics needs to be placed as close as possible to the SiPMs to preserve signal integrity. 
Crucially, in the context of this study, the electronics system is utilized within a laboratory characterization bench to evaluate sensors post-irradiation, rather than being subjected to the primary radiation field itself.
Therefore, here lies the necessity to design an amplifier that can work at a wide temperature range, between 300 K (ambient temperature) and cryogenic temperatures.
This amplifier must be suitable for SiPM characterization. 
Specifically, to preserve device rise times, which are typically $\sim1\ ns$ for devices such as the Hamamatsu S13360 SiPM series \cite{HamamatsuS13360Series2025}, and to enable precise timing measurements, an amplifier bandwidth exceeding $300\ \mathrm{MHz}$ is required.
This ensures the signal shape is not significantly distorted by the electronics.

In this paper, we first describe the circuit schematic and study the amplifier stability. 
Then we provide an overview of two different amplifier configurations. 
The two configurations are simulated and compared with real-world measurements at ambient temperature (300 K).  
After evaluating the pros and cons of the two solutions, we study the most promising one in more detail and measure its behavior at both ambient and cryogenic temperatures (liquid nitrogen).  
In the final section, we demonstrate the amplifier’s performance in detecting single-photon signals from a SiPM at the two extremes of the temperature range.

\section{Circuit analysis}
\label{sec:circuit_schematic}

For this work, a Transimpedance Amplifier (TIA) topology was selected for the front-end electronics to optimize the processing of SiPM current signals. 
The TIA provides a low input impedance interface, which is critical to minimize the impact of the SiPM’s internal capacitance on signal rise and fall times. 
By directly converting the instantaneous current into a voltage signal, the TIA also ensures a high-bandwidth response and high gain, preserving the fast temporal characteristics necessary for accurate signal detection. 

A two-stage TIA topology has been adopted in this design. 
The first stage utilizes a silicon-germanium Heterojunction Bipolar Transistor (HBT), which provides an ultra-fast, low-noise input interface with an exceptionally high transition frequency. 
The second stage employs a Current Feedback Operational Amplifier (CFOA) -— fabricated in a complementary bipolar process -— selected for its high slew rate and gain-independent bandwidth. This design is based on the topology presented in \cite{Dune_vecchio}, where the same HBT is used as the first stage and the conventional differential voltage operational amplifier is replaced with a CFOA to further enhance speed performance.

\subsection{Circuit description}
\label{sec:circuit_description}

\begin{figure}[htbp]

    \centering
    \tikzset{every node/.style={font=\Large}} 
    \ctikzset{amplifiers/thickness=2, transistors/thickness=3, amplifiers/scale=1.2}
    \begin{circuitikz}[scale=0.445 ,transform shape, american]       
          \draw (-1,0) node[circ]{} coordinate(in) to[short, i =$I_{in}$] ++(2,0) coordinate(V_B); 
          \draw (in) to[R,l_=$R_{s2}$] ++(-0.3,-2.5) -- ++(-1,0) to[transmission line, l^=$50\ \Omega$] ++(-2,0) to[generic,l^=$Z_{SiPM}$] ++(-1.5,0) -- coordinate(HVcompensation) ++(-1.25,0) to[R, l_=$R_{HV}$] ++(-1.25,0) -- ++(-1.75,0) node[ocirc,label=left:$V_{HV}$]{};
          \draw (HVcompensation) to[R, l_=$R_{s1}$] ++(0,-2) to[C,l_=$C_{HV}$] ++(0,-1.25) node[ground]{};
          \draw (in) to[short] ++(-0.3,2.5) to[C, l_=$C_t$, a^=\SI{1}{pF}, l2 halign=c] ++(-2,0) -- ++(-0.5,0) coordinate(pippo) to[R,l_=$R_{t}$] ++(0,-2) node[ground]{};
          \draw (pippo) -- ++(-1.5,0) to[transmission line, l^=$50\ \Omega$,a_=LEMO] ++(-3.5,0) node[ocirc,label=left:Test signal]{};
          \draw (V_B) node[npn,anchor=base](bjt){$\textbf{HBT}$};  
          \draw (bjt.emitter) to[short] ++(0,-0.5) coordinate(gnd);
          \draw (bjt.collector) to[short,-*] ++(0,0.5) node[left]{$V_c$} coordinate(coll) -- ++(0,0.3) to[R=$R_c$] ++(0,2.5)  to[battery2,name=myB,l=$V_{POS}$] ++(0,1) node[ground,rotate=180]{} coordinate(Vcc);
          \ctikztunablearrow{0.5}{1.3}{60}{myB}
          \draw (coll) -- ++(7.6,0) node[op amp, anchor=+,noinv input up](opamp){$\textbf{CFOA}$};
          \draw (coll) ++(2,0) node[circ]{} coordinate(parasitic) -- ++(0,0.3) to[R=$R_d$] ++(0,2) to[C=$C_d$] ++(0,1.5) coordinate(comp_coll);
          \draw (comp_coll) -- (Vcc -| comp_coll) node[ground,rotate=180]{};
          \draw (parasitic) ++(2,0) node[circ]{} coordinate(parasitic2) to[C=$C_x$] ++(0,3.8) coordinate(parasitc_gnd);
          \draw (parasitc_gnd) -- (Vcc -| parasitc_gnd) node[ground,rotate=180]{};
          \color{lightgray}
          \draw (parasitic2) ++(2,0) node[circ]{} to[C=$C_p$] ++(0,3.8) coordinate(parasitc_gnd2);
          \draw (parasitc_gnd2) -- (Vcc -| parasitc_gnd2) node[ground,rotate=180]{};
          \color{black}    
          \draw (opamp.-) to[short] ++(-2,0) -- ++(0,-0.3) to[R=$R_1$] ++(0,-1.2) to[C=$C_{1}$] ++ (0,-1.2) node[ground]{} coordinate(gnd2); 
          \draw (opamp.-) ++(-0.6,0) node[circ]{} coordinate(R2_node) to[short] ++(0,-2.2) to[R, l_=$R_2$]  ++(3.4,0) -| (opamp.out) node[circ]{};
          \draw (opamp.up) -- ++(0,0.2) node[vcc]{$V_{CC}$};
          \draw (opamp.down) -- ++(0,-0.2) node[vee]{$V_{EE}$};
          \draw (gnd) -- (gnd2 -| gnd) node[ground]{};
          \draw (opamp.out) -- ++(0.5, 0) coordinate(vout) node[circ,label=above:$V_{out}$]{} -- ++(0,-5) coordinate(FB);
          \draw (vout) to[R=$R_{out}$] ++(3,0) to[C=$C_{out}$] ++(1.5,0) coordinate(tooscilloscope);
          \draw (tooscilloscope) to[transmission line, l^=$50\ \Omega$,a_=LEMO] ++(2.5,0) coordinate(oscilloscope) to[R=$R_{osc}$] ++(0,-1.5) node[ground]{};
          \draw (oscilloscope) -- ++(1,0) node[ocirc,label=right:$V_{osc}$]{};
          \draw (FB) to[R=$R_f$] ++(-8,0) coordinate(FB2) -- (FB -| bjt.base) coordinate(FB3) -- (bjt.base) node[circ]{} node[above, xshift=-2pt]{$V_{be}$};
          \color{lightgray}
          \draw (FB) ++(-2, 0) -- ++(0,-2) to[C=$C_{p_{R_f}}$] ++(-4,0) -- ++(0, 2); 
          \draw (FB3) node[circ]{} to[C=$C_{pb}$] ++(0,-3.4) node[ground]{};
          \color{black}
          \draw (FB2) ++(-1.25,0) node[circ]{} -- ++(0,-0.4) to[R=$R_b$] ++(0,-1.5) to[C=$C_b$] ++(0,-1.5) node[ground]{};

          \draw[dashed,ultra thick, black] (-11.5,-6.75) rectangle (-0.5,4);
          \node at (-10.25,4.35) {\textcolor{black}{Input signals}};

          \draw[dashed,ultra thick, black] (4.5,2.25) rectangle (8.5,4.25);
          \node[rotate=0] at (8.9,3.25) {\textcolor{black}{$\mathbf{C_c}$}};

          \draw[dashed,ultra thick, black] (2.7,-8.5) rectangle (4.95,-4.5);
          \node[rotate=0] at (5.35,-6.5) {\textcolor{black}{$\mathbf{Z_b}$}};

          \draw[dashed,ultra thick, black] (6.4,-8) rectangle (11.15,-3.6);
          \node[rotate=0] at (11.55,-5.8) {\textcolor{black}{$\mathbf{Z_f}$}};

          \node[draw=black, thick, fill=white, align=center, inner sep=6pt] at (-8, 0) {

            Either HV or the test \\
            signal are connected, not \\
            both at the same time.
           };

    \end{circuitikz}

    \caption{Schematic of the amplifier circuit. The gray components represent the parasitic capacitances.}
    \label{fig:open_loop_and_closed_loop_configurations}
\end{figure}
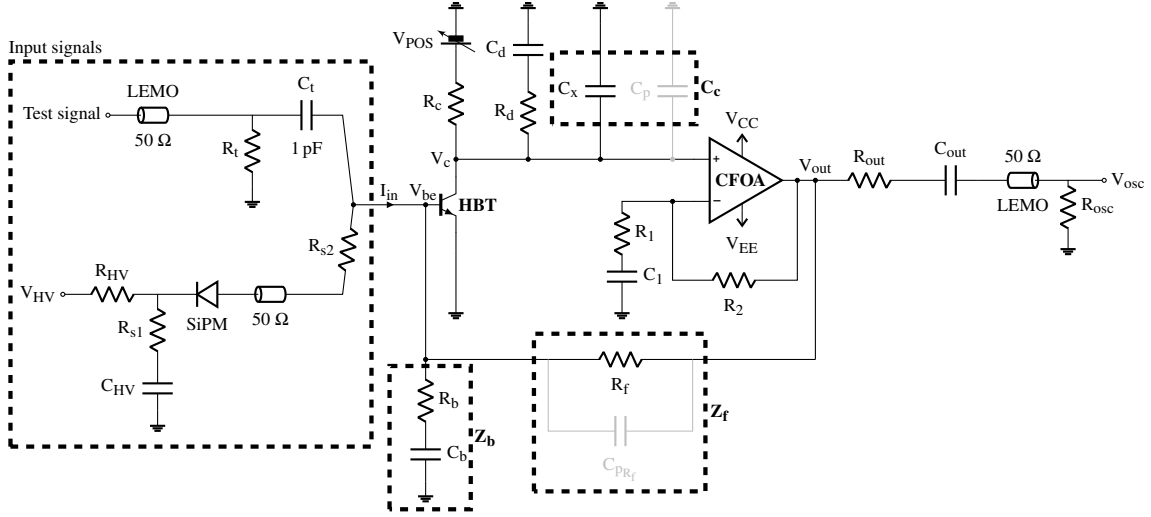

As illustrated in figure \ref{fig:open_loop_and_closed_loop_configurations}, the system is 
designed to support two distinct operating modes: a direct SiPM input or the injection of a known charge via a voltage step through a $1\ pF$ capacitor, $C_t$. 

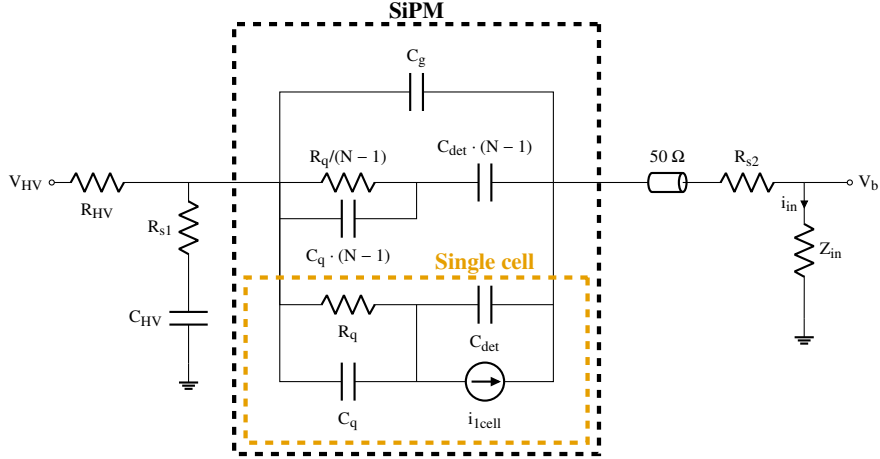
\begin{figure}[hbtp]
    \centering

    \begin{circuitikz}[american,scale=0.6 ,transform shape,]

        \draw (0,0) coordinate(start);
        \draw (start)  to[R=$R_q/(N-1)$] ++(3,0) coordinate(N-1cells) to[C=$C_{det}\cdot (N-1)$] ++(3,0);
        \draw (N-1cells) -- ++(0,-0.8
        ) to[C=$C_q\cdot(N-1)$] ++ (-3,0);
        \draw (start) -- ++(0,-2.7) to[R,l_=$R_q$] ++(3,0) coordinate(middlecell) to[C,l_=$C_{det}$] ++(3,0);
        \draw (start) -- ++(0,-4.4) to[C,l_=$C_q$] ++(3,0) to[I,l_=$i_{1cell}$] ++(3,0) -- ++(0,4.4);
        \draw (middlecell) -- ++(0,-1.7);
        \draw (start) -- ++(0,2) to[C=$C_g$] ++(6,0) -- ++(0,-2) coordinate(Vamp);
    
        \draw (Vamp) -- ++(2,0) to[transmission line, l^=$50\ \Omega$] ++(1,0) to[R=$R_{s2}$] ++(2.5,0) coordinate(Vb) to[R=$Z_{in}$,i>_=$i_{in}$] ++(0,-3) node[ground]{};
        \draw (Vb) -- ++(1,0) node[ocirc,label=right:$V_{b}$]{};
    
        \draw (start) ++ (-2,0) to[R, l_=$R_{s1}$] ++(0,-2) to[C,l_=$C_{HV}$] ++(0,-2) node[ground]{};
        \draw (start) -- ++(-3,0) to[R=$R_{HV}$] ++(-2,0) node[ocirc,label=left:$V_{HV}$]{};
    
        \draw[dashed,ultra thick, black] (7,3.5) rectangle (-1,-6);
        \node at (3,3.9) {\textcolor{black}{\Large{{$\mathbf{Z_{SiPM}}$}}}};

        \draw[dashed,ultra thick, darkorange] (6.75,-2.1) rectangle (-0.75,-5.75);
        \node[rotate=0] at (4.5,-1.8) {\textcolor{darkorange}{\Large{\textbf{Single cell}}}};
    
    \end{circuitikz}

    \caption{Focus on the SiPM structure, with N the number of SiPM cells, $R_q$ the quenching resistor, $C_d$ the single SiPM cell capacitor, $i_{1cell}$ the current signal of one fired cell and $C_g$ the grid parasitic capacitance. Outside of the SiPM structure, we have $R_{HV}$ and $C_{HV}$ which are used for the SiPM bias voltage filtering. $R_{s1}$ and $R_{s2}$ can be used for line termination or amplifier compensation. $Z_{in}$ is the amplifier input impedance and $i_{in}$ is the current flowing into the amplifying system.}
    \label{fig:SiPM_impedance}
\end{figure}

\noindent
For the latter, a voltage step is typically applied to $C_t$ to inject a high-frequency current impulse into the amplifier, enabling both characterization and system gain calibration. 
The passive elements $C_t$ and $R_{t}=50\ \Omega$ are both laid out in the closest proximity to the amplifier input to avoid reflections in the signal, and to minimize parasitic inductances. 
In the case of a SiPM input, a resistor $R_{HV}=10\ k\Omega$ is placed in series with the supply, while a capacitor $C_{HV}=100\ nF$ is connected to ground, in addition to $R_{s1}$, to filter the high-voltage SiPM power supply $V_{HV}$.
The resistors $R_{s1}$ and $R_{s2}$ also have two other purposes: they are used as termination for the SiPM signals and for the amplifier loop-gain compensation.
To reduce front-end effects that alter the SiPM pulse shape, $R_{s1}$ and $R_{s2}$ are replaced with short-circuits whenever feasible.
To quantify the impact of these resistors on the SiPM signal frequency spectrum, the structure of the SiPM pulse shape is analyzed in greater detail based on the model of a single fired cell shown in figure \ref{fig:SiPM_impedance} \cite{CorsiSiPM}.
The firing of a cell can be modeled by a $\delta(t)$-like current impulse that, in the Laplace transform domain, results $i_{1cell}(s)=Q_0$.
When a single cell fires, the current signal $i_{in}$ reaching the amplifier input is well approximated by:

\begin{equation}
    i_{in}(s)\simeq Q_0\frac{1+sC_qR_q}{\left[1+sC_qR_q\frac{(Z_{in}+R_s)}{(Z_{in}+R_s)+R_q/(N+1)}\right]\left[ 1+s(C_{det}(N+1)(Z_{in}+R_s)+(C_{det}+C_q)R_q)\right]}
    \label{eq:i_in_sipm_rs}
\end{equation}
\noindent
where $Z_{in}$ is the amplifier input impedance, $R_s=R_{s1}+R_{s2}$ and the bias filtering capacitor $C_{HV}$ is assumed a short circuit.
Equation (\ref{eq:i_in_sipm_rs}) exhibits two distinct negative real poles, responsible for the SiPM signal roll-off in the time domain.
Both poles depend on $R_s$. 
An increase in $R_s$, leads to a corresponding increase in the SiPM signal time constants.
For this reason, $R_s$ should be kept as small as possible.
Following the same reasoning, a higher amplifier input impedance $Z_{in}$ should also be avoided.
However, in cases where $C_g$ is sufficiently large or $R_q/N$ approaches a short circuit, the inclusion of a series resistor $R_s$ at the SiPM input becomes necessary to ensure amplifier stability (see section \ref{sec:stability_amp}).
In this paper, $R_{s1}$ and $R_{s2}$ are replaced with short circuits, as neither situation is relevant to the system under study.

As shown in figure \ref{fig:open_loop_and_closed_loop_configurations}, the collector current of the HBT is provided by the positive power supply $V_{POS}$ through the resistor $R_c$.   
The $V_{POS}$ power supply voltage can be changed to adjust the transistor transconductance ($g_m$) and tune the HBT gain; this is crucial for the circuit cryogenic/warm operation, as is explained later in section \ref{sec:hotvscold_operation}. 

The voltage signal at the HBT collector node ($V_c$) is amplified by a CFOA, which offers high output-current capability and a wide gain-independent bandwidth, provided the feedback resistor is appropriately selected; further details on the CFOA are given in appendix \ref{app:A}.
In principle, a CFOA can be used on its own to read out SiPMs with good performance \cite{CFOA_articolo}.
However, it exhibits worse signal-to-noise performance than by using the two-stage approach presented here. 
The CFOA is operated in a non-inverting configuration. 
At high frequencies, the closed-loop gain is given by: $\frac{V_{out}}{V_c} = 1 + \frac{R_2}{R_1}$.
At low frequencies, the gain reduces to unity due to the impedance of the $C_1$ capacitor. 
Consequently, the DC operating point is characterized by $V_{out} \simeq V_c \simeq V_{be}$, where $V_{be}$ is the base-emitter voltage of the HBT. 
The $V_{be}$ voltage of the HBT transistor should be approximately 0.7 V at room temperature and 1.0 V at 80 K. 
Associated with the other circuit elements, $V_{be}$, $V_{out}$, and $V_c$ actually exhibit slightly different voltages due to the HBT base current and the CFOA inverting-input bias current, both of which are of the order of $1-10\ \mu A$, depending on temperature.
In particular, $V_c\simeq1\ V$ at $300\ K$ because of these currents and large feedback resistors.
The CFOA is supplied with a negative voltage $V_{EE}$ and a positive voltage $V_{CC}$. 

The feedback loop of the two-stage amplifier is closed through $Z_f$, which comprises the resistance $R_f$ in parallel with its parasitic capacitance $C_{p_{Rf}}$. 
Amplifier stability is obtained via the compensation network $C_b$ and $R_b$ connected at the HBT base. 
These elements are in parallel with $C_{pb}$, a term that aggregates all parasitic capacitances at the amplifier input, as detailed in section \ref{sec:parasitics}.

Regarding the passive components attached to the HBT collector node, the $R_d$ resistor allows tuning of the open-loop gain, whilst the parasitic capacitor $C_p$ plays a crucial role in determining the circuit's frequency behavior.
The capacitor $C_x$ can be added in parallel to $C_p$ ($C_c = C_x+C_p$) to provide a dominant pole at the collector node, shaping the frequency behavior.

Finally, the output node $V_{out}$ is AC coupled to the scope through $C_{out}=100\  nF$, whose value was chosen to filter out DC and unnecessary low frequencies. 
In addition to the capacitor, an output resistance $R_{out}=50\  \Omega$, is present and is connected via a transmission line to the oscilloscope ($V_{osc}$), where the signal is terminated with $R_{osc}=50\ \Omega$.

\subsection{Open-loop gain}
If we do not consider the $Z_f$ feedback loop, the circuit from the HBT input to the CFOA output is an open-loop amplifier with a single-ended input at $V_{be}$ and the output at $V_{out}$. 

Assuming capacitances $C_d$ and $C_1$ act as short circuits over the frequency range of interest, and neglecting the output impedance, which represents the output impedance of a negative-feedback operational amplifier, the open-loop gain $A_{AMP}(s)$ given by the product of the HBT and CFOA stage transfer functions can be written as:

\begin{equation}
    A_{AMP}(s)\simeq \color{darkorange}g_m(R_d   \parallel   R_c) \frac{1}{1+sC_c(R_d  \parallel  R_c)}\color{black} \color{blue}\left[\left(1+\frac{R_2}{R_1}\right)\frac{1}{(1+sC_oR_2)(1+s\tau_{HF})} \right]\ \ \color{black}     
    \label{eq:open_loop_gain_ol_unified}
\end{equation}

\noindent
The open-loop DC gain is:
\begin{equation}
    A_{{AMP}_0}\simeq g_m(R_d  \parallel  R_c) \left(1+\frac{R_2}{R_1}\right)
\end{equation}

\noindent
where the HBT and CFOA transfer functions are respectively highlighted in orange and blue.

The CFOA transfer function shows two poles, as discussed in appendix \ref{app:A}.
The time constant $C_oR_2$ is determined by the local CFOA feedback, where $C_o$ is an internal CFOA capacitor that depends both on the type of CFOA and temperature, whereas $\tau_{HF}$ represents the CFOA second pole.
$A_{{AMP}_0}$ is the low-frequency open-loop gain.

\subsection{Loop gain}
\label{sec:loop_gain_and_feedback_stability}

The open-loop amplifier is integrated in a closed-loop system, as shown in figure \ref{fig:open_loop_and_closed_loop_configurations}. 
The feedback loop is closed through impedance $Z_f$, while loop gain compensation is provided by the network $Z_b = R_b + 1/sC_b$.
Additional compensation can be provided by $R_s = R_{s1}+R_{s2}$ if the SiPM impedance is almost purely capacitive.

To derive the full expression of the loop-gain, we make some assumptions on the passive components in the feedback loop.
We define the input impedance $Z_i$ as the parallel combination of the parasitic capacitance $C_{pb}$ with the branch containing the SiPM impedance $Z_{SiPM}$ and series resistance $R_s$: $Z_i = (Z_{SiPM} + R_s) \parallel \left(1/sC_{pb}\right)$.
$Z_i$ is in parallel with $Z_b$, hence we have $Z_b'=Z_b  \parallel  Z_i$. 
Owing to their usual pF or sub-pF magnitude, the SiPM parasitic capacitances $C_g$ and $Cq\cdot N$ have a marginal effect on stability and are thus neglected (see figure \ref{fig:SiPM_impedance}). 
This yields a simplified SiPM impedance $Z_{SiPM}\simeq\frac{1}{sNC_{det}}+R_q/N$, hence $Z_i\simeq (R_s+\frac{1}{sNC_{det}}+R_q/N) \parallel (1/sC_{pb})$.

Regarding the feedback path, the $R_f$ resistor is in parallel with the parasitic capacitor $C_{p_{Rf}}$, forming $Z_f=\frac{R_f}{1+sC_{p_{Rf}}R_f}$.
We also assume that $C_b$ is a short-circuit at the frequencies of interest, meaning that $Z_b\simeq R_b$.
Finally, we say that $NC_{det}\gg C_{p_{Rf}}$ and $NC_{det}\gg C_{pb}$, which is generally true since $NC_{det}$ is usually of the order of tens or hundreds of pF, whereas the others do not exceed a few pF.
If we call $R_{qN}=R_q/N$, $C_{detN}=NC_{det}$ and by using equation (\ref{eq:open_loop_gain_ol_unified}), we obtain the loop gain:

\begin{equation}
    \begin{aligned}
        T_{AMP}(s)&= -A_{AMP}(s)\cdot\frac{Z_b'}{Z_b'+Z_f} \simeq\\[2ex]
        &\simeq -g_m(R_c \parallel R_d)
        \left(1+\frac{R_2}{R_1}\right)\frac{1}{1+sC_c(R_c \parallel R_d)}\frac{1}{(1+sC_oR_2)(1+s\tau_{HF})} \cdot \\[2ex]
        &  \cdot \frac{R_b}{R_b+R_f}\color{bluishgreen}\frac{1+sC_{detN}(R_s+R_{qN})}{1+sC_{detN}[R_{qN}+R_s+(R_b \parallel Z_f)]} \color{reddishpurple}\frac{1+sC_{p_{Rf}}R_f}{1+s(C_{p_{Rf}}+C_{pb})[(R_b \parallel Z_f) \parallel (R_{qN}+R_s)]}
        \label{eq:loop_gain_pure}
    \end{aligned}
\end{equation}

As we can observe in equation (\ref{eq:loop_gain_pure}), the SiPM contributes with an additional pole to the loop-gain transfer function, the bluish green term. 
This pole is compensated by a zero, whose frequency can be adjusted with $R_s$. 

For a large-area SiPM device, $R_{qN}$ and $C_{detN}$ present relatively low impedances. 
In this case, the inclusion of $R_s$ suppresses ringing in the amplifier response by properly actuating the pole-zero.

The parasitic capacitances $C_{pb}$ and $C_{p_{Rf}}$ also add a zero-pole compensation (pink term), which, as is discussed in section \ref{sec:parasitics}, is negligible when evaluating the amplifier’s stability to a first approximation.

Assuming the SiPM zero is properly set to compensate the SiPM pole (green term), and the parasitic term is negligible because it lies at high frequency (pink term), equation (\ref{eq:loop_gain_pure}) simplifies to:
\begin{equation}
    \begin{aligned}
        T_{AMP}(s)&\simeq-g_m(R_c \parallel R_d)\left(1+\frac{R_2}{R_1}\right)\frac{1}{1+sC_c(R_c \parallel R_d)}\frac{1}{(1+sC_oR_2)(1+s\tau_{HF})} \cdot\\[2ex]
        &\qquad\qquad\qquad\qquad\qquad\qquad\qquad\qquad\qquad\qquad \cdot\frac{[R_b \parallel (R_s+R_{qN})]}{R_f+[R_b \parallel (R_s+R_{qN})]}=\\
        &=- \frac{T_{AMP_0}}{(1+sC_c(R_c \parallel R_d))(1+sC_oR_2)(1+s\tau_{HF})}
        \label{eq:loop_gain_unified_no_simplifications}
    \end{aligned}
\end{equation}
where $Z_f \simeq R_f$ ($C_{p_{R_f}}$ is neglected), $T_{AMP_0}=A_{AMP_0}\cdot\beta$, with $A_{AMP_0}=g_m(R_c \parallel R_d)\left(1+\frac{R_2}{R_1}\right)$ and $\beta = \frac{[R_b \parallel (R_s+R_{qN})]}{R_f+[R_b \parallel (R_s+R_{qN})]}$.
The $\beta$ term corresponds to the high-frequency value of the bluish-green term in equation (\ref{eq:loop_gain_pure}) multiplied by $\frac{R_b}{R_b+R_f}$.

\subsection{Feedback stability}
\label{sec:stability_amp}

To derive an expression for stability, we approximate the loop gain as a two-pole expression. 
As is discussed in sections \ref{sec:stab_cond_ODP} and \ref{sec:stab_cond_TDP}, this is a valid approximation for our specific case. 
Using equation (\ref{eq:loop_gain_unified_no_simplifications}) and neglecting $\tau_{HF}$ in the first approximation, the loop gain can be expressed as:

\begin{equation}
    \begin{aligned}
        T_{AMP}(s)\simeq - \frac{T_{AMP_0}}{(1+s\tau_{1_{LG}})(1+s\tau_{2_{LG}})}
        \label{eq:loop_gain_unified_due_poli}
    \end{aligned}
\end{equation}
\noindent
where $\tau_{1_{LG}}>\tau_{2_{LG}}$.
The closed-loop gain can be derived from the loop gain and, neglecting the direct transmission term, is given by \cite{articolo_reti_reazionate}:
\begin{equation}
\begin{aligned}
    G_{CL}(s)=G_\infty\ \cdot \frac{-T_{AMP}(s)}{1-T_{AMP}(s)}&= -Z_f \cdot \frac{T_{AMP_0}}{s^2\tau_{1_{LG}}\tau_{2_{LG}}+s(\tau_{1_{LG}}+\tau_{2_{LG}})+T_{AMP_0}+1}=\\[2ex]
    &\simeq -Z_f \cdot \frac{1}{(1+s\tau_{1_{CL}})(1+s\tau_{2_{CL}})}
    \label{eq:generic_closed_loop}
\end{aligned}
\end{equation}
\noindent
where $G_\infty=-Z_f$ is the closed-loop gain when the loop gain is considered infinite, $\frac{T_{AMP_0}}{T_{AMP_0}+1}\rightarrow1$ and $\tau_{1_{CL}}>\tau_{2_{CL}}$.

As can be seen from equation (\ref{eq:generic_closed_loop}), the condition $1 - T_{AMP}(s) = 0$ must be avoided, as it leads to instability. 
To ensure this, the loop gain phase must be less than $|-180^\circ|$ when $|T_{AMP}(s)| = 1$. 
To avoid ringing in the output signal, a minimum phase margin (PM) of $45^\circ$ is typically required, which corresponds to a phase of $-135^\circ$ at the unity-gain frequency. 
Under this condition, the closed-loop gain poles are generally complex conjugates, and the time-domain response exhibits ringing, with an overshoot of more than $20\%$.
To achieve a flat, ring-free response and to reduce sensitivity to additional parasitic effects, a more conservative design is adopted, unconventionally targeting real and distinct poles to achieve stability. 

The closed-loop gain in equation (\ref{eq:generic_closed_loop}) has two poles $| -1/\tau_{1_{CL}} |$ and $|-1/\tau_{2_{CL}}|$, and their time constants can be expressed as:

\begin{equation}
    \tau_{1,2_{CL}} = \frac{\tau_{1_{LG}}+\tau_{2_{LG}}}{2T_0}\left( 1 \pm \sqrt{1-\frac{4T_0\tau_{1_{LG}}\tau_{2_{LG}}}{(\tau_{1_{LG}}+\tau_{2_{LG}})^2}} \right)
    \label{eq:poles_closed_loop}
\end{equation}
\noindent
where the two poles are distinct and real only if the quadratic denominator has a non-negative discriminant. Hence:

\begin{equation}
    \Delta=1-4\frac{\tau_{1_{LG}}\tau_{2_{LG}}}{(\tau_{1_{LG}}+\tau_{2_{LG}})^2}(T_{AMP_0}+1)\geq 0\quad \longrightarrow \quad T_{AMP_0} \leq \frac{(\tau_{1_{LG}} - \tau_{2_{LG}})^2}{4\tau_{1_{LG}}\tau_{2_{LG}}}
    \label{eq:discriminant_closed-loop_gain}
\end{equation}

Since $T_{AMP_0} = A_{AMP_0}\cdot\beta$, with $\beta = \frac{[R_b \parallel (R_s + R_{qN})]}{R_f + [R_b \parallel (R_s + R_{qN})]}$, an upper bound on $R_b$ can be derived by choosing $R_f \gg R_b$. We also consider that $[R_b \parallel (R_s + R_{qN})]\simeq R_b$, meaning that we assume that $R_{qN}\gg R_b$ and $R_s$ is not needed to improve stability.
Under this approximation, $\beta \simeq R_b / R_f$, and the upper bound follows from equation (\ref{eq:discriminant_closed-loop_gain}):

\begin{equation}
    R_b\leq R_f\cdot \frac{(\tau_{1_{LG}} - \tau_{2_{LG}})^2}{4A_{AMP_0}\tau_{1_{LG}}\tau_{2_{LG}}}
    \label{eq:rb_upper_limit}
\end{equation}

\noindent
Equation (\ref{eq:rb_upper_limit}) does not take into account the presence of third poles and their impact on the closed-loop phase.

Finally, to justify the approximation $Z_b\simeq R_b$ that we first used to derive equation (\ref{eq:loop_gain_pure}), we need to make some simplifications.
We also need to consider that the impedance $Z_b$ in the loop gain calculation (see equation (\ref{eq:loop_gain_pure})) is in parallel with $Z_i$, which, however, can be approximated as $Z_b$ when $Z_i$ is much larger than $Z_b$. 
This situation occurs either when no SiPM is attached, or when $R_q/N \ (+R_s) \gg R_b$, neglecting $N C_q$, and $C_g$ is very small, as shown in figure \ref{fig:SiPM_impedance}.

These conditions are typically satisfied for small single SiPMs with areas of $1$–$9\ mm^2$, which have a low number of pixels. 
As a result, $R_q/N$ is generally high and $C_g$ is negligible. 
When the SiPM impedance is relatively low, it can negatively affect amplifier stability. 
To mitigate this effect, it is generally preferable to place $R_s$ in series, as shown in equation (\ref{eq:loop_gain_pure}).

For this dissertation, however, $R_s$ is set to $0\ \Omega$, since the SiPM impedance is assumed to be sufficiently high.
This assumption is typically valid and is supported by the measurements presented in section \ref{sec:sipm_measurement}.

Then, $R_b$ can be selected such that, in series with $C_b$ (see figure \ref{fig:open_loop_and_closed_loop_configurations}), the high-pass frequency $|-1/(C_b R_b)|$ is approximately an order of magnitude smaller than the angular frequency $\omega_{T1}$ at which $|T(s)| = 1$. 
Under this condition, the impedance can be approximated as $Z_b \simeq R_b$.

To determine $\omega_{T1}$, $T_{AMP}(s)$ is approximated as a single-pole transfer function up to the frequency where its magnitude reaches unity, with $|-1/\tau_{1_{LG}}|$ representing the dominant pole. 
This approximation is appropriate when the primary design objective is to ensure stability and the two loop gain poles are sufficiently well separated.
If, in addition, $\omega_{T1}\tau_{1_{LG}} \gg 1$, then:

\begin{equation}
    \omega_{T1}\simeq T_{AMP_0}/\tau_{1_{LG}}\simeq \frac{A_{{AMP}_0} R_b}{\tau_{1_{LG}} R_f}
    \label{eq:omega_t1}
\end{equation}
\noindent
where $\beta\simeq R_b/R_f$. 
Therefore, we can extract a lower bound for $R_b$, which is: 

\begin{equation} 
    R_b\geq \sqrt{10 \cdot\frac{\tau_{1_{LG}}}{A_{{AMP}_0}}\cdot \frac{R_f}{C_b}}
    \label{eq:lower_rb_limit}
\end{equation}

\noindent    
where we can use $C_b$ to adjust this lower limit, 
Combining the results shown in equations (\ref{eq:rb_upper_limit}) and (\ref{eq:lower_rb_limit}), the stability constraints on $R_b$ can be expressed as:

\begin{equation}
    \sqrt{10 \cdot\frac{\tau_{1_{LG}}}{A_{{AMP}_0}}\cdot \frac{R_f}{C_b}}\leq R_b\leq R_f\cdot \frac{(\tau_{1_{LG}} - \tau_{2_{LG}})^2}{4A_{AMP_0}\tau_{1_{LG}}\tau_{2_{LG}}}
    \label{eq:rb_limits}
\end{equation}

\subsection{Closed-loop transfer function}

Using equation (\ref{eq:generic_closed_loop}), the transfer function of the closed-loop amplifier can be expressed as:

\begin{equation}
    \frac{V_{out}(s)}{i_{in}(s)} = G_\infty \cdot \frac{-T_{AMP}(s)}{1 - T_{AMP}(s)} = -Z_f \cdot \frac{-T_{AMP}(s)}{1 - T_{AMP}(s)} 
    \label{eq:transfer_function_general}
\end{equation}

A similar transfer function can be found for the signal gain between the test input and the oscilloscope ($V_{osc}$, see figure \ref{fig:open_loop_and_closed_loop_configurations}). 
If the input is the one coming from the pulse generator, a step voltage with amplitude $V_t$ and a rise time constant $\tau_{gen}$ charges the test capacitor $C_t$, generating a current impulse $i_{in}(t)\simeq V_tC_t/\tau_{gen}\cdot e^{-t/{\tau_{gen}}}$ which flows into the amplifier. 

The response may be smoothed by the two low-pass filters at $\tau_{cable}=\frac{1}{2\pi\cdot BW_{cable}}$ that account for the non-ideal propagation of the test and output signals through the LEMO cables (assumed to be of the same length). 
Also, the oscilloscope adds a low-pass filter at $\tau_{osc}=\frac{1}{2\pi\cdot BW_{osc}}$.
Finally, we consider the high-pass filtering and attenuation by a factor two due to $C_{out}$ and the line termination which is accomplished by two $R_{out}$ resistors ($R_{osc}=R_{out}$).
The output signal is then expected to be:

\begin{equation}
    \begin{split}
        V_{osc}(s) &= \frac{V_{out}(s)}{i_{in}(s)}\cdot \left[\frac{1}{(1+s\tau_{cable})^2}\frac{1}{1+s\tau_{osc}}\frac{sC_{out}R_{out}}{1+2sC_{out}R_{out}}\right] \cdot i_{in}(s)\\[2ex] 
        & \simeq -Z_f \cdot \frac{-T_{AMP}(s)}{1-T_{AMP}(s)} \cdot \left[\frac{1}{(1+s\tau_{cable})^2}\frac{1}{1+s\tau_{osc}}\frac{sC_{out}R_{out}}{1+2sC_{out}R_{out}}\right] \cdot \frac{ V_tC_t}{1+s\tau_{gen}}
        \label{eq:osc_transf_function}
    \end{split}
\end{equation}

If the input signal comes from the SiPM, the transfer function is the same as in equation (\ref{eq:osc_transf_function}), but without the contribution of one LEMO cable, the one connected to the pulse generator. 
In addition, the transfer function of the SiPM, shown in equation (\ref{eq:i_in_sipm_rs}) replaces that of the pulser.

In this work, two configurations of the same topology are analyzed: the \enquote{OA Dominant Pole} (ODP, OA stands for Operational Amplifier) and the \enquote{Transistor Dominant Pole} (TDP).
The schematic, shown in figure \ref{fig:open_loop_and_closed_loop_configurations}, is the same for the two configurations. 
The main difference between them lies in the operating mode of the (CFOA): in the ODP configuration, a high-value local feedback resistor $R_2$ is used, whereas in the TDP configuration a low-value $R_2$ is employed, shifting the dominant pole of the CFOA to a higher frequency.
Moreover, each configuration employs different passive components and CFOA models in order to optimize speed.
The two amplifiers are very similar, so we can make a general explanation of how the circuit works, and later elaborate on the two separately.

\subsection{TDP and ODP shared considerations}

Given the evaluations presented in section \ref{sec:stability_amp} on loop stability, the two amplifier configurations were designed by accurately selecting passive and active elements. 
Most passive/active elements are shared between the two distinct configurations.
The passive components are selected to ensure the amplifier operates reliably at cryogenic temperatures: capacitors with C0G/NP0 dielectric and thin film resistors. 

The HBT used is the Infineon BFP640, which is a NPN heterojunction bipolar transistor with a high transition frequency $f_T$ of several GHz which depends on the collector current \cite{infineon_bfp640_datasheet}. 
This transistor was already used in a previous design where it was operated down to cryogenic temperatures \cite{Si-GeTransistors} \cite{DUNE_amplifier}. 
At ambient temperature the circuit works at $V_{POS}=5\ V$, whereas at lower temperatures $V_{POS}$ needs to be reduced, as is explained in section \ref{sec:hotvscold_operation}. 
With $V_c\simeq 1\  V$ at ambient temperature and $R_c=1- 4\ k\Omega$, the current flowing in the collector is of the order of a few mA, corresponding to a transition frequency of about 10 GHz \cite{infineon_bfp640_datasheet}.

For both configurations the amplifier feedback resistor is $R_f=7.5\ k\Omega$, so that the two amplifier configurations can be compared with the same closed-loop gain.

Regarding the CFOA, two different devices are used: LMH6702 \cite{lmh6702} and LMH6703 \cite{lmh6703}, both supplied with $V_{CC}=5\ V$ and $V_{EE}=-5\ V$. 
Their performance is slightly different, but they are both CFOAs with high unity gain bandwidth ($>1\ GHz$), high slew rate ($>3000 \  V/\mu s$) and low noise ($<2.5\ nV/\sqrt{Hz}$).
However, even if very similar, one is more suitable for certain applications and vice versa. 

The constants differ across the two configurations, reflecting variations in both the local feedback loop (namely $R_2$) and the specific CFOA utilized.
A comparison of the technical specifications for these two CFOAs \cite{lmh6702} \cite{lmh6703} reveals that the LMH6703 is more aggressively optimized for high-bandwidth performance at low closed-loop gains ($4-10\ V/V$).
Further analysis of the device parameters suggests that the output capacitance $C_o$ (see equation (\ref{eq:closed_loop_CFOA})) of the LMH6702 is larger than that of the LMH6703.
This is consistent with the observation that both devices exhibit similar phase margins at a gain of ($2\ V/V$), indicating that their dominant poles occur at approximately the same frequency (see equation (\ref{eq:T_CFOA_new})). 
Given that $R_{2,LMH6702} < R_{2,LMH6703}$, it follows that $C_{o,LMH6702} > C_{o,LMH6703}$. 
Consequently, the higher output capacitance of the LMH6702 allows it to maintain a larger phase margin for higher values of $R_2$, making it better suited for applications that require higher feedback resistance.

\subsection{ODP configuration analysis}
This subsection analyzes the ODP configuration, where the dominant pole is placed at the CFOA stage. We first derive the stability conditions for the compensation network, followed by the resulting closed-loop transfer function. Finally, we detail the choice of components and the specific parameters used to implement this design.

\subsubsection{Stability}
\label{sec:stab_cond_ODP}
In this configuration, we set $R_2$ to tens of $k\Omega$, a sufficiently high value to limit the CFOA’s bandwidth and place the dominant pole of the loop gain at $|- 1/(C_oR_2)|$. 
The CFOA gain, adjustable via $R_1$, is maximized to achieve the highest possible amplifier open-loop gain.
The capacitor $C_x$ is omitted ($C_c = C_p$) so that the collector node does not limit the overall loop-gain bandwidth. 
Additionally, to push the collector pole to even higher frequencies, we take $R_d \ll R_c$.

To determine the stability condition (equation (\ref{eq:rb_limits})), we continue analyzing $T_{AMP}(s)$ under simplifying assumptions. 
As discussed already in section \ref{sec:stability_amp}, we use that $\beta\simeq R_b/R_f$.
In addition, the high-frequency term $\tau_{HF}$ has little effect on $T_{AMP}(s)$; as shown in section \ref{sec:components_choice_ODP}, it nearly coincides with the pole defined by $|-1/(C_p R_d)|$, allowing the two high-frequency poles to be treated as coincident.
We combine these two poles into a single equivalent pole, for which the -3 dB attenuation occurs at a lower frequency: $\left|-\sqrt{\sqrt{2}-1}/(C_p R_d)\right|$.
The value of this equivalent pole is derived by finding the frequency at which the transfer function with two coincident poles drops by 3 dB: $\frac{1}{\sqrt{(1+\omega^2(C_pR_d)^2)^2}}=-3\ dB=1/\sqrt{2}$.
This shift is justified by the fact that, when we derive the $R_b$ upper bound in equation (\ref{eq:rb_limits}), we are primarily interested in how the modulus of $T_{AMP}(s)$ evolves between the first and second poles. 
We are less concerned with the accuracy of the behavior beyond the second pole. 

Equation (\ref{eq:loop_gain_unified_no_simplifications}) is approximated as:
\begin{equation}
    T_{ODP}(s)\simeq -g_mR_d \left(1+\frac{R_2}{R_1}\right) \frac{1}{1+sC_oR_2 }\frac{1}{1+s\frac{C_pR_d}{\sqrt{\sqrt{2}-1}}} \cdot \frac{R_b}{R_f}
    \label{eq:loop_gain_open_loop_configuration}
\end{equation}

To determine the bounds on $R_b$, we use equation (\ref{eq:rb_limits}) with $A_{AMP_0}=g_m R_d\left(1+\frac{R_2}{R_1}\right)$, $\tau_{1_{LG}} = C_oR_2$, and $\tau_{2_{LG}} = \frac{C_p R_d}{\sqrt{\sqrt{2}-1}}$. 
Putting together the lower and upper bounds, we obtain:

\begin{equation}
     \sqrt{10 \cdot\frac{C_oR_2}{g_mR_d(1+R_2/R_1)}\cdot \frac{R_f}{C_b}}\le {R_b} \leq R_f\cdot \frac{\left(C_oR_2 - \frac{C_pR_d}{\sqrt{\sqrt{2}-1}}\right)^2 }{4g_m R_d (1+R_2/R_1) C_oR_2 \frac{C_pR_d}{\sqrt{\sqrt{2}-1}}}
    \label{eq:stability_rb_open_loop}
\end{equation}

\noindent
If the upper limit is too low and the condition cannot be satisfied, we need to act on the open-loop gain. 
We can either lower the low-frequency gain ($A_{{AMP}_0}$), which means lowering $g_m$ or $(1+R_2/R_1)$, or increase the second dominant pole frequency (lower $C_p R_d$, specifically by reducing $R_d$, since $C_p$ cannot be further reduced). 
If this is still not enough, then $R_f$ must be increased. 
The trade-off of implementing these stability-enhancing modifications is a reduction in the amplifier bandwidth.
We can also act on the lower limit of these range of values by changing $C_b$.

\subsubsection{Transfer function}

The ODP configuration transfer function can be found using equation (\ref{eq:transfer_function_general}):
\begin{equation}
    \frac{V_{out}(s)}{i_{in}(s)} \simeq -Z_f \cdot \frac{1}{1 + s \tau_{1_{ODP}}} \cdot \frac{1}{1 + s \tau_{2_{ODP}}}\\
\end{equation}
The pole time constants $\tau_{1,2_{ODP}}$ can be obtained from equation (\ref{eq:poles_closed_loop}) by substituting the loop-gain parameters expressed in equation (\ref{eq:loop_gain_open_loop_configuration}).

As shown in section \ref{sec:components_choice_ODP}, these two time constants are real and distinct.
Therefore, the feedback amplifier can be regarded as a two-pole system. 
The output response to a current impulse $i_{in}=Q_0\delta(t)$ from the test capacitor $C_t$ ($\tau_{gen}$ is assumed negligible) is defined as:
\[
v_{out}=Q_0\cdot\left(\frac{e^{-t/\tau_{1_{ODP}}}}{\tau_{1_{ODP}}-\tau_{2_{ODP}}}-\frac{e^{-t/\tau_{2_{ODP}}}}{\tau_{1_{ODP}}-\tau_{2_{ODP}}}\right)
\]

The simulated and measured output signals for this configuration are shown in section \ref{sec:output_signal_simulation}.

\subsubsection{Components choice}
\label{sec:components_choice_ODP}

As previously mentioned when deriving equation (\ref{eq:loop_gain_open_loop_configuration}), in this configuration the CFOA needs to work with high values of $R_2$, about tens of $k\Omega$.
For this application, the LMH6702 is the most suitable option of the two.
This CFOA has the second pole $|-1/\tau_{HF}|$ at about $1.7\ GHz$.
The use of high-value resistors in the feedback path is not a common practice for the operation of this CFOA.
For this reason the device was studied with an ad-hoc test bench, which is shown in appendix \ref{app:B}. 
The used resistors are $R_2=9.1\ k\Omega$ and $R_1=20\ \Omega$. 
The measurements reported in the test described in appendix \ref{app:B} indicate that the gain is slightly lower than expected, which can be accounted for by assuming a slightly larger resistance, namely $R_1 = 22\ \Omega$. 
The closed-loop bandwidth of the CFOA is $|-1/(2\pi C_oR_2)| \simeq 4.5\ MHz$, which implies $C_oR_2 \simeq 35\ ns$. 
This measured value is consistent with the approximate value one would extract from the CFOA manual \cite{lmh6702}.

There are other passive elements in the amplifier network. The capacitor $C_1=150\ nF$, together with $R_1$, determines the frequency at which the CFOA works at full closed-loop gain, which is $\sim 50\  kHz$.
The HBT current is provided by $R_c = 1\ k\Omega$. At ambient temperature ($T=300\ K$) $V_c\simeq 1\ V$ and $V_{POS}=5\ V$, resulting in $g_m=I_c/V_t=\frac{V_{POS}-V_c}{R_c}\cdot\frac{q}{KT}\simeq 154\ mA/V$.
Then, $C_p=1.95\ pF$ --- measured in appendix \ref{app:C} --- and $R_d=51\  \Omega$ produce a time constant $C_pR_d\simeq 0.1 \ ns$ ($\sim 1.6\ GHz$).
This pole at $1.6\ GHz$ is very close to the second pole of the CFOA ($1.7\ GHz$).
Then, as previously stated in section \ref{sec:stab_cond_ODP}, we are treating the two poles as coincident, with the -3 dB attenuation being shifted at a larger time constant: $C_pR_d\  / \sqrt{\sqrt{2}-1} \simeq 0.15\ ns$.

Finally, equation (\ref{eq:stability_rb_open_loop}) can be used to find a value for the compensating resistor $R_b$ (and $C_b$ consequently).
Using all values used for this configuration and equation (\ref{eq:stability_rb_open_loop}), the condition for stability becomes $R_b\lesssim 130\ \Omega$. 
The chosen value is $R_b=100\ \Omega$, which lies within the range of usable $R_b$ values. 
Paired with $C_b=330\ pF$, the selected $R_b$ stays above the lower limit of equation (\ref{eq:stability_rb_open_loop}) of $49\ \Omega$.

Using equation (\ref{eq:poles_closed_loop}) with this $R_b$ value and knowing that $\tau_{1_{LG}}=C_cR_c$, $\tau_{2_{LG}}=C_oR_2$, $A_{AMP_0} = g_m R_c \left(1 + R_2/R_1\right)$ and $\beta\simeq R_b/R_f$, the two pole time constants can be calculated, yielding: $\tau_{1_{ODP}}\simeq 0.59\ ns$, corresponding to $\sim 270 \ MHz$, and $\tau_{2_{ODP}}\simeq 0.21\ ns$, corresponding to $\sim 760 \ MHz$.

\subsection{TDP configuration}
Following the same analytical structure as the ODP configuration, this subsection evaluates the TDP design. We first establish the stability condition under these revised dynamics, then derive the corresponding transfer function, and conclude with the choice of components.

\subsubsection{Stability}

\label{sec:stab_cond_TDP}
In this alternative configuration, the CFOA is operated with a low $R_2$ resistor (hundreds of $\Omega$), causing the $|-1/(C_oR_2)|$ pole to shift to a higher frequency. 
$R_2$ is chosen close to the value suggested in the datasheet to maximize the bandwidth.
The collector node pole --- $|-1/(C_cR_c)|$ --- is made the dominant pole in the loop gain, while $R_d = \infty$ and $C_d = 0$, as they are not needed (see figure \ref{fig:open_loop_and_closed_loop_configurations}).
In this case, the $|-1/\tau_{HF}|$ pole is neglected, since it lies at a higher frequency than $|-1/(C_oR_2)|$. 
Also in this case, $\beta \simeq R_b/R_f$.

Equation (\ref{eq:loop_gain_unified_no_simplifications}), with all these simplifications, becomes:

\begin{equation}
    T_{TDP}(s)\simeq- g_mR_c \left(1+\frac{R_2}{R_1}\right)\frac{1}{1+sC_cR_c} \frac{1}{(1+sC_oR_2) }\cdot \frac{R_b}{R_f} 
    \label{eq:loop_gain_closed_loop_configuration}
\end{equation}

\noindent
Applying equation (\ref{eq:rb_limits}) and setting $\tau_{1_{LG}}=C_cR_c$, $\tau_{2_{LG}}=C_oR_2$, and $A_{AMP_0}=g_mR_c\left(1+\frac{R_2}{R_1}\right)$, we obtain the upper and lower limits for $R_b$:

\begin{equation}
    \sqrt{10 \cdot\frac{C_CR_c}{g_mR_c(1+R_2/R_1)}\cdot \frac{R_f}{C_b}} \le {R_b}\le R_f\cdot \frac{\left(C_cR_c - C_oR_2\right)^2}{4g_m R_c (1+R_2/R_1) C_cR_c C_oR_2}
    \label{eq:stability_rb_closed_loop}
\end{equation}

\noindent
If the range of $R_b$ is not large enough, we need to act on $A_{AMP}(s)$ and ultimately on $R_f$ to change the upper limit. If we want to change the former, one way is to diminish the low-frequency gain (lower $g_m$ or $(1+R_2/R_1)$) or increase the second pole frequency (lower $C_oR_2$). 
Also in this case, the lower limit of these range of values can be modified by changing $C_b$.

\subsubsection{Transfer function}

Applying the general model from equation (\ref{eq:transfer_function_general}) to this configuration yields the closed-loop transfer function:
\begin{equation}
    \frac{V_{out}(s)}{i_{in}(s)} \simeq -Z_f \cdot \frac{1}{1 + s \tau_{1_{TDP}}} \cdot \frac{1}{1 + s \tau_{2_{TDP}}}\\
\end{equation}
Here, the pole constant $\tau_{1,2_{TDP}}$ can be obtained from equation (\ref{eq:poles_closed_loop}) by substituting the loop-gain parameters shown in equation (\ref{eq:loop_gain_closed_loop_configuration}).

As we will see in section \ref{sec:closed_loop_design}, these two poles can be considered coincident.
The amplifier’s time-domain response to an ideal current pulse $i_{in}=Q_0\delta(t)$ from the test capacitor $C_t$ ($\tau_{gen}$ is considered negligible) is then characterized by:
\[
v_{out}=Q_0\cdot\frac{t}{\tau_{TDP}^2}e^{-t/\tau_{TDP}}
\]

Also for this configuration, the simulated and measured output signals are presented in section \ref{sec:output_signal_simulation}.

\subsubsection{Components choice}
\label{sec:closed_loop_design}
For this configuration, we select the LMH6703, as it offers the highest closed-loop bandwidth between the two chosen models.
We select $R_2 = 390\ \Omega$ and $R_1 = 91\ \Omega$ as a compromise between achieving high loop gain, wide bandwidth and adequate stability. 
These values are recommended by the manufacturer in the LMH6703 manual \cite{lmh6703}.
With these resistors, the CFOA closed-loop bandwidth is $\sim 700\ MHz$, which gives $C_oR_2\simeq0.23\ ns$ (see equation (\ref{eq:loop_gain_closed_loop_configuration})). 

Regarding $C_c$, the chosen value is $(2.2+1.69)\ pF=3.89\  pF$, where $C_p\simeq1.69\ pF$ is the extracted value for the parasitic capacitance on the collector node (see appendix \ref{app:C}) and $C_x=2.2\ pF$.
The combination of $R_c=2.4\ k\Omega$ and $C_c$ gives the dominant pole of the open-loop gain of the amplifier, resulting in a time constant $C_cR_c\simeq9.3\  ns$.
As previously mentioned, at ambient temperature ($T=300\ K$) $V_c\simeq 1\ V$ and $V_{POS}=5\ V$, meaning that $g_m=I_c/V_t=\frac{V_{POS}-V_c}{R_c}\cdot\frac{q}{KT}\simeq 64\ mA/V$.

The last passive elements to be determined are the compensating elements $R_b$ and $C_b$. 
Employing equation (\ref{eq:stability_rb_closed_loop}), the $R_b$ upper bound can be extracted: $R_b\lesssim 85 \  \Omega$. 
We select $R_b=100\ \Omega$ — as in the ODP configuration — in series with a capacitor $C_b=330\ pF$.
The chosen value of $R_b$ is higher than the upper bound computed above.
This choice is made to limit the noise, since $\overline{i^2_{in}}$ decreases as $R_b$ increases (see section \ref{sec:noise_equations}).
The capacitor $C_b$ is selected such that the resistor remains above the lower limit imposed by equation (\ref{eq:stability_rb_closed_loop}), which yields $32\ \Omega$, leaving some margin in case the SiPM impedance is not negligible and appears in parallel with $R_b$.

Because of this $R_b$ resistor, the two poles are complex conjugates and their magnitude can be computed using equation (\ref{eq:poles_closed_loop}), with $\tau_{1_{LG}}=C_cR_c$, $\tau_{2_{LG}}=C_oR_2$, $A_{AMP_0} = g_m R_c \left(1 + R_2/R_1\right)$ and $\beta\simeq R_b/R_f$. 
Since the chosen $R_b$ is very close to the computed upper limit, the imaginary part is minimal and we therefore neglect it and we consider them coincident real time constants: $\tau_{TDP}=\tau_{1_{TDP}}=\overline{\tau_{2_{TDP}}}\simeq \frac{\tau_{1_{LG}}+\tau_{2_{LG}}}{2 A_{AMP_0}\beta+1}\simeq 0.42\ ns$, corresponding to a cut-off frequency of $\sim 380 \ MHz$.
The assumption of neglecting the imaginary part is strengthened by the fact that in the time-domain, the simulated output signals present zero to negligible overshoot (see section \ref{sec:output_signal_simulation}).

Table \ref{tab:configurations_network_elements} shows a summary of all the components and characteristics of both amplifier configurations. 

\begin{table}[tbp]
    \centering
    \caption{Summary of all passive and active components used in the two amplifier configurations: ODP and TDP configurations.}
    \vspace{2mm}
    
    \begin{tabular}{c|c|c}
        \toprule
         \textbf{Network elements} & \textbf{TDP configuration} & \textbf{ODP configuration} \\
        \midrule
        \textbf{$\mathbf{HBT}$} & BFP640 & BFP640 \\
        \textbf{$\mathbf{R_c}$} & $2.4\ k\Omega$ & $1\ k\Omega$ \\
        \textbf{$\mathbf{R_d}$} & $\infty$ & $51\ \Omega$ \\
        \textbf{$\mathbf{C_d}$} & 0 & $150\  nF$ \\
        \textbf{$\mathbf{C_p}$} & $1.69\ pF$ & $1.95\ pF$ \\
        \textbf{$\mathbf{C_x}$} & $2.2\ pF$ & 0\\
        \textbf{$\mathbf{CFOA}$} & LMH6703 & LMH6702 \\
        \textbf{$\mathbf{R_1}$} & $91\ \Omega$ & $22\ \Omega$ \\
        \textbf{$\mathbf{C_1}$} & $150\  nF$ & $150\  nF$ \\
        \textbf{$\mathbf{R_2}$} & $390\ \Omega$ & $9.1\ k\Omega$ \\
        \textbf{$\mathbf{R_f}$} & $7.5\ k\Omega$ & $7.5\ k\Omega$ \\
        \textbf{$\mathbf{R_b}$} & $100\ \Omega$ & $100\ \Omega$ \\
        \textbf{$\mathbf{C_b}$} & $330\ pF$ & $330\ pF$ \\
        \textbf{$\mathbf{\tau_{1_{TDP}}}/\mathbf{\tau_{1_{ODP}}}$} & $0.42\ ns$ & $0.59\ ns$\\
        \textbf{$\mathbf{\tau_{2_{TDP}}}/\mathbf{\tau_{2_{ODP}}}$} & $0.42\ ns$ & $0.21\ ns$ \\
        \bottomrule
    \end{tabular}
    
    \label{tab:configurations_network_elements}
\end{table}

\subsection{Input impedance}
As already discussed in \ref{sec:circuit_description}, another important parameter of an amplifier for SiPM signals is the input impedance. 
In fact, if the amplifier input impedance is not low enough, the faster signal components may be lost, degrading the timing performance of the sensors (see equation (\ref{eq:i_in_sipm_rs})). 

The input impedance of the amplifier can be computed as follows:
\begin{equation}
    Z_{in}(s)=\frac{Z_{open}}{1-T_{AMP}(s)}
    \label{eq:full_input_imp}
\end{equation}

\noindent
where $Z_{open}=Z_f \parallel  Z_b$ and $T_{AMP}(s)$ is the amplifier loop gain of the chosen configuration. 
Note that this is the amplifier impedance without anything connected at its input.
When a SiPM is connected at the input, the impedance becomes $Z_{in}' = Z_{in}  \parallel  Z_{SiPM}$ if $R_s=0$ (see figure \ref{fig:open_loop_and_closed_loop_configurations}).

Since $R_f$ is very large at the frequencies of interest (amplifier bandwidth) if compared to $Z_b$ (see table \ref{tab:configurations_network_elements}), equation (\ref{eq:full_input_imp}) can be simplified, leading to the input impedance being:
\begin{equation}
    Z_{in}(s)\simeq\frac{Z_{b}}{1-T_{AMP}(s)}
    \label{eq:simple_input_imp}
\end{equation}
\noindent

The amplifier input impedance simulations are shown in section \ref{sec:input_imp_results}.

\subsection{Noise}
\label{sec:noise_equations}

Both configurations have three main noise contributions: $R_b$, $R_f$ and the HBT, series and parallel noise sources. 
The HBT can also contribute to noise with its base spreading resistance ($R_{bb}$), whose value is estimated to be up to a few tens of Ohms, depending on the collector current \cite{Dune_vecchio}. 
There are other contributions related to the CFOA and resistors at the HBT collector nodes, but these are generally negligible because they need to be divided by the HBT gain when referred to the amplifier input.

Therefore, the noise contributions at the output can be computed by assuming $g_m\rightarrow\infty$, which allows us to disregard all second order noise contributions.
This additional approximation renders the noise expression for the two configurations equivalent, since most of the differences arise from the open-loop gain topology.
The noise at the output is:

\begin{equation}
    \begin{split}
        \overline{v_{on}^2}=\left | \frac{-T_{AMP}(s)}{1-T_{AMP}(s)}\right |^2
        \Biggl\{\left| \frac{Z_b+Z_f}{Z_b} \right|^2 \cdot\left[4k_BTR_{bb}+\overline{e^2_{HBT}}\right] & +\left|\frac{Z_f}{Z_b}\right|^2\cdot4k_BTR_b +
        \\[1ex]
        &
        +R_f^2\cdot\overline{i_{HBT}^2} + 4k_BTR_f \Biggr\}
        \label{eq:noise}
    \end{split}
\end{equation}

\noindent
where $T_{AMP}(s)$ is the loop gain of either one of the two configurations, $\overline{e_{HBT}^2}=4k_BT\frac{1}{2g_m}$, $\overline{i_{HBT}^2}=2qI_b$, $k_B=1.38\cdot 10^{-23} \ J/K$ (Boltzmann constant), $q=1.6\cdot10^{-19}\  C$ (electron charge), $T$ is the amplifier operating temperature expressed in degrees Kelvin, $g_m=\frac{I_c}{V_T}$, $I_c=\frac{V_{POS}-V_c}{R_c}$ is the HBT collector current, $V_T=\frac{k_BT}{q}$, $I_b=\frac{I_c}{h_{FE}}$ is the HBT base current and $h_{FE}$ is the HBT current gain.
To evaluate the RMS noise at the amplifier output, equation (\ref{eq:noise}) must be integrated over the amplifier’s bandwidth.

The equivalent input current noise source can be calculated by dividing the output noise (equation (\ref{eq:noise})) with the amplifier closed-loop gain (equation (\ref{eq:loop_gain_unified_no_simplifications})): 

\begin{equation}
\begin{aligned}
    \overline{i_{in}^2}&=\frac{\overline{v_{on}^2}}{\left|Z_f\cdot \frac{-T_{AMP}(s)}{1-T_{AMP}(s)}\right |^2} =\\[2ex]
    &= \left| \frac{Z_b+Z_f}{Z_bZ_f} \right|^2 \cdot\left[4k_BTR_{bb}+\overline{e^2_{HBT}}\right] +\left|\frac{1}{Z_b}\right|^2\cdot4k_BTR_b 
        +\frac{R_f^2}{|Z_f|^2}\cdot\overline{i_{HBT}^2} + \frac{4k_BTR_f}{|Z_f|^2}
    \label{eq:noise_input_referred}
\end{aligned}
\end{equation}

The noise measurements and simulations are presented in section \ref{sec:input_and_output_noise}.

\section{Circuit board layout and cables}
\label{sec:layout}
\begin{figure}[hbtp]
    \centering
    \includegraphics[width=0.6\linewidth]{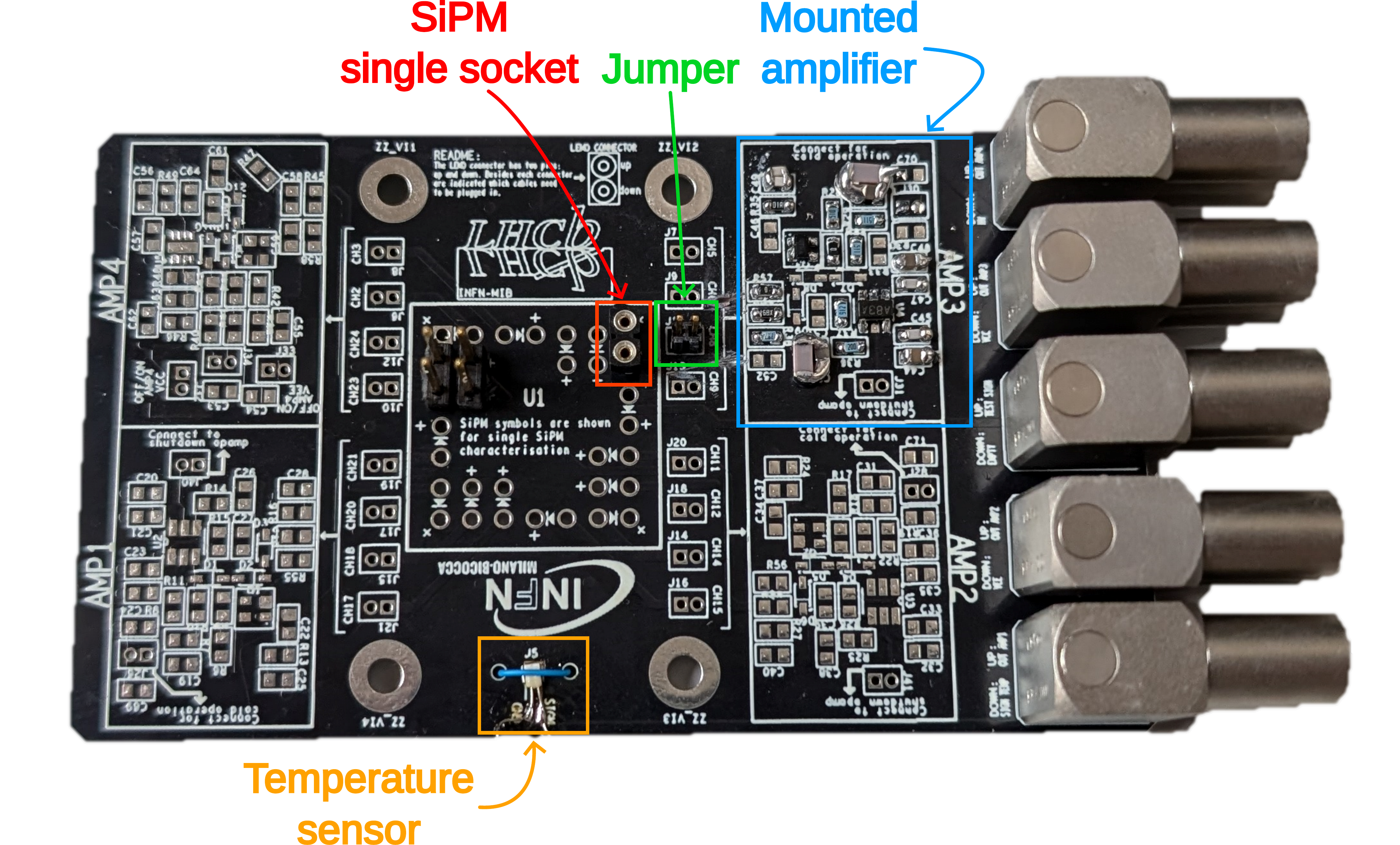}
    \caption{Amplifier layout. It features four amplifiers around a SiPM test structure (white square at the center). There is one PT1000 temperature sensor coupled to the PCB ground pad for more precise temperature measurements. In this picture only one amplifier and one SiPM socket are mounted.}
    \label{fig:amplifier_layout}
\end{figure}

A prototype of the amplifier board is shown in figure \ref{fig:amplifier_layout}.
This Printed Circuit Board (PCB) is used to test different configurations and identify the optimal option.
The layout centers around a square SiPM test socket located at its midpoint.
At the corners are four different amplifiers, each taking four different inputs. 
Each SiPM can be connected to the amplifier through a jumper, allowing the characterization of one SiPM at a time.
To measure the operating temperature of the amplifier and the SiPMs, there is a PT1000 temperature sensor on the top layer of the PCB, with the sensor being in contact to an uncovered ground pad to optimize thermal coupling between the temperature sensor and the PCB.
Finally, all signals and voltages are brought through coaxial cables with LEMO connectors. Five double-stacked LEMO connectors are used for the amplifier outputs, the temperature sensor signal and to provide the PCB with the test signal, $V_{CC}$, $V_{EE}$, $V_{POS}$ and $V_{HV}$.

Several different cables and connectors were, trying out both their cryogenic temperature endurance, bandwidth and characteristic impedance. 
To test their bandwidth and impedance we performed time domain reflection/time domain transmission measurements with a 18 GHz sampling oscilloscope (Agilent Technologies DCA-X 86100D).
To test the cable cryogenic reliability, the cable was dipped and extracted from liquid nitrogen several times and the signal integrity between the two cable ends was constantly checked.
Among the tested cables, commercial CAT8 Ethernet and USB-C 3.2 cables did not have sufficient bandwidth for analog signals, probably due to dielectric loss. 
Coaxial cables with MCX connectors were also tested (RG174 dielectric), but after a few thermal cycles the connector central pin would be sucked in and the cable would become unusable. 
Finally, we tested both SMA and LEMO cables, and both proved to be viable solutions. 
However, we ultimately chose to use the latter.
We selected the 80 cm (4 ns delay) LEMO cables
These cables have a maximum bandwidth of $|-1/(2\pi \tau_{cable})|\simeq3\ GHz$, where $\tau_{cable}=53\ ps$ is the time constant corresponding to a $-3$ dB attenuation of the cable.

Regarding the PCB traces, to distribute the test signal to four amplifiers while minimizing reflections, all transmission lines are terminated with $50\ \Omega$ (see figure \ref{fig:test_signal_path}). The signal generator drives a $50\ \Omega$ line through a series $50\ \Omega$ resistor, is split into four branches via $30\ \Omega$ resistors, and each branch reaches a transmission line terminated to ground with $50\ \Omega$.

\begin{figure}[hbtp]
    \centering
    \begin{circuitikz}[scale=0.5 ,transform shape, american] 
        \draw (0,0) node[ground]{} to[sinusoidal voltage source, v_=$V_s$] ++(0,2) to[short] ++(1,0)
                     to[R, l_=$50\ \Omega$] ++(2,0)
                     to[transmission line, l_=$50\ \Omega$] ++(2,0)
                     to[R, l_=$30\ \Omega$] ++(2,0) coordinate(centro);
        \draw (centro) to[R,l^=$30\ \Omega$] ++(75:2.5) 
                       to[transmission line, l^=$50\ \Omega$] ++(75:1) 
                       -- ++(75:0.5) coordinate(restognd1)
                       -- ++(75:1) node[ocirc,label=above:Test signal 1]{} ;
                       \draw (restognd1) to[R, l_=$50\ \Omega$] ++(165:2) node[ground,rotate=-105]{};
        \draw (centro) to[R,l^=$30\ \Omega$] ++(30:3) 
                       to[transmission line, l^=$50\ \Omega$] ++(30:1) 
                       -- ++(30:0.5) coordinate(restognd2)
                       -- ++(30:1) node[ocirc,label=right:Test signal 2]{} ;
                       \draw (restognd2) to[R, l_=$50\ \Omega$] ++(120:2) node[ground,rotate=-150]{};
        \draw (centro) to[R,l^=$30\ \Omega$] ++(-30:3) 
                       to[transmission line, l^=$50\ \Omega$] ++(-30:1) 
                       -- ++(-30:0.5) coordinate(restognd3)
                       -- ++(-30:1) node[ocirc,label=right:Test signal 3]{} ;
                       \draw (restognd3) to[R, l_=$50\ \Omega$] ++(60:2) node[ground,rotate=-210]{};
        \draw (centro) to[R,l_=$30\ \Omega$] ++(-75:3) 
                       to[transmission line, l^=$50\ \Omega$] ++(-75:1) 
                       -- ++(-75:0.5) coordinate(restognd4)
                       -- ++(-75:1) node[ocirc,label=below:Test signal 4]{} ;
                       \draw (restognd4) to[R, l_=$50\ \Omega$] ++(-165:2) node[ground,rotate=-75]{};
       \draw[dashed,ultra thick, blue] (-1,-1) rectangle (3,3);
       \node at (0.2,3.25) {\textcolor{blue}{Signal generator}};
    \end{circuitikz}

    \caption{Test signal path from the signal generator to the "Test signal X" input for the four different amplifying channels.}
    \label{fig:test_signal_path}
\end{figure}

This PCB prototype has four layers. To properly transport high-frequency signals, the PCB is manufactured using the Rogers RO4350B dielectric. 
This material has a low dissipation factor up to tens of $GHz$, as well as a low dielectric thermal coefficient \cite{rogers}. 
Each layer has this high frequency laminate, except for the core one, placed in between the two internal layers, which is made of a more common FR4 laminate. Because of this, all traces transporting high-frequency signals are drawn on the top or bottom of the PCB.

The aim for this amplifier is to characterize SiPMs with minimum signal degradation, which means that the amplifier needs to work at the highest possible frequencies. 
Also, the signal needs to be stable and not present ringing due to either amplifier instability or signal reflections at high frequencies. 
For this reason, it is necessary to take care of the parasitic capacitances and inductances in the layout. 

\subsection{Parasitics}
\label{sec:parasitics}

Figure \ref{fig:schematics_with_parasitics_new} illustrates the complete amplifier schematic, highlighting the relevant parasitic components that are examined in this section.

\begin{figure}[hbtp]

    \centering
    \ctikzset{amplifiers/thickness=2, transistors/thickness=3, amplifiers/scale=1.2}
    
    \begin{circuitikz}[scale=0.6 ,transform shape, american]   
    
          \draw (0,0) to[R=$R_s$] ++(2,0) to[L=$L_{p_{in}}$] ++(+2,0) -- ++(5.5,0) coordinate(V_B);
          \draw (V_B) node[npn,anchor=base](bjt){$\textbf{HBT}$};  
          \draw (bjt.emitter) to[short] ++(0,-0.5) coordinate(gnd);
          \draw (bjt.collector) to[short,-*] ++(0,0.5) node[left]{$V_c$} coordinate(coll) -- ++(0,0.3) to[R=$R_c$] ++(0,2.5)  to[battery2,name=myB,l=$V_{POS}$] ++(0,1) node[ground,rotate=180]{} coordinate(Vcc);
          \ctikztunablearrow{0.5}{1.3}{60}{myB}
          \draw (coll) -- ++(7,0) node[op amp, anchor=+,noinv input up](opamp){$\textbf{CFOA}$};
          \draw (coll) ++(2,0) node[circ]{} coordinate(parasitic) -- ++(0,0.3) to[R=$R_d$] ++(0,2) to[C=$C_d$] ++(0,1.5) coordinate(comp_coll);
          \draw (comp_coll) -- (Vcc -| comp_coll) node[ground,rotate=180]{};
          \draw (parasitic) ++(2,0) node[circ]{} coordinate(parasitic2) to[L=$L_x$] ++(0,2) to[C=$C_x$] ++(0,1.5) coordinate(parasitc_gnd);
          \draw (parasitc_gnd) -- (Vcc -| parasitc_gnd) node[ground,rotate=180]{};
          \color{black}
          \draw (parasitic2) ++(2,0) node[circ]{} to[C=$C_p$] ++(0,3.8) coordinate(parasitc_gnd2);
          \draw (parasitc_gnd2) -- (Vcc -| parasitc_gnd2) node[ground,rotate=180]{};
          \draw (opamp.-) to[short] ++(-2,0) -- ++(0,-0.3) to[R,l_=$R_1$] ++(0,-1.2) to[C, l_=$C_1$] ++ (0,-1.2) node[ground]{} coordinate(gnd2); 
          \draw (opamp.-) ++(-0.6,0) node[circ]{} coordinate(R2_node) to[short] ++(0,-2.2) coordinate(R2) to[R=$R_2$]  ++(3.4,0) -| (opamp.out) node[circ]{};
          \draw (R2) -- ++(0,-1.5) to[C=$C_{p_{R_2}}$]  ++(3.4,0) -| (opamp.out);
          \draw (opamp.up) -- ++(0,0.2) node[vcc]{$V_{CC}$};
          \draw (opamp.down) -- ++(0,-0.2) node[vee]{$V_{EE}$};
          \draw (gnd) -- (gnd2 -| gnd) node[ground]{};
          \draw (opamp.out) -- ++(0.5, 0) node[circ]{} -- ++(0,-5) coordinate(FB);
          \draw (opamp.out) ++(0.5,0) -- ++(0.5,0) node[ocirc,label=right:$V_{out}$]{} ; 
          \draw (FB) to[R=$R_f$] (FB -| bjt.base) coordinate(FB2) to[L=$L_{p_{R_f}}$] (bjt.base) node[circ]{} node[above]{$V_b$};
          \draw (FB) -- ++(-4.1,0) -- ++(0,-1.5) to[C=$C_{p_{R_f}}$] ++(-3,0) -- ++(0,1.5) ;
          \draw (V_B) ++(-2.5,0) node[circ]{} coordinate(Zb) to[generic=$Z_b$] ++(0,-2) node[ground]{};
          \draw (Zb) ++(-2.5,0) node[circ]{} coordinate(C_pb) to[C=$C_{pb}$] ++(0,-2) node[ground]{};
          \draw (C_pb) ++(-4.5,0) coordinate(ZSiPM) to[generic=$Z_{SiPM}$] ++(0,-2) node[ground]{};

          \draw[dashed,ultra thick, darkorange] (-0.75,-3.25) rectangle (6,1.25);
          \node at (-0.25,1.6) {\textcolor{darkorange}{\Large $Z_i$}};
          \draw[dashed,ultra thick, blue] (-1,-3.5) rectangle (8,2.25);
          \node at (-0.25,2.6) {\textcolor{blue}{\Large $Z_b'$}};
          
    \end{circuitikz}

    \caption{Schematic of the amplifier configurations with all parasitics included. The orange dashed box contains the impedances denoted by $Z_i$, while the blue dashed box highlights those denoted by $Z_b'$.}
    \label{fig:schematics_with_parasitics_new}
\end{figure}

As far as parasitic inductances are concerned, the goal during the layout phase was to minimize trace length and select a low thickness dielectric to minimize their value. 
Cadence Sigrity simulations carried out on parts of the PCB layout indicate a general rule of thumb: each 4 mm of trace length introduces approximately 1 nH of parasitic inductance.
Parasitic inductances in the feedback ($L_{p_{Rf}}$) and input ($L_{p_{in}}$) paths can potentially compromise amplifier stability or signal integrity. 
To better understand their impact, these two contributions are analyzed separately.

For the $L_{p_{Rf}}$ case, we consider the term highlighted in pink in the loop-gain expression of equation (\ref{eq:loop_gain_pure}). 
This term originates from the parasitic capacitances associated with the feedback and input traces. 
In the high-frequency limit, this term can be approximated as $\frac{C_{p_{Rf}}}{C_{p_{Rf}}+C_{pb}} \simeq \frac{C_{p_{Rf}}}{C_{pb}}$, with $C_{pb} \gg C_{p_{Rf}}$, as discussed below. 
Introducing $L_{p_{Rf}}$ in series with $Z_f=R_f \parallel C_{p_{Rf}}$ (see figure \ref{fig:schematics_with_parasitics_new}) modifies this factor to $\frac{C_{p_{Rf}}}{C_{pb}\left(1+sL_{p_{Rf}}C_{p_{Rf}}\right)}$, thereby introducing an additional pole in the loop gain. 
This extra pole reduces the phase margin and may eventually lead to instability. 
Simulations indicate that the effect becomes significant only for parasitic inductances on the order of several hundred nHs.

The input parasitic inductance $L_{p_{in}}$, on the other hand, is particularly relevant when the signal originates from the SiPM. 
When a microcell fires a current $i_{in}$, the generated current divides between the amplifier input and the remaining cells, as illustrated in figure \ref{fig:SiPM_impedance}. 
Under these conditions, the amplifier output can be expressed as $v_o = -Z_f\cdot \frac{i_{in}}{1+sL_{p_{in}}/Z_{cells}}$, where $Z_{cells}$ represents the impedance of the unfired cells, and the input impedance of the amplifier is assumed to be negligible.
This parasitic inductance introduces a frequency-dependent factor, $\frac{1}{1+sL_{p_{in}}/Z_{cells}}$, which attenuates and distorts the SiPM signal. 
According to simulations, this effect becomes noticeable for inductance values exceeding a few tens of nHs.

Cadence Sigrity simulations estimate parasitic inductances of $L_{p_{Rf}} \lesssim 5\ nH$ and $L_{p_{in}} \lesssim 1\ nH$. 
Since both values are well below the thresholds at which stability degradation becomes significant, the impact of these parasitic inductances can be considered negligible for the present application.

In addition to reducing the traces length, to reduce the risk of high-frequency ringing due to parasitic inductance between the HBT emitter and ground, the transistor emitter is connected to the ground planes with short traces and many vias, as found out in \cite{Dune_vecchio}.

Ground planes are removed under critical nodes like the HBT collector and CFOA to reduce stray capacitance. 
While effective, this interruption creates a parasitic inductor $L_x$ that induces signal oscillations, further detailed in section \ref{sec:output_signal_simulation}.
To further reduce parasitic capacitances at the collector node, the original single resistor in the network is replaced by two equal resistors connected in series.
This arrangement effectively halves the parasitic capacitance in parallel to the resistor.

One of the most critical parasitic capacitances for amplifier stability is that present at the HBT collector, denoted $C_p$.
With careful layout, the dominant and unavoidable contribution to $C_p$ becomes the input capacitance of the CFOA. 
Here, $C_p$ refers to the total parasitic capacitance at the collector node with either the LMH6702 or LMH6703 connected. 
Its value was estimated to be $1.95\ pF$ for the LMH6702 and $1.69\ pF$ for the LMH6703, based on a dedicated measurement described in appendix \ref{app:C}.

The base node of the HBT exhibits the parasitic capacitance, $C_{pb}$, which include the track capacitances, estimated at $1.1\ pF$, and the HBT's $C_{be}$ and $C_{bc}$ capacitances, estimated at $0.5\ pF$, as reported in the HBT's technical data sheet \cite{infineon_bfp640_datasheet}. 
The capacitance $C_{pb}$ also includes a contribution from $C_t$, which is used for amplifier characterization or gain calibration. 
While $C_t$ could be removed during SiPM measurements to minimize parasitics, doing so would require an impractical and repetitive desoldering process whenever calibration is required. 
Consequently, $C_t$ remains permanently on the PCB, resulting in 
a total $C_{pb} \simeq 2.6\ pF$.

Finally, we should also consider the parasitic capacitances in parallel with $R_2$ and $R_f$ ($C_{p_{R_2}}$ and $C_{p_{Rf}}$). 
Since these resistors have values on the order of $\sim k\Omega$, even a small parasitic capacitance of $\sim 0.1\ pF$ or lower in parallel can affect high-frequency performance. 
The total parasitic capacitance in parallel with a resistor soldered on the PCB pads is the sum of the resistor package and PCB contributions. 
The first contribution is not provided by the manufacturer and is difficult to measure at high frequencies. 
The second contribution is likewise challenging to measure, and the Cadence Sigrity simulator does not come in handy since it is not designed for this type of parasitic simulation.
To get an estimation of the total value of these parasitic capacitances, we exploited simulations of the entire circuit, and estimated the magnitude of these parasitics by comparing the measured and simulated signals.
We expect these parasitic capacitances to be of the order of tens of $fF$ (more details in section \ref{sec:output_signal_simulation}).

These parasitic capacitances, although necessary for a more accurate simulation of the output signal, have little effect on stability in a first-order approximation.
The time constants due to the parasitics $C_{pb}$ and $C_{p_{Rf}}$ in the loop gain in equation (\ref{eq:loop_gain_pure}) are about $\tau_z = C_{p_{Rf}}R_f\simeq 0.6\ ns$ and $\tau_p \simeq C_{pb}R_b\simeq 0.3\ ns$, with $\tau_z$ the parasitic zero and $\tau_p$ the parasitic pole.
Since both the ODP and TDP closed-loop gain dominant poles have a time constant of $<0.6\ ns$ (see table \ref{tab:configurations_network_elements}), this zero-pole pair adds a small phase increase  that slightly improves stability. 
Therefore, neglecting this pair in a first-order approximation represents a conservative design approach that ensures baseline stability. 

\section{Amplifier measurements and performance}
\label{sec:amplifier_performance}
The two configurations were simulated using a MATLAB script and subsequently implemented on PCB.
The amplifier performance was simulated by applying the inverse Laplace transform on the equations found in section \ref{sec:circuit_schematic}. 
Following the PCB implementation, the simulations were compared with measurements. 
They were then adjusted to account for the presence of second-order parasitic components ($C_{pb}$ and $C_{p_{Rf}}$). 
In particular, $C_{p_{Rf}}$ was unknown, and it was estimated by adjusting its value to match the simulation to the measured response.

To test the amplifier, the signal generator used was the Active Technologies Pulse Rider PG-1072. 
This is a very fast pulse generator with a rise time of about $95\  ps$. 
The amplifier output and the input test signal were each connected through $4\ ns$ cables, which, as previously mentioned, add a double low-pass filter at $BW_{cable}\simeq3\ GHz$ ($\tau_{cable} \simeq 52\ ps$).
To sample the electrical signals, a high-bandwidth oscilloscope was used, namely the Rhode \& Schwarz RTO64, featuring a 6 GHz bandwidth and a sampling rate of 20 GS/s.

After deciding which of the two configuration gives the best results, a thorough study was performed on only one of those configurations with additional testing at cryogenic temperatures.

\subsection{Output signal}
\label{sec:output_signal_simulation}
All simulations were carried out assuming $h_{FE} \simeq 300\ A/A$, measured with collector-to-emitter voltage $V_{be} \simeq 1\ V$ and collector current $I_c \simeq 4\ mA$ at 300 K, no resistive attenuation from the cables, $T \simeq 300\ K$, and $V_{CC}=-V_{EE}=5\ V$.
For the output signal simulations, equation (\ref{eq:osc_transf_function}) was employed.  
The collector node voltage depends on $V_{be}$, $I_b$, $R_2$, and $R_1$.  
Because $R_2$ differs between the two configurations, the collector voltage was approximately $V_C \simeq 1\ V$ for the ODP configuration and $V_C \simeq 0.8\ V$ for the TDP configuration.

Simulations were then compared to the measured signals, both obtained using an input step voltage of $S = 50\ mV$ across the test capacitor $C_t$, corresponding to a charge signal of $\sim 300\ ke^-$.

\begin{figure}[hbtp]
    \centering
    \includegraphics[width=0.55\linewidth]{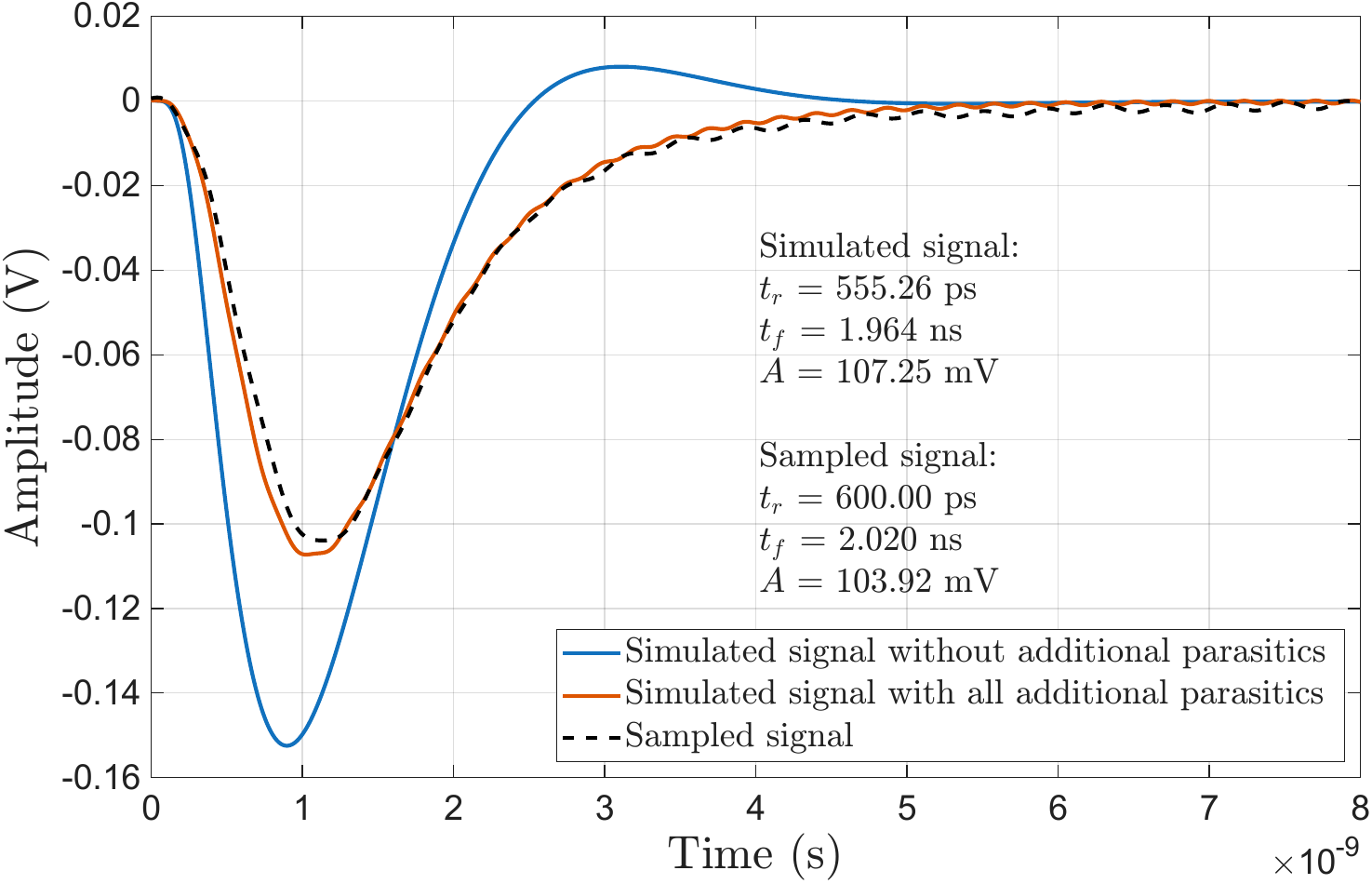}
    \qquad
    \includegraphics[width=0.55\linewidth]{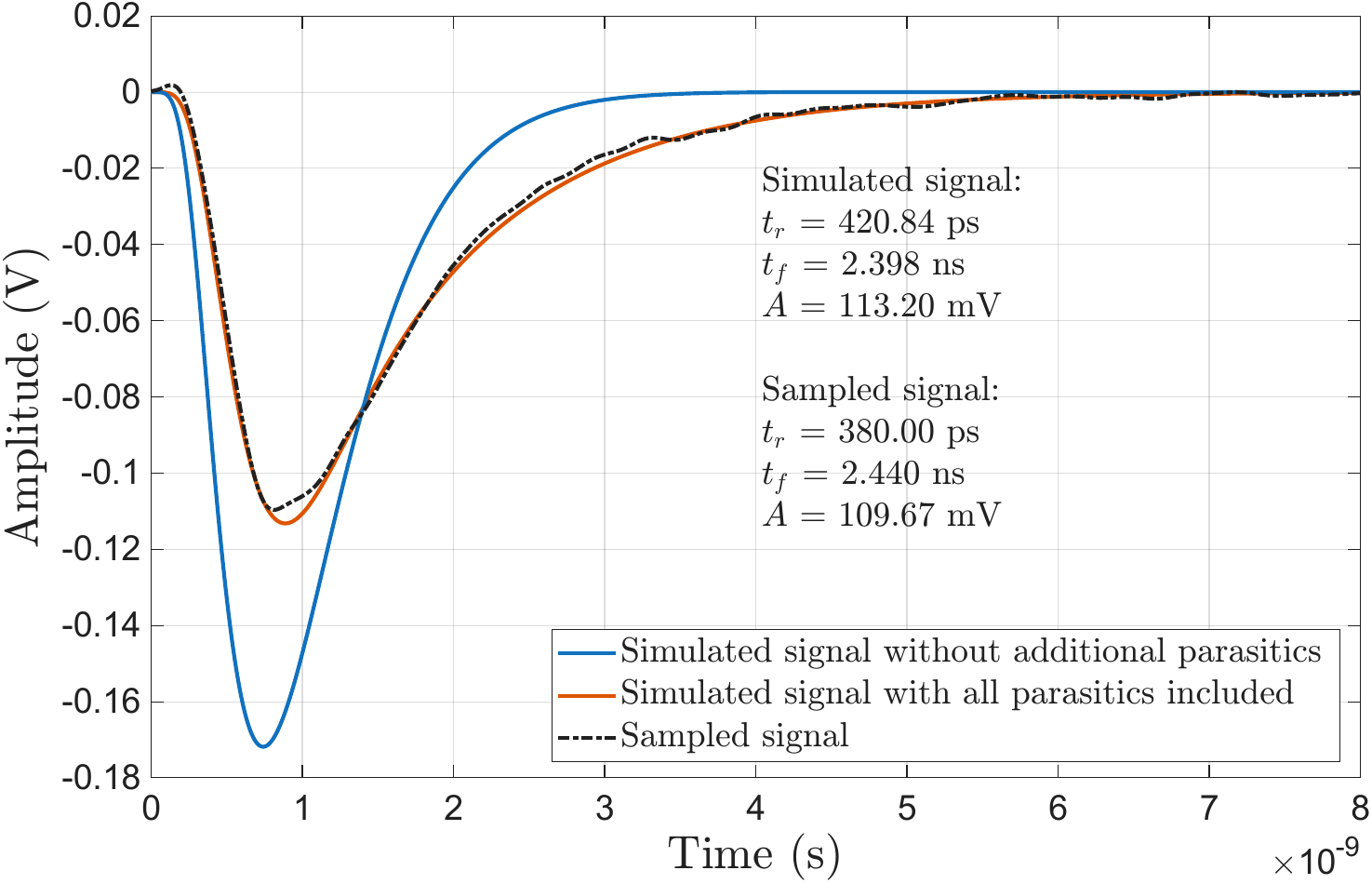}
    \caption{Simulation of the TDP (up) and ODP (down) configurations response to an impulse ($T=300\ K$). The dashed line represents the sampled signal, while the solid lines are the outputs of the simulation. The input signal is a step voltage with an amplitude $S=50\ mV$ and fast rise time ($95\ ps$) that charges a test capacitor of $1\ pF$. The resulting current signal carries a charge of $\sim 300\ ke^-$.}
    \label{fig:signal_simulations}
\end{figure}

Figure \ref{fig:signal_simulations} shows the output signals for both configurations. Each plot has three different curves. One is the measured output signal. 
The other two are simulations: one is the inverse Laplace transform of equation (\ref{eq:osc_transf_function}) --- which contemplates the presence of only $C_p$ as a parasitic capacitance --- while the other results from the same equation, but considers the effect of the other parasitics mentioned in section \ref{sec:parasitics}. 

The additional parasitic capacitances considered were $C_{pb}$, whose value was discussed in section \ref{sec:parasitics}, $C_{R_2}$, and $C_{pRf}$.
The parasitic capacitances in parallel to $R_2$ was considered negligible because the ground plane was removed below its PCB footprint. 
For the feedback resistor $R_f$, a parallel parasitic capacitance of $C_{pRf}\simeq 80\  fF$ was estimated by matching the simulated curve to the measured signal.

The curves in figure \ref{fig:signal_simulations} that consider more parasitics are reasonably close to representing the measured signal behavior. 
One important effect of the parasitic capacitances on the output signal is a reduction in signal amplitude. 
This occurs because the increased system time constants, while keeping the total input charge the same, spread the signal over a longer time, thereby lowering its peak amplitude.

The TDP real-time measurements exhibit high-frequency oscillations. These may be attributed to the parasitic inductance $L_x$, which produces an effect when $C_x$ is added to the circuit, as occurs in the TDP configuration.
The TDP simulation without additional parasitic effects exhibits a small overshoot, indicating that the two poles have a slight imaginary component. 
However, as clearly shown in the plot, this effect is negligible.

Table \ref{tab:summary_both_values} summarizes the computed value for the amplitude and rise and fall times for both sampled and simulated waveforms.
In addition to the parameters extracted from the waveforms, the Gain-Bandwidth Product (GBW) was determined using the time-domain system response. 
First, the dominant pole frequency was estimated from the measured  signal fall time via the first-order approximation.
The GBW was then calculated by multiplying this dominant pole frequency by the closed-loop gain, which is set by the feedback resistance ($R_f = 7500\ \Omega$).
This approach neglects the contribution of the rise time, which leads to an overestimation of the dominant pole frequency and, consequently, the GBW.

\begin{table}[htbp]
    \centering
    \caption{Comparison of measured and simulated signal parameters for TDP and ODP configurations at 300 K, including the Gain-Bandwidth Product (GBW) computed from the measured fall-time.}
    \label{tab:summary_both_values}
    \vspace{2mm}
    \begin{tabular}{l cc} 
        \toprule
        \textbf{Parameter} & \textbf{TDP Config.} & \textbf{ODP Config.} \\
        \midrule
        \textbf{Rise time (ps)} & & \\
        \quad Measured & 600.00 & 380.00 \\
        \quad Simulated & 555.26 & 420.84 \\
        \textbf{Fall time (ns)} & & \\
        \quad Measured & 2.02   & 2.44   \\
        \quad Simulated & 1.96   & 2.40   \\
        \textbf{Amplitude (mV)} & & \\
        \quad Measured & 103.92 & 109.67 \\
        \quad Simulated & 107.25 & 113.20 \\
        \textbf{GBW (GHz)} & & \\
        \quad Computed & 1300 & 1076 \\
        \bottomrule
    \end{tabular}
\end{table}

\subsection{Input and output noise}
\label{sec:input_and_output_noise}
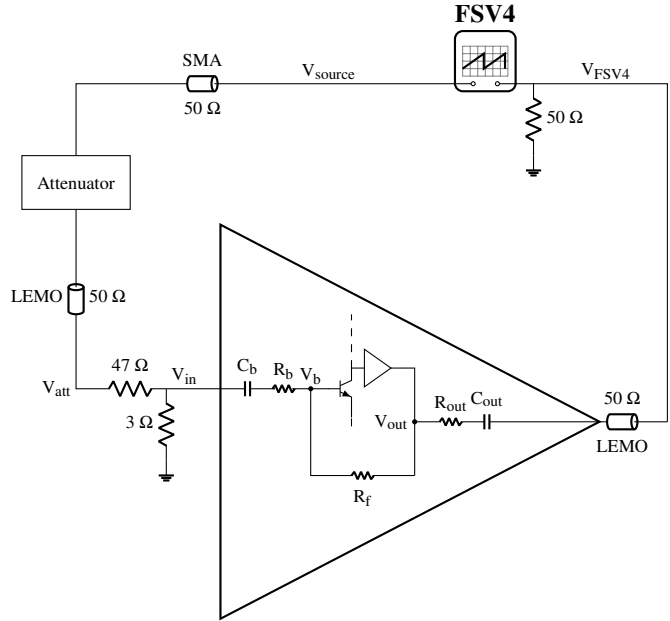
\begin{figure}[hbtp]

    \centering
    \tikzset{every node/.style={font=\Large}} 
    \ctikzset{amplifiers/thickness=1, transistors/thickness=1, amplifiers/scale=0.75, transistors/scale=0.75}
    
    \begin{circuitikz}[scale=0.475 ,transform shape, american]       
          \ctikzset{resistors/scale=0.5, capacitors/scale=0.5}  
          \draw (0,0) coordinate(VB) node[npn,anchor=base](bjt){};
          \draw (VB) -- ++(-0.5,0) coordinate(vblabel) -- ++(-0.5,0) to[R,l_=$R_b$] ++(-0.5,0) to[C,l_=$C_b$] ++(-1.5,0) -- ++(-1,0) node[above]{$V_{in}$} -- ++(-0.5,0) coordinate(Vsourcein); 
          \ctikzset{resistors/scale=1, capacitors/scale=1}
          \draw (Vsourcein) to[R,l_=$47\ \Omega$] ++(-2,0) -- ++(-0.5,0) node[left]{$V_{att}$} to[transmission line, l_=$50\ \Omega$,a^=LEMO] ++(0,5) coordinate(attenuator);
          \draw (Vsourcein) to[R,l_=$3\ \Omega$] ++(0,-2) node[ground]{};
          \node[draw,minimum width=3cm,minimum height=1.5cm,anchor=south](attenu) at (attenuator){Attenuator};
          \draw (attenu.north) -- ++(0,2) to[transmission line, l_=$50\ \Omega$,a^=SMA] ++(7,0) node[above]{$V_{source}$} -- ++(4,0) to[short,o-] ++(0,0) node[oscopeshape, anchor=in 1,scale=2](signAnalyzer){{\textbf{\small FSV4}}};
          \draw[dashed] (bjt.emitter) to[short] ++(0,-0.5);
          \draw[dashed] (bjt.collector) -- ++(0,1.5) node[left]{};
          \draw (bjt.collector) to[short] ++(0.2,0) coordinate(opampnode);
          \draw (opampnode) node[buffer ,anchor=in](opamp){};   
          \draw (opamp.out) -- ++(0.5, 0) -- ++(0,-1.5) coordinate(vout) -- ++(0,-1.5)  coordinate(FB);
          \ctikzset{resistors/scale=0.5, capacitors/scale=0.5}
          \draw (vout) node[circ,label=left:$V_{out}$]{} -- ++(0.5,0) to[R=$R_{out}$] ++(1,0) to[C=$C_{out}$] ++(1,0) coordinate(tooscilloscope);
          \ctikzset{resistors/scale=1, capacitors/scale=1}
          \draw (tooscilloscope) -- ++(2,0) to[transmission line, l^=$50\ \Omega$,a_=LEMO] ++(2.5,0) coordinate(ACout) ;
          \draw (signAnalyzer.in 2)  to[short,o-] ++(0,0) -- ++(1,0) coordinate(rfsv4) to[R=$50\ \Omega$] ++(0,-2) node[ground]{};
          \draw (rfsv4) -- ++(2,0) node[above]{$V_{FSV4}$} -- (signAnalyzer.in 2 -| ACout) -- (ACout); 

          \ctikzset{resistors/scale=0.5}
          \draw (FB) to[R=$R_f$] (FB -| vblabel) coordinate(FB2) -- (vblabel) node[circ]{} node[above]{$V_b$};

          \draw[thick] (-3,-5.5-0.92) coordinate(A) -- (-3,5.5-0.92) coordinate(B) -- (7.5,-0.92) coordinate(C) -- cycle;

    \end{circuitikz}

    \caption{Setup for the amplifier noise amplitude spectral density measurement.}
    \label{fig:fsv4_testbench}
\end{figure}

Using equation (\ref{eq:noise_input_referred}), the input-referred noise can be simulated and compared to real measurements of the same quantity.
The noise was measured using a signal analyzer: Rohde $\&$ Schwarz FSV4. 
The network used for this measurement is shown in figure \ref{fig:fsv4_testbench}. 
The equivalent input current noise was measured following four steps:

\begin{enumerate}
    \item The signal sourced by the FSV4 was attenuated (with the passive resistive attenuator Wavetek Model 5080) and characterized by measuring the transfer function between $V_{source}$ and $V_{att}$ ($\mathbf{H_{cal}}$). 
    Note that The $47\ \Omega$ and $3\ \Omega$ resistors were used to terminate the transmission line at the input. 
    These resistors further attenuate the $V_{source}$ signal by $V_{in}=V_{att}\cdot 3\ \Omega/(47\ \Omega + 3\ \Omega)$. 
    \item The amplifier transfer function between $V_{FSV4}$ and $V_{in}$ ($\mathbf{H_{amp}}$) was measured by sampling $V_{FSV4}$ and subsequently dividing it by $H_{cal}$. 
    \item The output noise spectral density ($\sqrt{\overline{\mathbf{v_{on}^2}}}$) was measured by disconnecting the cable at $V_{att}$ and sampling $V_{FSV4}$ over the frequency range from $10\ kHz$ to $1\ GHz$.
    \item The input-referred voltage noise was obtained by dividing the measured output spectral density by the measured amplifier transfer function: $\overline{v_{in}^2}=\overline{v_{on}^2}/\left|H_{amp}\right|^2$.
    To obtain the input-referred current noise $\sqrt{\overline{\mathbf{i_{in}^2}}}$, we divided the measured input-referred voltage noise by the computed impedance $Z_b=1/sC_b+R_b$.
    The expression used to estimate the input-referred current noise from the measured $\overline{v_{on}^2}$ is then:
    \begin{equation}
        \sqrt{\overline{i_{in}^2}}=\sqrt{\frac{\overline{v_{on}^2}}{|H_{amp}|^2\cdot |Z_b|^2}}
    \end{equation}
\end{enumerate} 

\begin{figure}[hbtp]
    \centering
    \includegraphics[width=0.7\linewidth]{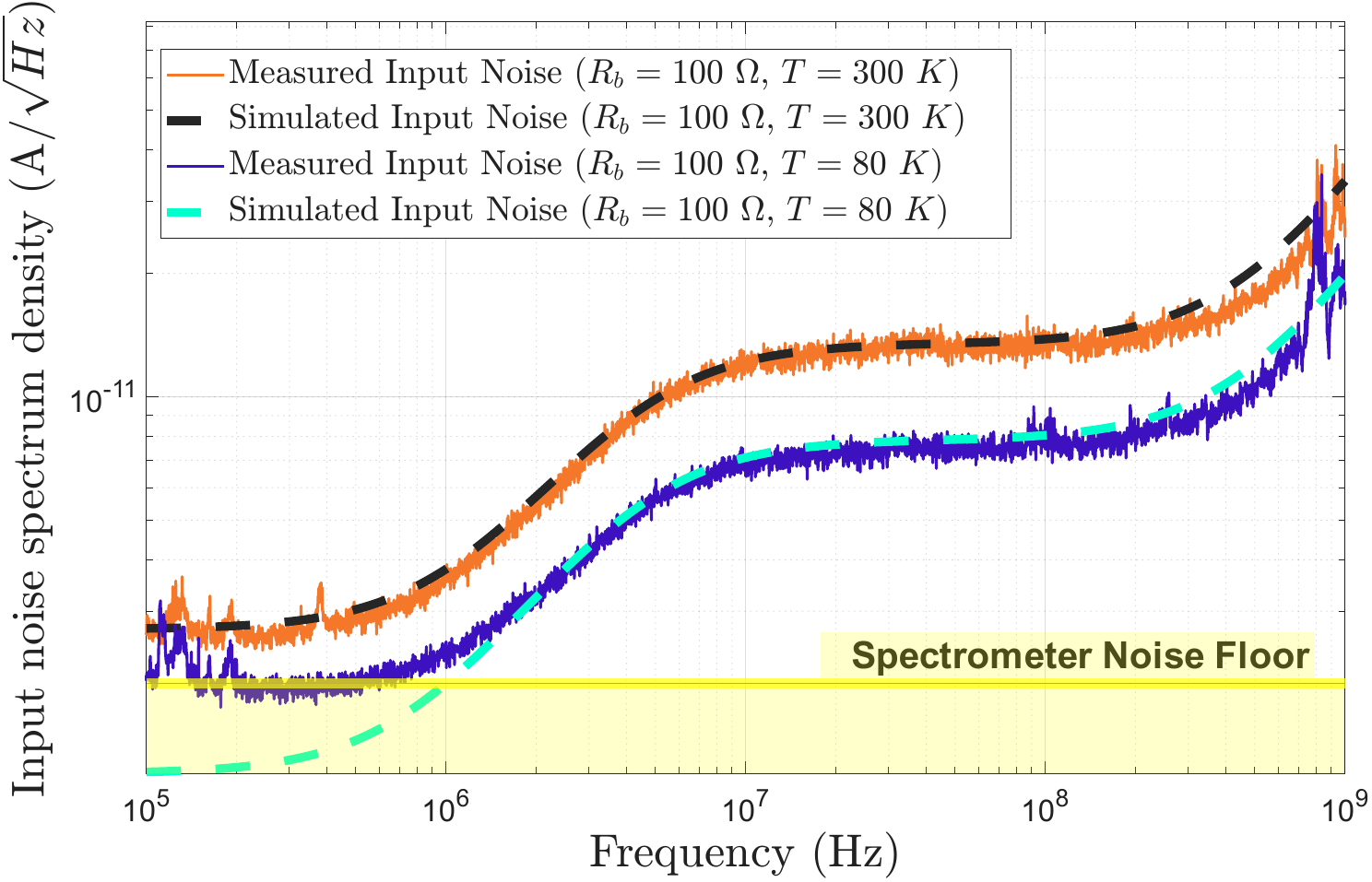}
    \caption{Plot of the measured (solid lines) and simulated (dashed lines) input-referred noise spectra at 300 K and 80 K of the ODP configuration.}
    \label{fig:input_referred_noise}
\end{figure}

The measured and simulated noise spectra for the ODP configuration at ambient and cryogenic temperatures are plotted in figure \ref{fig:input_referred_noise}.
At ambient temperature and on the flat part of the spectrum, the input noise is $\sim 13\ pA/\sqrt{Hz}$, whilst at 80 K is $\sim 8\ pA/\sqrt{Hz}$.
Most of the noise comes from just the $R_b$ resistor, meaning that the two configurations feature about the same input noise.
At cryogenic temperatures, the noise sensitivity is limited by the spectrometer noise floor to approximately $2\ pA/\sqrt{{Hz}}$.

\begin{figure}[hbtp]
    \centering
    \includegraphics[width=0.6\linewidth]{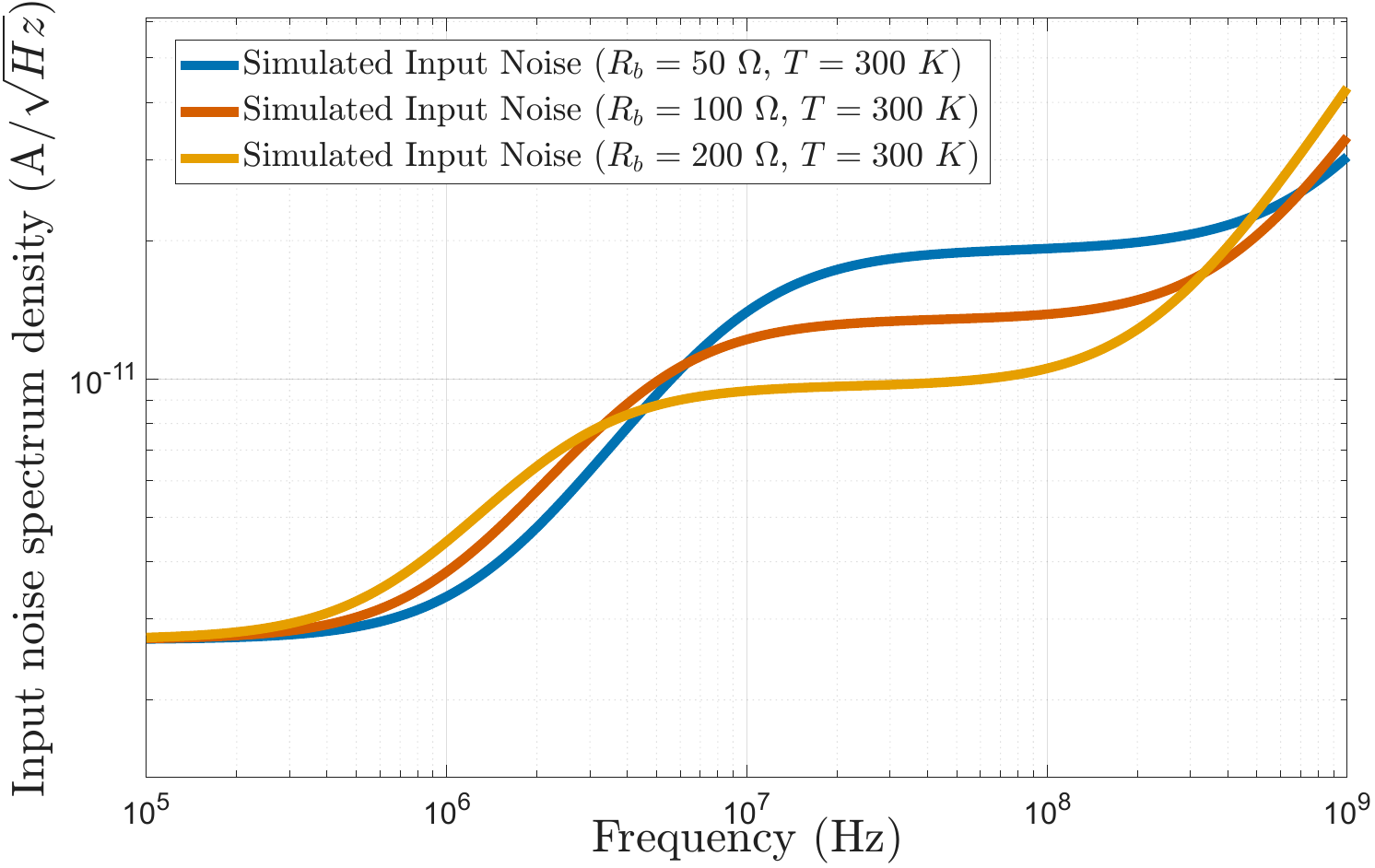}
    \caption{Input noise spectrum density as a function of different $R_b$ values at $T=300\ K$.}
    \label{fig:input_referred_noise_vs_Rb}
\end{figure}

Figure \ref{fig:input_referred_noise_vs_Rb} shows the effect of changing $R_b$ on the total input-referred noise. 
As can be seen, $R_b$ is the most impactful parameter on the noise, as the total current noise magnitude follows the value of this resistor.
The ideal case would be when $R_b\rightarrow\infty$ (note that $R_b$ is connected to ground through $C_b$ as in figure \ref{fig:SiPM_impedance}), but this would produce instability in the amplifier, as shown in equations (\ref{eq:stability_rb_open_loop}) and (\ref{eq:stability_rb_closed_loop}).
If we wanted to use this amplifier for a specific SiPM, we could remove $R_b$ and $C_b$ and optimize the loop gain --- namely adjusting the feedback resistor --- considering a specific SiPM impedance.

The output integrated noise is the integral of the output noise measured with the FSV4. 
For the two configurations, the output integrated noise is $\sigma_{N_{out_{TDP}}}=710\ \mu V_{RMS}$ and $\sigma_{N_{out_{ODP}}}=740\ \mu V_{RMS}$. 
Both configurations yield comparable noise levels, largely due to $R_b$ being the primary noise source. The small difference is caused by the mismatch in amplifier bandwidths.

The Signal-to-Noise Ratio (SNR) can be calculated as:

\begin{equation}
    SNR(dB)=10log_{10} \left( \frac{S_{out}^2}{\sigma_{N_{out}}^2} \right)
\end{equation}
where $S_{out}$ is the amplifier output signal amplitude and $\sigma_{N_{out}}$ is the integrated noise at the output of the amplifier (expressed in $V_{RMS}$).

Using the signal amplitudes of the sampled signals shown in figure \ref{fig:signal_simulations}, $S_{out_{TDP}} = 103.92\ mV$ and $S_{out_{ODP}} = 109.67\ mV$ for $S_{in} = 50\ mV$, corresponding to a signal of $\sim 300\ ke^-$, and the sampled noise values obtained above, the two SNRs were computed. 
They are: $SNR_{TDP} = 2.14 \cdot 10^4 = 43.3\ dB$ and $SNR_{ODP} = 2.20 \cdot 10^4 = 43.4\ dB$.

\subsection{Electronics jitter}

To find an expression for the electronics jitter, we make a simplification by stating that the output signal rise time in response to a delta signal is:

\begin{equation}
    f(t)=S\cdot (1-e^{-t/\tau})
    \label{eq:simple_signal_jitter}
\end{equation}

\noindent
where $\tau$ is the rise time constant.
The jitter can be defined as the fluctuation of the time when the signal crosses a given voltage threshold, conventionally set to half the signal amplitude. By setting equation (\ref{eq:simple_signal_jitter}) equal to $S/2$ and solving for t, one finds $t_{1/2}=\tau ln(2)$.

The electronics jitter can be expressed using the relation $dt = dv / f'(t)$:

\begin{equation}
    \sigma_t^2= \left. \left(  \frac{d f(t)}{d t} \right)^{-2}\right|_{t=t_{1/2}} \cdot\sigma_N^2=\frac{4\tau^2\sigma_N^2}{S^2}\simeq0.83\cdot\frac{t_r^2}{SNR} \longrightarrow \sigma_t\simeq 0.91\cdot\frac{t_r}{\sqrt{SNR}}
    \label{eq:jitter}
\end{equation}

where $t_r=\tau ln(9)$ is the rise time.
Finally, by computing the jitter of the same input signal, the TDP and ODP configurations contribute respectively with $\sigma_{t_{TDP}}\simeq3.7\ ps_{RMS}$ and $\sigma_{t_{ODP}}\simeq2.3\ ps_{RMS}$, where the $SNR$ used are those calculated in section \ref{sec:input_and_output_noise} and $t_r$ are those of the sampled signals in figure \ref{fig:signal_simulations}, corresponding to $FWHM_{t_{TDP}}=\sigma_{t_{TDP}}\cdot 2\sqrt{2ln(2)}=8.8\ ps$ and $FWHM_{t_{ODP}}=5.5\ ps$, a result that favors the latter configuration for SiPM characterization. 
This difference between the two is mainly due to the ODP configuration exhibiting a larger bandwidth (lower rise time), while the two $SNR$ values are almost identical, as summarized in table \ref{tab:summary_configurations}.

To verify these computations, we can perform measurements of these jitter values using the test signal.
However, to better evaluate the amplifier’s jitter contribution for specific input signals, an equivalent input signal with non-ideal rise time and defined amplitude should be used, as is shown in section \ref{sec:jitter_measurements}.
This is necessary because real signals, despite carrying the same total charge as the ideal test signal, have lower amplitude due to their limited bandwidth.
Both these factors impact the jitter measurement and need to be taken into account.

\subsection{Input impedance}
\label{sec:input_imp_results}

\begin{figure}[hbt]
    \centering
    \includegraphics[width=0.75\linewidth]{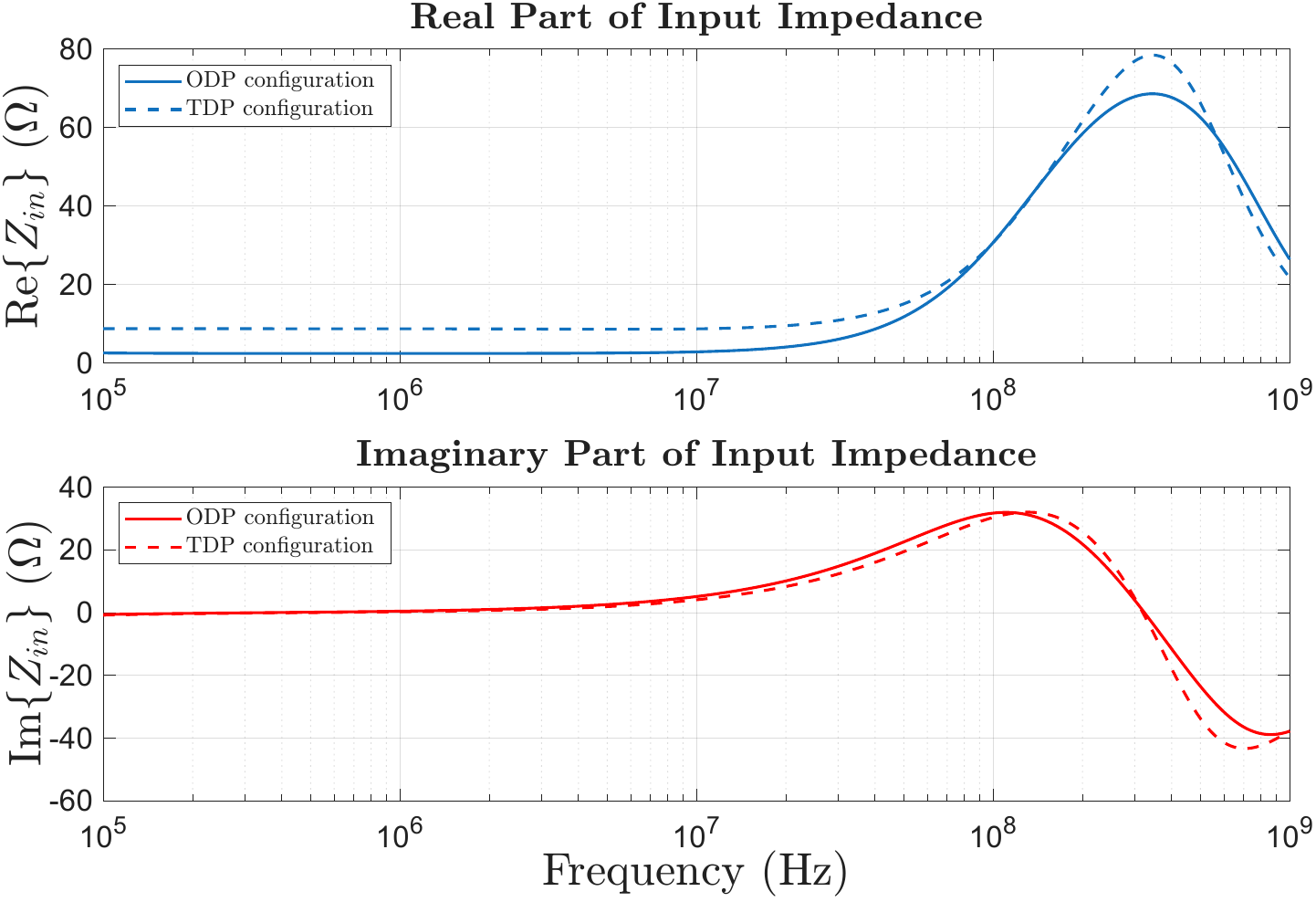}
    \caption{Simulated input impedance (real and imaginary parts) for the TDP and ODP configurations.}
    \label{fig:input_impedance_sim_both}
\end{figure}

Since the amplifier topology is a TIA, by design the input impedance is very low, close to $0\ \Omega$.
The input impedances for both configurations were simulated using equation (\ref{eq:simple_input_imp}) and the plots are shown in figure \ref{fig:input_impedance_sim_both}.

The low-frequency ($< 10$ MHz) input impedance real part values are $Z_{in_{LF, TDP}} \simeq 8.7~\Omega$ for the TDP configuration and $Z_{in_{LF, ODP}} \simeq 2.4~\Omega$ for the ODP one. 
This difference originates from the higher loop gain in the ODP configuration since the other parameters in equation (\ref{eq:simple_input_imp}) remain constant. 
With increasing frequency, the loop gain diminishes, causing a corresponding rise in input impedance. At several hundred MHz, the impedance reaches approximately $70\ \Omega$ and $80\ \Omega$ for the ODP and TDP configurations, respectively.

Regarding input signal integrity, the SiPM is placed only a few millimeters from the amplifier input. This distance is short enough that the connection is considered electrically short at the frequencies of interest, which effectively mitigates the signal reflections. 
Although a 50 $\Omega$ series resistor could be added to provide standard matching, this option was omitted because it would reduce the input signal bandwidth, as shown in equation (\ref{eq:i_in_sipm_rs}).

\subsection{OA dominant pole configuration: hot vs cold temperatures operation}
\label{sec:hotvscold_operation}

Table \ref{tab:summary_configurations} summarizes the ambient-temperature characteristics of the two configurations calculated in the previous section. Additionally, it introduces power dissipation.
This is mostly due to the CFOA and measures about 175 mW for the TDP configuration and 165 mW for the ODP.

Between the two, the ODP configuration is faster, which is a most crucial parameter for jitter measurements. 
Therefore, we continued the in-depth study of the ODP configuration, as it provides better overall performance, and proceeded to analyze its behavior at different temperatures.

\begin{table}[tbp]
    \centering
    
    \caption{TDP and ODP configuration characteristics at 300 K. The third column shows the ODP characteristics at 80 K with $V_{POS}$ tuned for stable amplifier operation. All these values were computed in section \ref{sec:amplifier_performance} using the amplifier measured response to a fast ($95\ ps$ rise time) step signal of $50\ mV$ ($\sim 300 \ ke^-$) which charges a test capacitor of $1\ pF$ and sends to the amplifier input a $\delta(t)$-like current signal. For all three columns, the amplifier gain is 7500 $\Omega$.}
    \vspace{2mm}
    
    \resizebox{\textwidth}{!}{%
    \begin{tabular}{c|c|c|c}
        \toprule
         \textbf{Characteristics} & \textbf{TDP configuration (300 K)} & \textbf{ODP configuration (300 K)} & \textbf{ODP configuration (80 K)} \\
        \midrule
        \textbf{Rise time: $\mathbf{t_r}$} & $600\ ps$ & $380\ ps$ & $540\ ps$  \\
        \textbf{Fall time: $\mathbf{t_f}$} & $2.02\ ns$ & $2.44\ ns$ & $1.41\ ns$\\
        \textbf{Signal amplitude: $\mathbf{S_{out}}$} & $103.92\ mV$ & $109.67\ mV$ & $147.00\ mV$ \\
        \textbf{Gain-Bandwidth Product: GBW}  & 1300 & 1076 & 1862\\
        \textbf{Input noise: $\mathbf{\sqrt{\overline{i_{in}^2}}}$: } & $\sim 13\ pA/\sqrt{Hz}$ & $\sim 13\ pA/\sqrt{Hz}$ & $\sim 8\ pA/\sqrt{Hz}$ \\  
        \textbf{Integrated output noise: $\mathbf{\sigma_{N_{out}}}$} & $710\ \mu V_{RMS}$ & $740\ \mu V_{RMS}$ & $470\ \mu V_{RMS}$ \\
        \textbf{Signal-to-noise ratio: SNR} & $43.3\ dB\simeq2.14\cdot10^4$ & $43.4\ dB\simeq2.20\cdot10^4$ & $49.9\ dB\simeq 9.8\cdot 10^4$\\
        \textbf{Electronics jitter: $\mathbf{\sigma_t}$} & $3.7\ ps_{RMS}$ & $2.3\ ps_{RMS}$ & $1.6\ ps_{RMS}$\\        
        \textbf{Input impedance ($<$10 MHz): Re($Z_{in}$)} & $8.7\ \Omega$ & $2.4\ \Omega$  & $2.4\ \Omega$\\
        \textbf{Power dissipation per channel} & 175 mW & 165 mW & 100 mW\\
        \bottomrule
    \end{tabular}
    }
    
    \label{tab:summary_configurations}
\end{table}

At cryogenic temperatures, the active components can behave differently. The HBT is known to work well at lower temperatures because of its heterogeneous structure, which allows the transistor to have a larger gain at lower temperatures \cite{Si-GeTransistors}. 
On the other hand, the CFOA needs to be tested at cryogenic temperatures to verify its behavior. 

Measurements of the local CFOA closed-loop gain and rise time were performed in liquid nitrogen ($\sim 80\ K$); the experimental setup and methodology are described in appendix \ref{app:B}. 
From these measurements, we obtained a closed-loop gain of $G_{CFOA\ real} \simeq 345\ V/V$ at $80\ K$, which corresponds to a corrected value of $R_1 \simeq 26.4\ \Omega$. 
In addition, the measured rise time yields $C_o R_2 \simeq 25\ ns$ (see equation (\ref{eq:loop_gain_open_loop_configuration})).
These values are to be confronted with their counterparts at 300 K, which are $G_{CFOA_{real \ 300K}}\simeq 411\ V/V$ and $C_oR_2\simeq 35\  ns$ (see appendix \ref{app:B}).
The amplifier gain diminishes as temperature decreases.
Although the gain measured at $80\ K$ is lower than that measured at $300\ K$, the higher bandwidth yields a comparable gain–bandwidth product at both temperatures.

Among all of these temperature dependent parameters, the most critical for the amplifier stability is the HBT gain, which increases significantly with decreasing temperatures. 
To counteract this effect, a variable power supply voltage $V_{POS}$ is used (see figure \ref{fig:open_loop_and_closed_loop_configurations}). 
Since the HBT gain is directly proportional to $g_m = I_c / V_T = I_c / (k_B T / q)$, a practical way to maintain a stable amplifier bandwidth across varying operating temperatures is to adjust the HBT bias current by tuning $V_{POS}$.

$V_{POS}$ is manually adjusted with temperature, decreasing from about $5\ V$ at room temperature to $2.6\ V$ at 80 K, matching the behavior observed at ambient conditions. 
Table \ref{tab:summary_configurations} summarizes the differences in amplifier operation at ambient and cold temperatures, with all values computed and measured as shown in the previous sections. 

\begin{figure}[hbtp]
    \centering
    \includegraphics[width=0.6\linewidth]{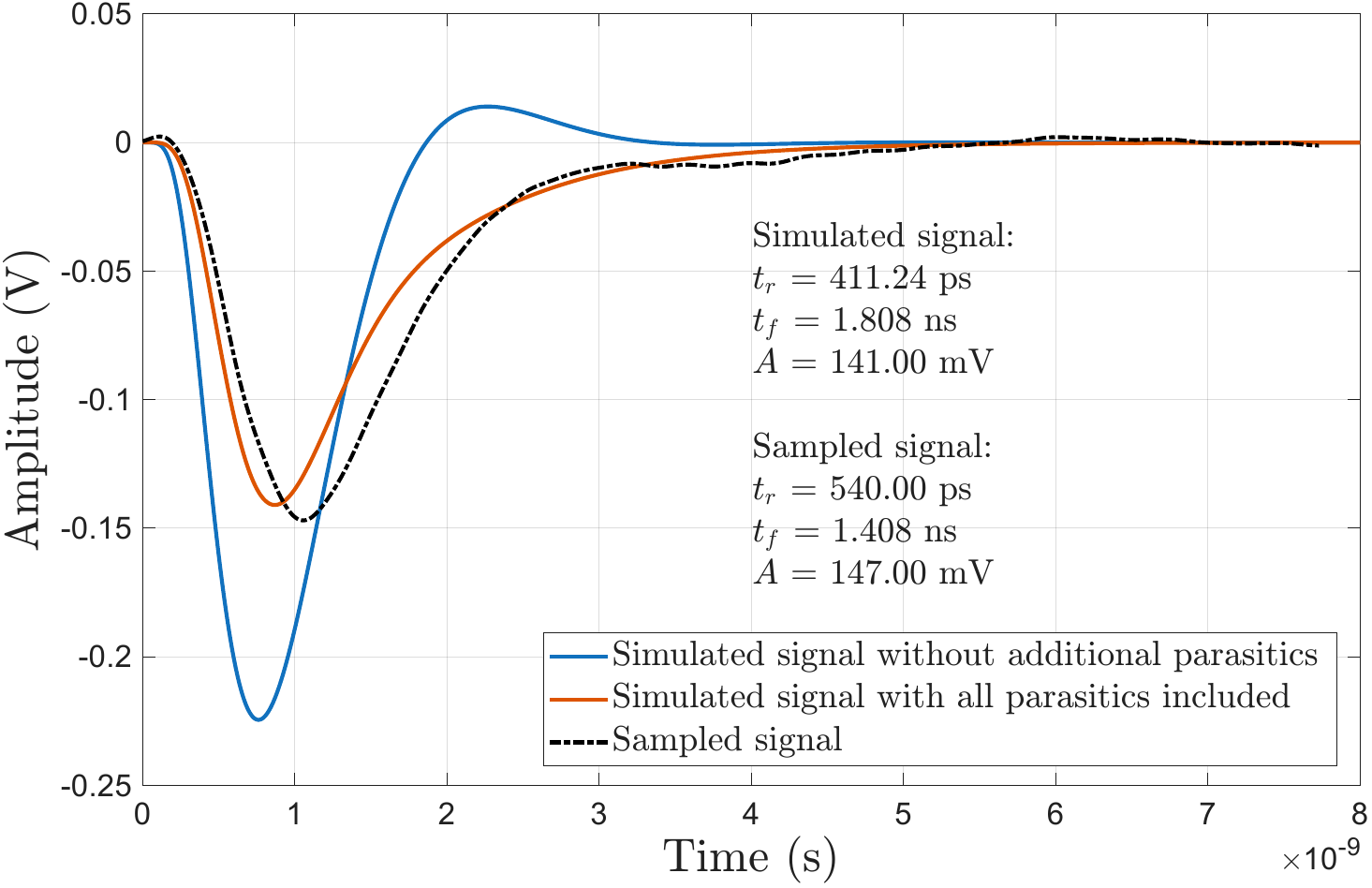}
    \includegraphics[width=0.6\linewidth]{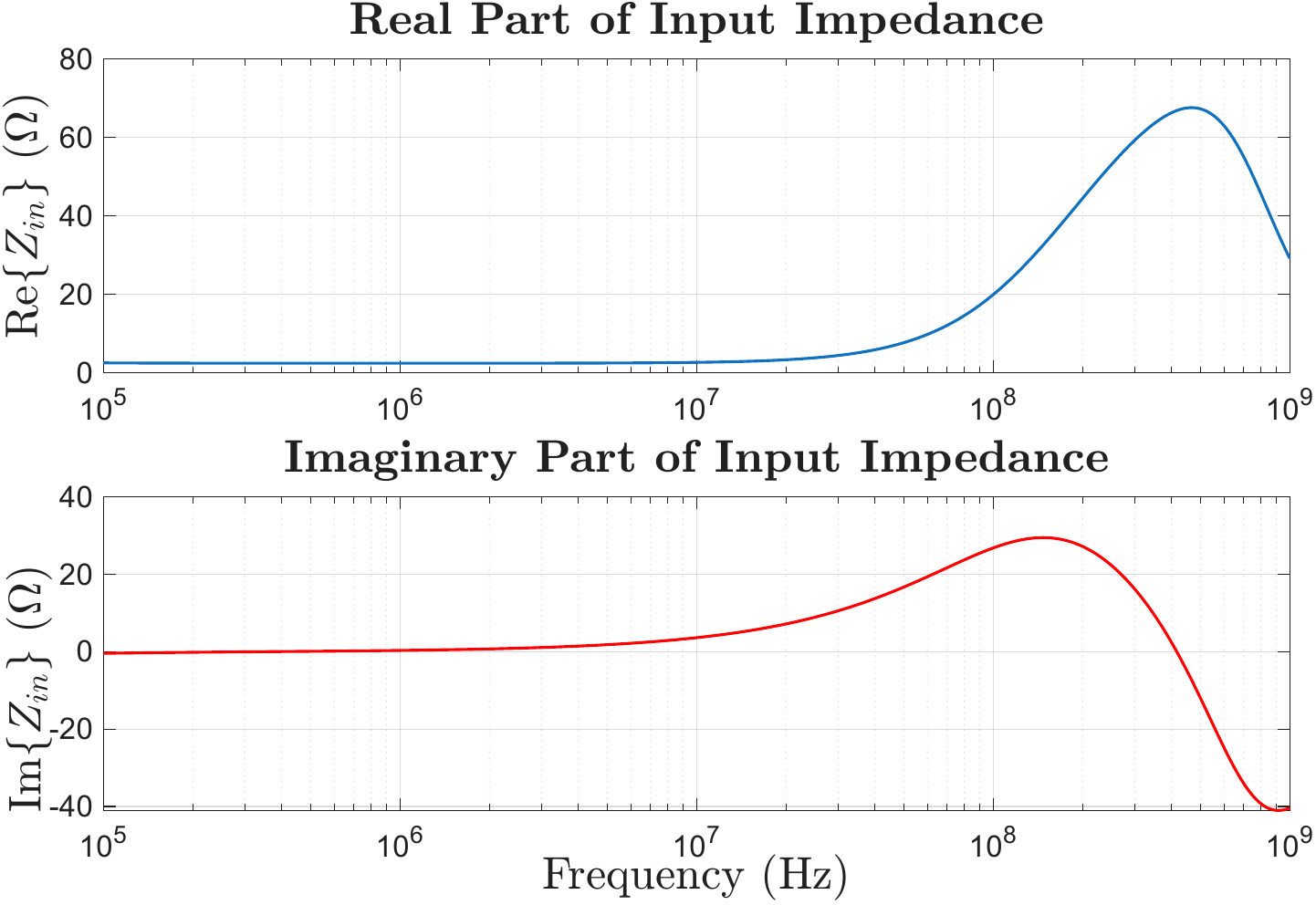}
    \caption{On the top is shown the simulation of the ODP configurations response to an impulse ($T=80\ K$). The dashed line represents the sampled signal, while the solid lines are the outputs of the simulation. The input signal is a step voltage with an amplitude $S=50\ mV$ and fast rise time ($95\ ps$) that charges a test capacitor of $1\ pF$. The resulting current signal carries a charge of $\sim 300\ ke^-$. On the bottom is shown the simulated input impedance (real and imaginary parts) for the ODP configurations at $T=80\ K$.}
    \label{fig:80K_sim_output_and_input_impedance}
\end{figure}

Figure \ref{fig:80K_sim_output_and_input_impedance} shows the amplifier's response to a test signal at both temperatures. 
As with the earlier comparison between the TDP and ODP configurations, the measured signal is compared to the simulations. 
In this case, even the model that accounts for all known parasitics does not perfectly match the measured waveform. 
This discrepancy likely arises from unmodeled temperature-dependent effects. 
Consequently, because the simulated transfer function deviates from the actual behavior, the simulated input impedance may also differ from the real input impedance.

\begin{figure}[hbtp]
    \centering
    \includegraphics[width=0.7\linewidth]{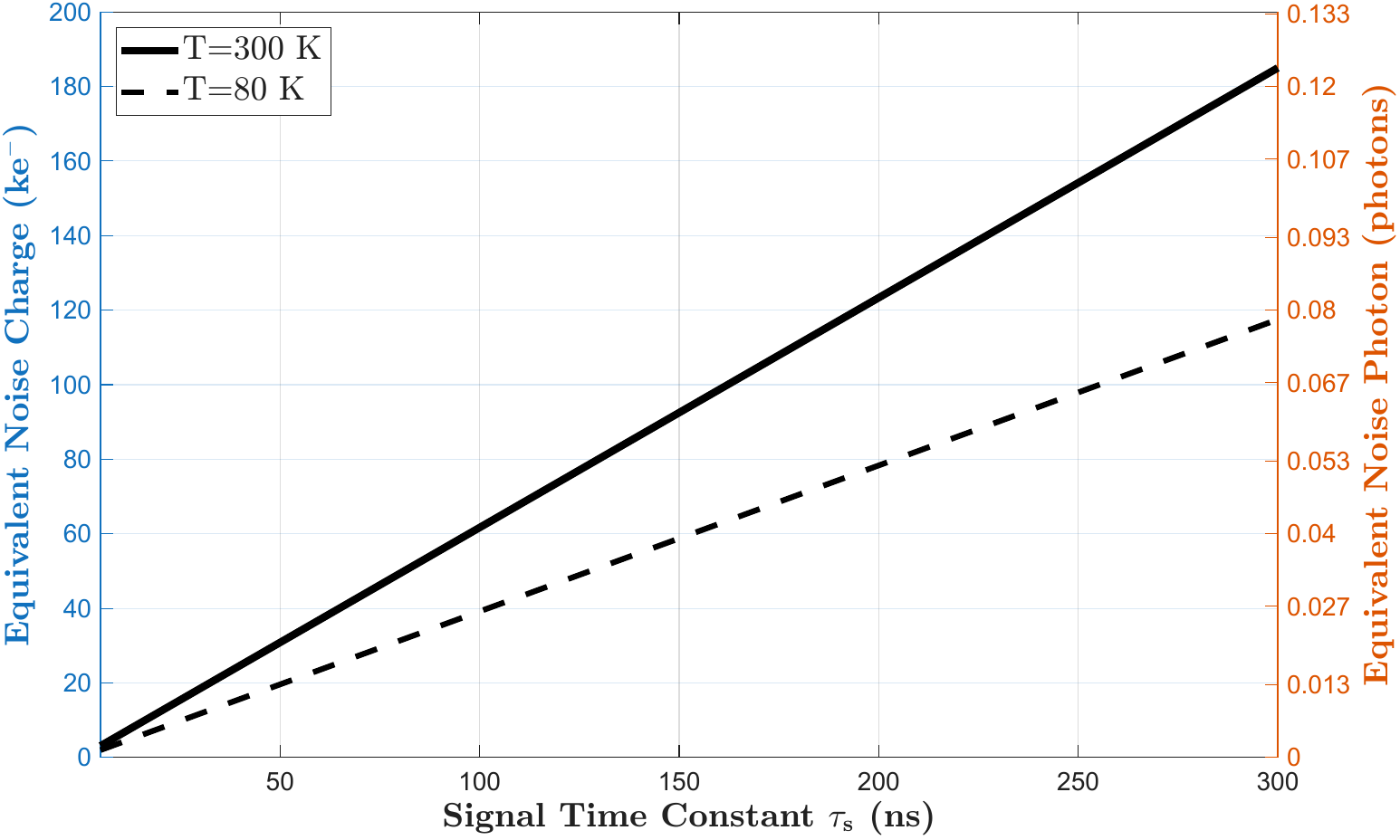}
    \caption{Equivalent noise charge and equivalent noise photon as functions of the fall-time constant of an input signal emulating a single-cell SiPM discharge. The two curves correspond to amplifier operation at ambient temperature and in liquid nitrogen.}
    \label{fig:ENC_ENP}
\end{figure}

The equivalent noise charge (ENC) can also be evaluated for the amplifier when driven by a SiPM signal.
Based on equation (\ref{eq:i_in_sipm_rs}), when the product $C_q R_q$ is negligible, the SiPM pulse can be approximated by a single-exponential current $i_{SiPM} = \frac{Q}{\tau_s} e^{-t/\tau_s}$, where $Q$ is the charge released by a single cell discharge and $\tau_s$ is the SiPM cell recharge time constant.
Under this assumption, the corresponding amplifier output is $v_{out} = R_f \cdot i_{SiPM}$.
The resulting signal-to-noise ratio (SNR) is therefore given by:

\begin{equation}
    SNR=\frac{V_{out_{MAX}}}{\sigma_{N_{OUT}}}=\frac{R_f}{Q\tau_s}\cdot \frac{1}{\sigma_{N_{OUT}}}
\end{equation}
The equivalent noise charge is therefore:
\begin{equation}
    ENC=\frac{\sigma_{N_{OUT}}\tau_s}{qR_f}
\end{equation}
\noindent
where $q$ is the electron charge.
Assuming a SiPM electron gain of $G_{SiPM}=1.5\ {Me^-}$, typical for SiPMs operated at $3\ V_{OV}$ with large cell sizes ($50\ \mu m$–$75\ \mu m$) \cite{HamamatsuS13360Series2025}, we can also derive the equivalent noise photon (ENP) as $ENP = ENC / G_{SiPM}$.
By fixing the total charge produced by a single photon to $G_{SiPM}$ and varying the time constant over which this charge is delivered to the amplifier, the plot shown in figure \ref{fig:ENC_ENP} is obtained. 
The figure illustrates the dependence of the amplifier ENC and ENP on the bandwidth of the input signal.

In Figure \ref{fig:ENC_ENP}, the 80 K curve exhibits a flatter slope, since the $\sigma_{N_{OUT}}$ decreases at cryogenic temperatures.
However, the SiPMs used in this work exhibit longer fall-time constants at low temperatures, since the SiPM cell recharge time depends on $R_q$, which has a negative temperature coefficient.
Hence, for the same amount of SiPM-generated charge, we generally obtain similar results in terms of ENC/ENP.

Finally, the amplifier power consumption per channel is about 100 mW at cryogenic temperatures (with adjusted $V_{POS}$).
Because the amplifier is designed for single-channel operation rather than high-density parallel arrays, the primary thermal constraint is preventing localized film boiling in the liquid nitrogen, as vapor bubbles would interfere with the SiPM characterization. 
The critical heat flux limit for liquid nitrogen is roughly 150 to 200 $mW/mm^2$ \cite{critical_heat_flux_LN2}. 
Even under the conservative assumption that all 100 mW is dissipated entirely by the CFOA package, which has a surface area of about 5 $mm^2$, the resulting heat flux is only 20 $mW/mm^2$, which is well below the critical threshold. 
Consequently, the amplifier safely avoids localized film boiling and maintains a stable thermal equilibrium at 77 K.

\section{SiPM measurements}
\label{sec:sipm_measurement}
To assess how well the amplifier works in a real-world scenario, the amplifier was tested with SiPM signals. We mostly used  an Hamamatsu S13360-1350 sensor ($1.3 \times 1.3\ mm^2$), especially for jitter measurements, but we also used the Hamamatsu S13360-3050 ($3 \times 3\ mm^2$), to validate the amplifier operation with different and bigger sensors. 

Figure \ref{fig:SiPM_impedance} shows an equivalent circuit for the SiPM.
Using the Hamamatsu S13360-1350 SiPM as a reference, values for the SiPM passive elements can be extracted.
If we use the values of gain ($G=1.7\cdot 10^6\ $ at $V_{ov}=3\ V$), terminal capacitance ($C_{term}=60\ pF$) and the number of cells ($N=667$) provided by Hamamatsu \cite{HamamatsuS13360Series2025}, we can find that $C_{det} \simeq \frac{q\cdot G} {V_{ov}}\simeq 91\ fF$ \cite{HamamatsuSiPM2018}. 
This represents an overestimation of $C_{det}$, since the calculation includes the contribution of $C_q$, which is therefore implicitly neglected.
Since $N\cdot C_{det}\simeq C_{term}$, the capacitance $C_g$, which is less than $1\ pF$, can be considered negligible. 
The same calculation can be performed with other SiPMs, yielding similar conclusions.
Regarding the quenching resistor, we measured the IV-curves of SiPMs with areas ranging from $1.3 \times 1.3\ mm^2$ to $3 \times 3\ mm^2$ and we extrapolated that $R_q/N\gtrsim 100\ \Omega$.
Hence, in these cases, we should not have instability problems since $C_g$ is negligible and $(R_q/N)  \parallel R_b$ remains above the lower bound of $49\ \Omega$ for $R_b$, as derived in equation (\ref{eq:loop_gain_open_loop_configuration}).
Furthermore, computing the pole–zero pairs in equation (\ref{eq:loop_gain_pure}) by using these values of $C_{det}$, $R_q/N$, and $R_b$ (with $R_s = 0$) places the zero and the pole in the tens of MHz range, which is sufficiently low to classify them as low-frequency poles.

If $C_g$ is on the order of several tens of picofarads, one way to mitigate amplifier instability is to place $R_s$ in series with the preamplifier.
Alternatively, the amplifier bandwidth can be reduced by decreasing $V_{POS}$, which in turn lowers the amplifier’s gain–bandwidth product.
As previously anticipated, for our next tests, $R_{s1}$ and $R_{s2}$ were both short circuited ($R_s=R_{s1}+R_{s2}=0$), as the SiPMs we tested did not produce any instabilities when attached at the amplifier inputs.

The initial objective of this section is to test and observe the signals generated by the SiPM after shaping by our amplifier, in order to verify that the SiPM impedance does not induce ringing or instability in the amplifier response. The remainder of the section focuses on jitter measurements, using both SiPM signals and test pulses, to evaluate the amplifier’s influence on the overall SiPM timing performance.

For these measurements we used the Rhode \& Schwartz RTO64 oscilloscope, which offers $6\ GHz$ bandwidth and a $20\ GS/s$ sampling rate.

\subsection{Signal shape}
\label{sec:sipm_signal_shape}
\begin{figure}[hbtp]
    \centering
    \includegraphics[width=0.48\linewidth]{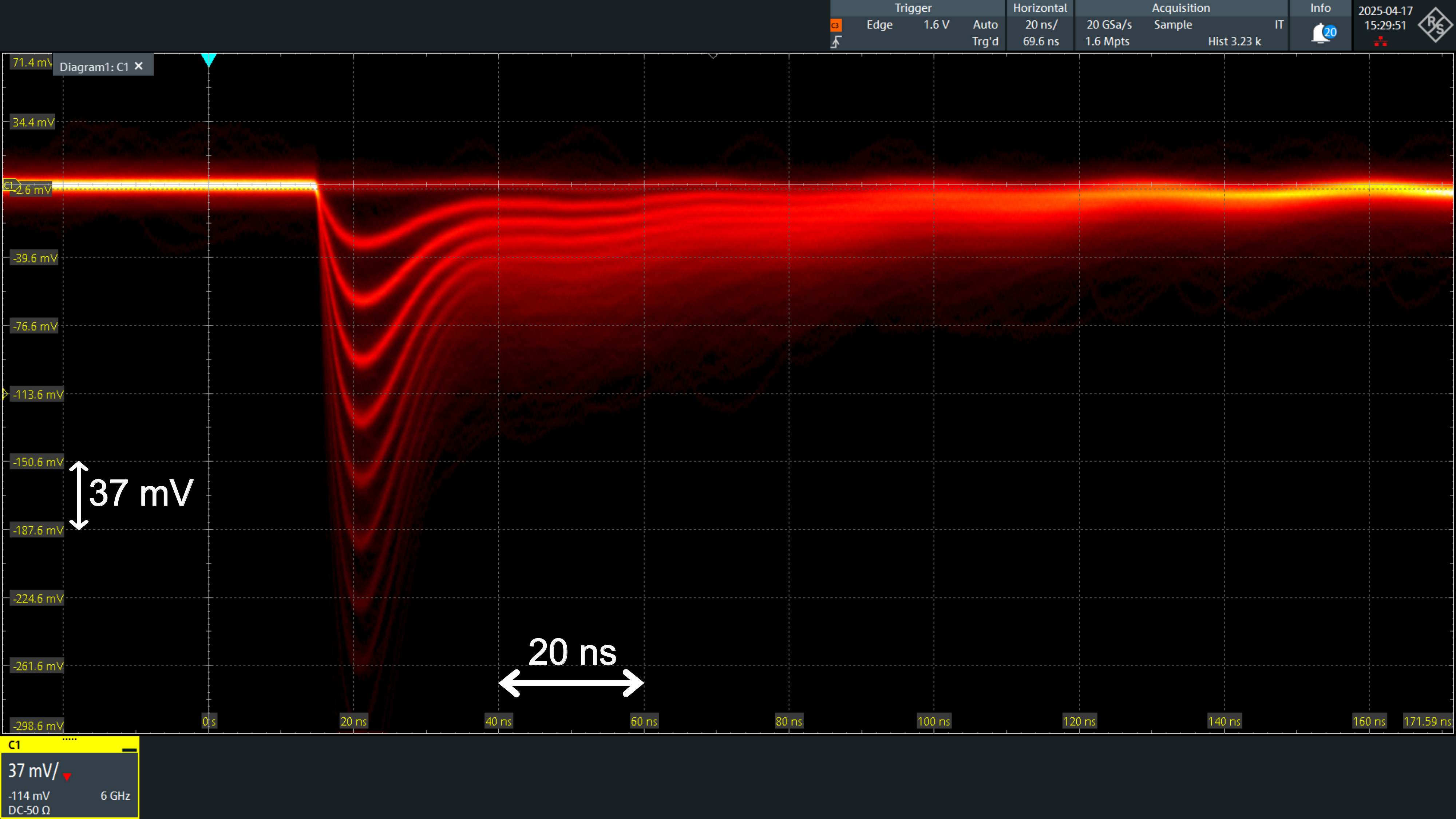}
    \includegraphics[width=0.48\linewidth]{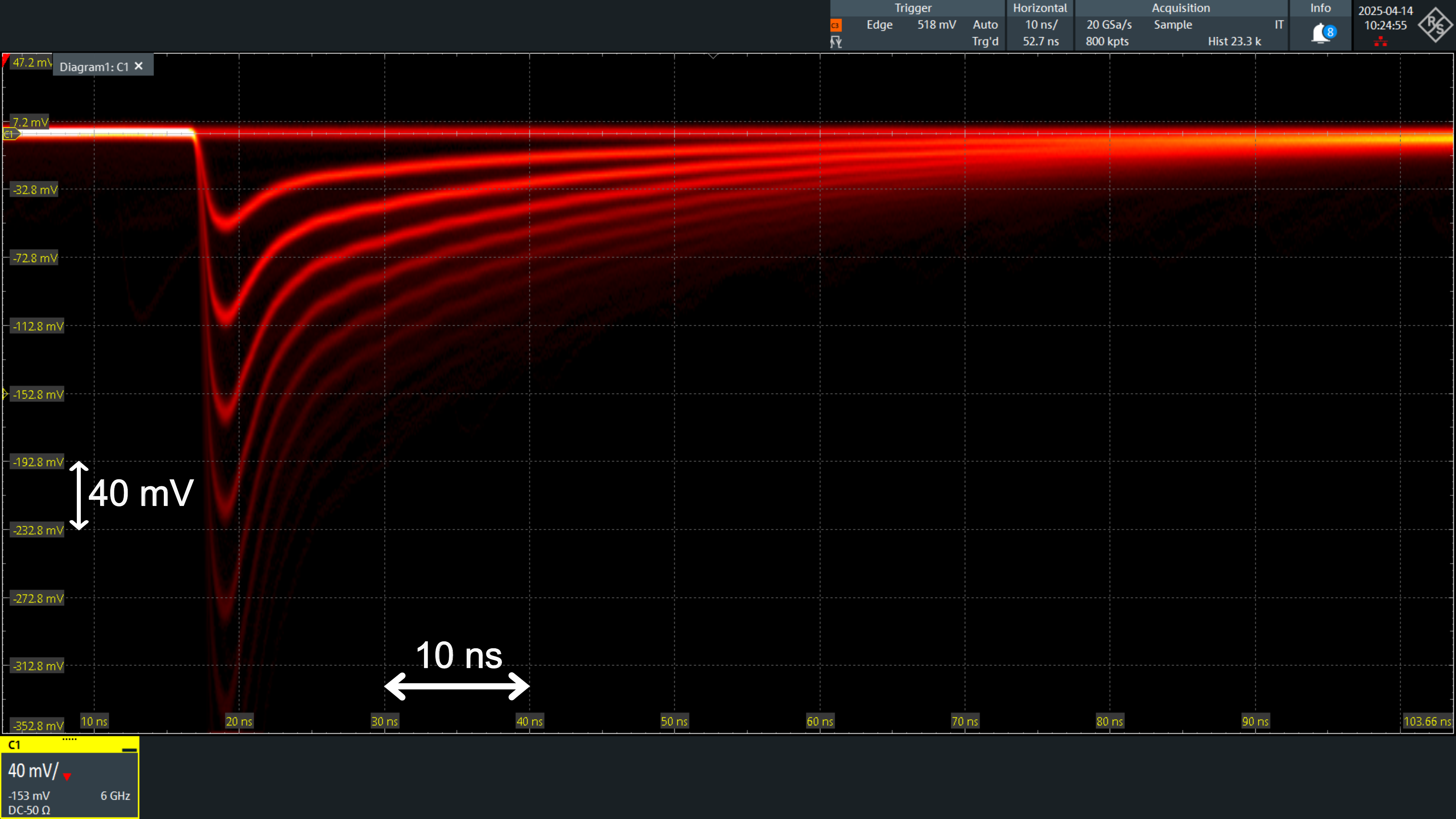}
    \caption{Photon counting at $T=80\ K$ (left) and $T=300\ K$ (right) using the Hamamatsu S13360-1350 sensor.}
    \label{fig:80_300_K_photoncounting}
\end{figure}

To illuminate the SiPM, we used an optical fiber cable which transmits 405 nm laser signals controlled by the Hamamatsu PLP-10 pulser (70 ps FWHM pulses). 
The end of the optical fiber was fixed close ($\sim 1 \ cm$ distance) to the SiPM active area. 
The PLP-10 pulser output trigger signal was employed as time reference. 
Figure \ref{fig:80_300_K_photoncounting} shows the amplifier capability of distinguishing the different amount of impinging photons signals coming from the S13360-1350, both at ambient and cryogenic temperature. 

Using equation (\ref{eq:i_in_sipm_rs}), the SiPM signal can be expressed in simplified form when ($Z_{in}+R_s$) is negligible \cite{DUNE_amplifier}:

\begin{equation}
    i_{SiPM}\simeq V_{OV}\left(C_q\delta(t)+C_d\frac{e^{-t/\tau_s}}{\tau_s}\right)
    \label{eq:sipm_signal}
\end{equation}

\noindent
where $V_{OV}$ is the over-voltage applied to the SiPM and $\tau_s=(C_d+C_d)R_q$.
The signal described in equation (\ref{eq:sipm_signal}) is made of two parts: the first is a fast current spike, the second is an exponential decay due to the SiPM cell recharge. 
The first part can be better described by the inverse Laplace of equation (\ref{eq:i_in_sipm_rs}), where a series impedance to the SiPM slows down this signal.
The response to the $\delta (t)$ changes based on how much impedance is in series to the SiPM. 
The larger the series impedance, the longer the signal fall time becomes, which also causes the pulse peak to broaden and decrease in amplitude.

Figure \ref{fig:80_300_K_photoncounting} shows an oscilloscope trace of an amplified SiPM signal, which is the convolution of the SiPM input signal and the amplifier time response: there is the slow fall slope due to the cell recharge and the peak due to the amplification of the $\delta(t)$ part of the input signal. 
Between $80\ K$ and $300\ K$, the main difference in signal shape is the tail duration. At cold temperatures the signal tail is longer (tens of ns of difference), due to the quenching resistor technology, whose temperature gradient makes the resistor increase at lower temperatures.

\begin{figure}[hbtp]
    \centering
    \includegraphics[width=0.48\linewidth]{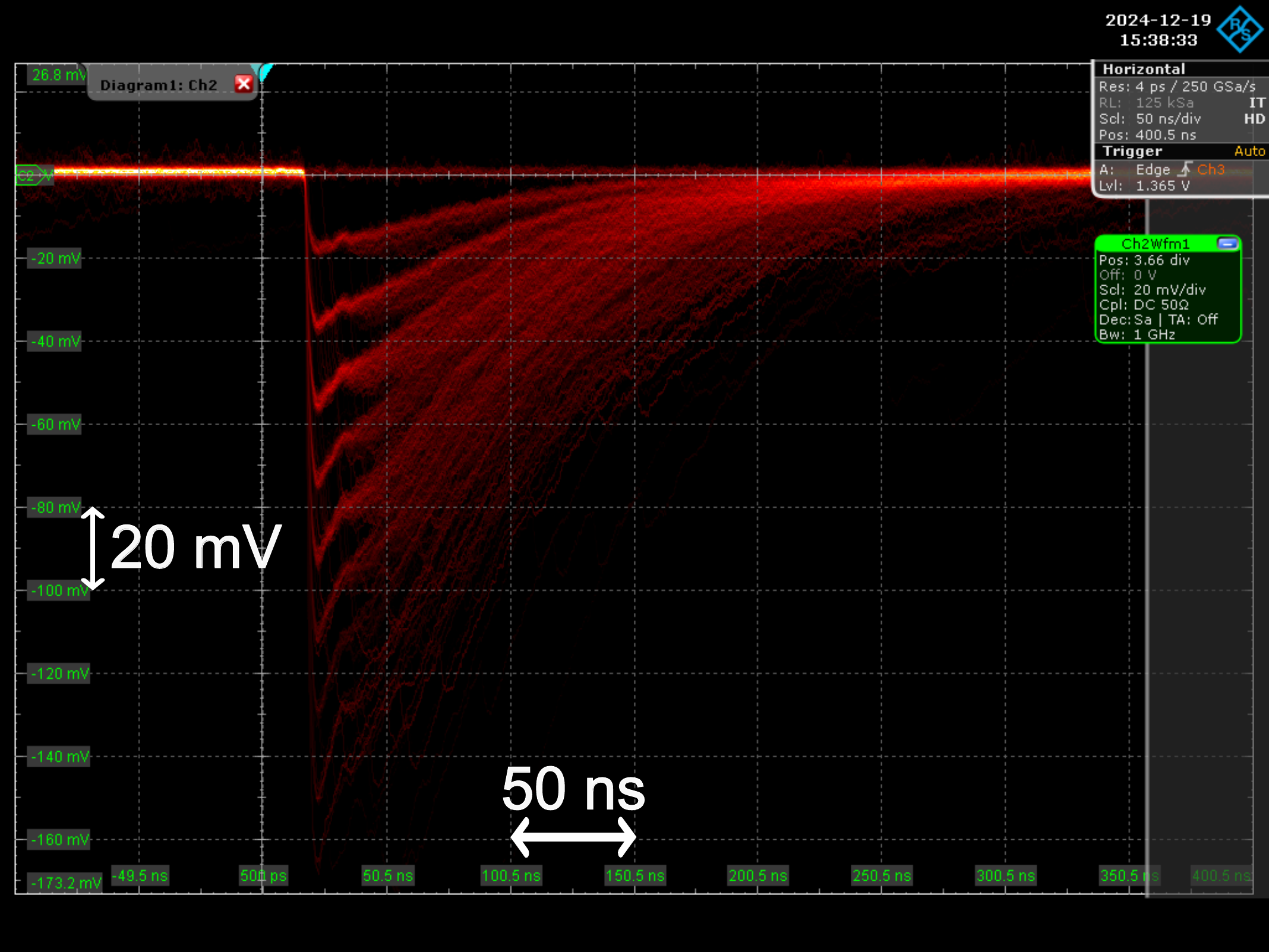}
    \caption{Photon counting at $T=80\ K$ using the Hamamatsu S13360-3050 sensor.}
    \label{fig:80K_photoncounting_3x3}
\end{figure}

The Hamamatsu S13360-3050 was tested only at cryogenic temperatures, since its large area makes the noise due to DCR high enough to make the single photon signals barely visible at room temperature (see figure \ref{fig:80K_photoncounting_3x3}).

\subsection{Jitter measurements}
\label{sec:jitter_measurements}
The SiPM was operated with a range of $V_{OV}$ values.

\begin{figure}[hbtp]
    \centering
    \includegraphics[width=0.48\linewidth]{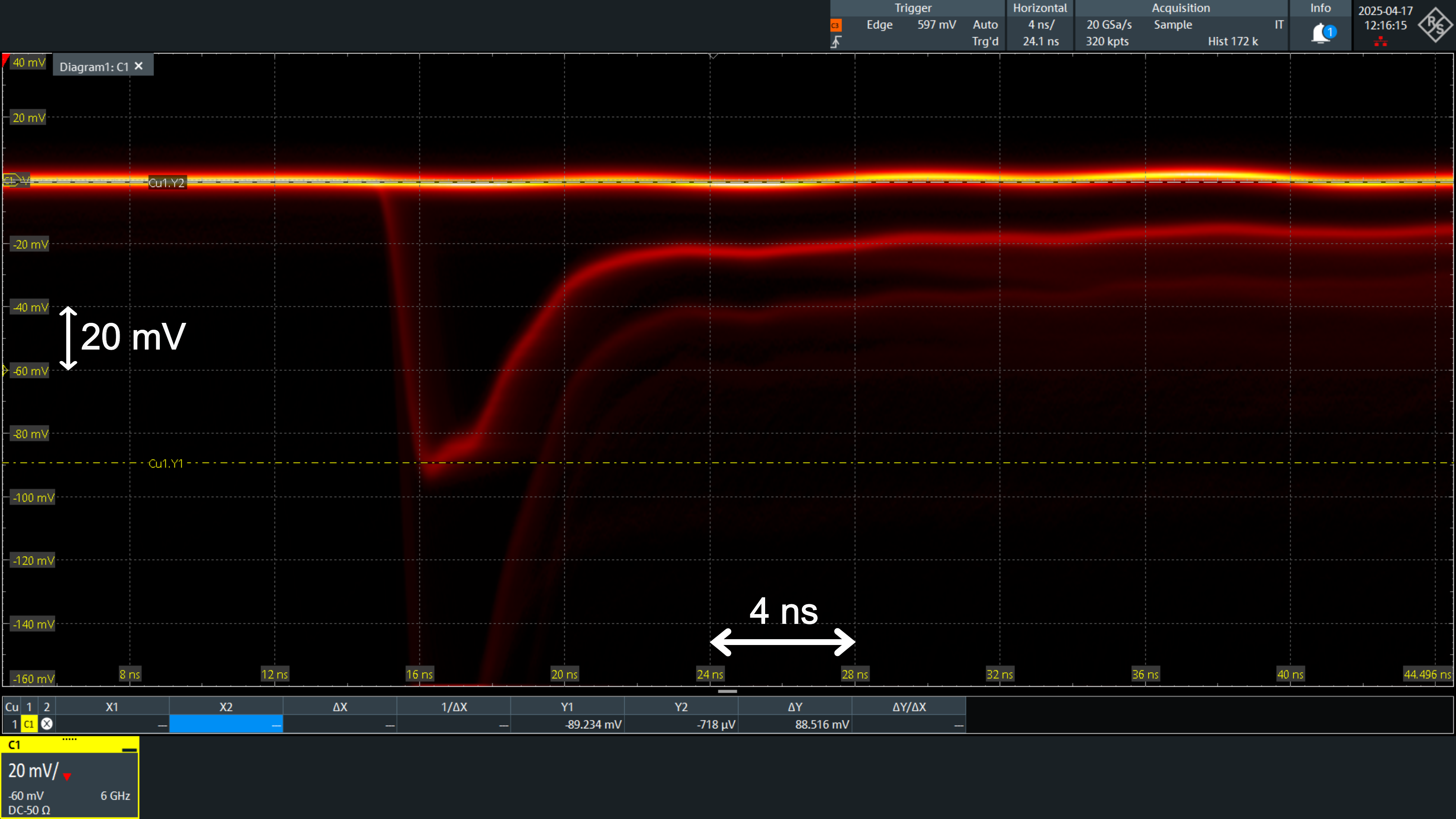}
    \includegraphics[width=0.48\linewidth]{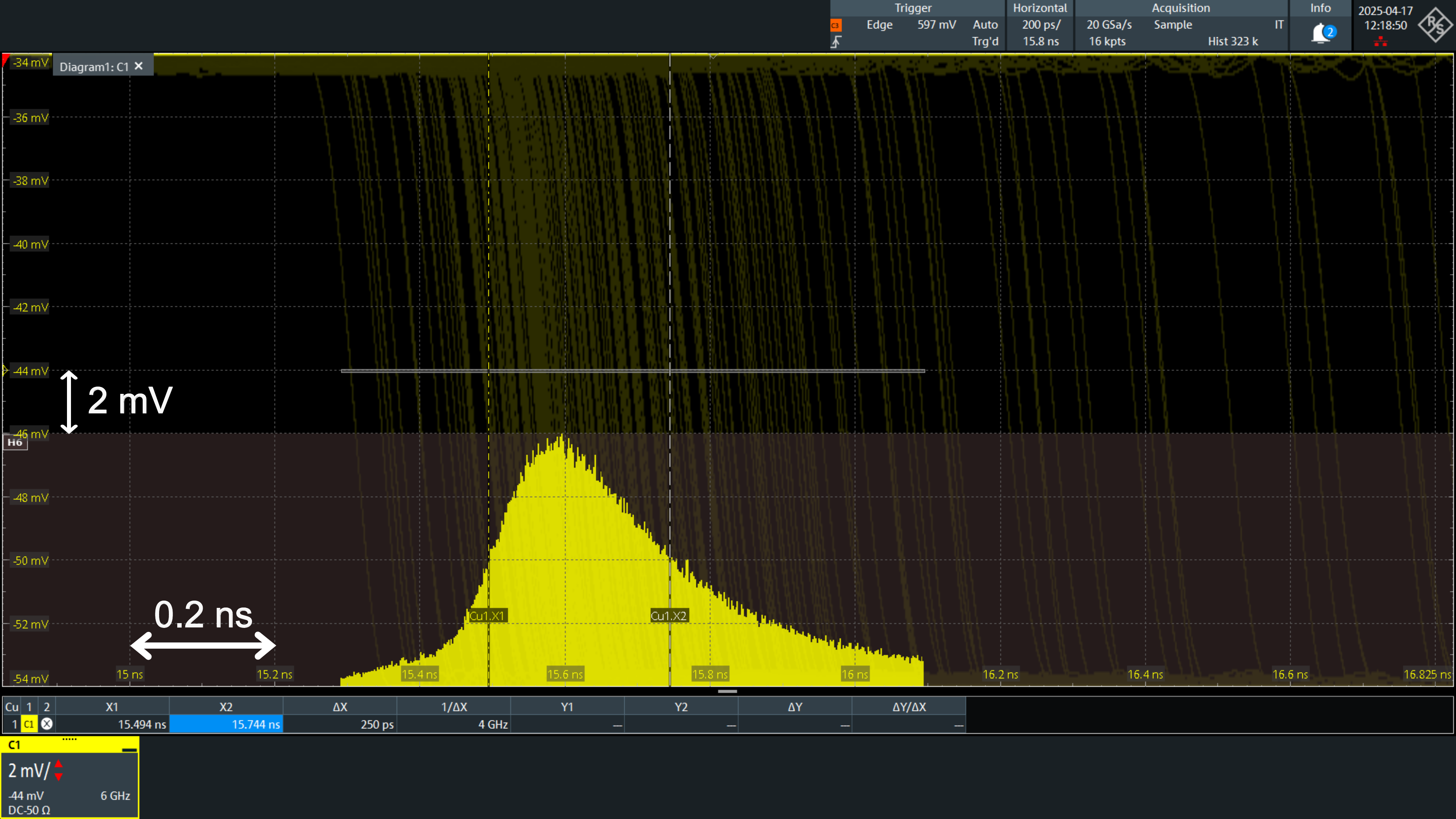}
    \caption{Output signal (left) and jitter measurement (right) of SiPM signals (Hamamatsu S13350-1350). These measurements were performed at $80\ K$.}
    \label{fig:SiPM_jitter_measurement}
\end{figure}

For both ambient temperature and 80 K, a histogram was generated from the distribution of signal arrival times, where the arrival time -- measured with respect to the trigger output of the laser -- was defined as the instant the signal reached half of its amplitude.
The shape of this histogram gives information on the jitter. 
Figure \ref{fig:SiPM_jitter_measurement} shows the process of measuring the signal amplitude and formation of the histogram. 
It can be observed that the histogram follows a Gaussian-like distribution with a rightward skew, indicated by the longer right tail.
This effect is more prominent at low temperatures and could be associated to the formation of charges near the surface of the active region, located farther from the depletion region, which take more time to reach the high-electric field region and start the avalanche \cite{histogram_long_tail}. 

Because of this tail, to evaluate jitter we could either compute the standard deviation value of the timestamps that create the histogram or measure the FWHM of the histogram. We chose the latter, as it provides a better representation of the dispersion in the signal arrival time.

We also measured the jitter contribution due to the amplifier alone, so that we could see the impact of the electronics on this measurement. To do so, the laser was turned off and the test signal was applied to the amplifier input.
In this case, there is no SiPM attached at the input, as we verified that it does not affect the jitter measurements with $R_q/N \gg R_b$ (as discussed in section \ref{sec:stability_amp}). 
We omitted the SiPM at the input to avoid signal distortion, which would arise from part of the signal flowing into the SiPM and part into the amplifier’s feedback network.
The rise time is however similar with or without the SiPM at the input, hence the jitter measurements are unaffected.

The test signal amplitude was tuned to match the amplitude of SiPM signals at different over-voltages, despite the much shorter tail.
This was done because jitter also depends on the signal amplitude, as it can be seen in equation (\ref{eq:jitter}). 
A larger signal amplitude results in reduced jitter.
Therefore, a non-constant jitter contribution from the amplifier is expected when using different test signals.
The rise time of the SiPM signal also needs to be considered, as it is slower than the test signal that was used for the previous measurements. 
Its rise time slightly depends on temperature, and it was measured to be about $\sim1 \ ns$, higher than the $300-500 \ ps$ of the amplified test signal.
To emulate the SiPM signal rise time we put in series to $C_t$ a resistor of $910\ \Omega$, that integrates the test signal, making it slower.
The jitter of the amplified test signal is only due to electronics noise, hence a Gaussian distribution is observed.

\begin{figure}[hbtp]
    \centering
    \includegraphics[width=0.75\linewidth]{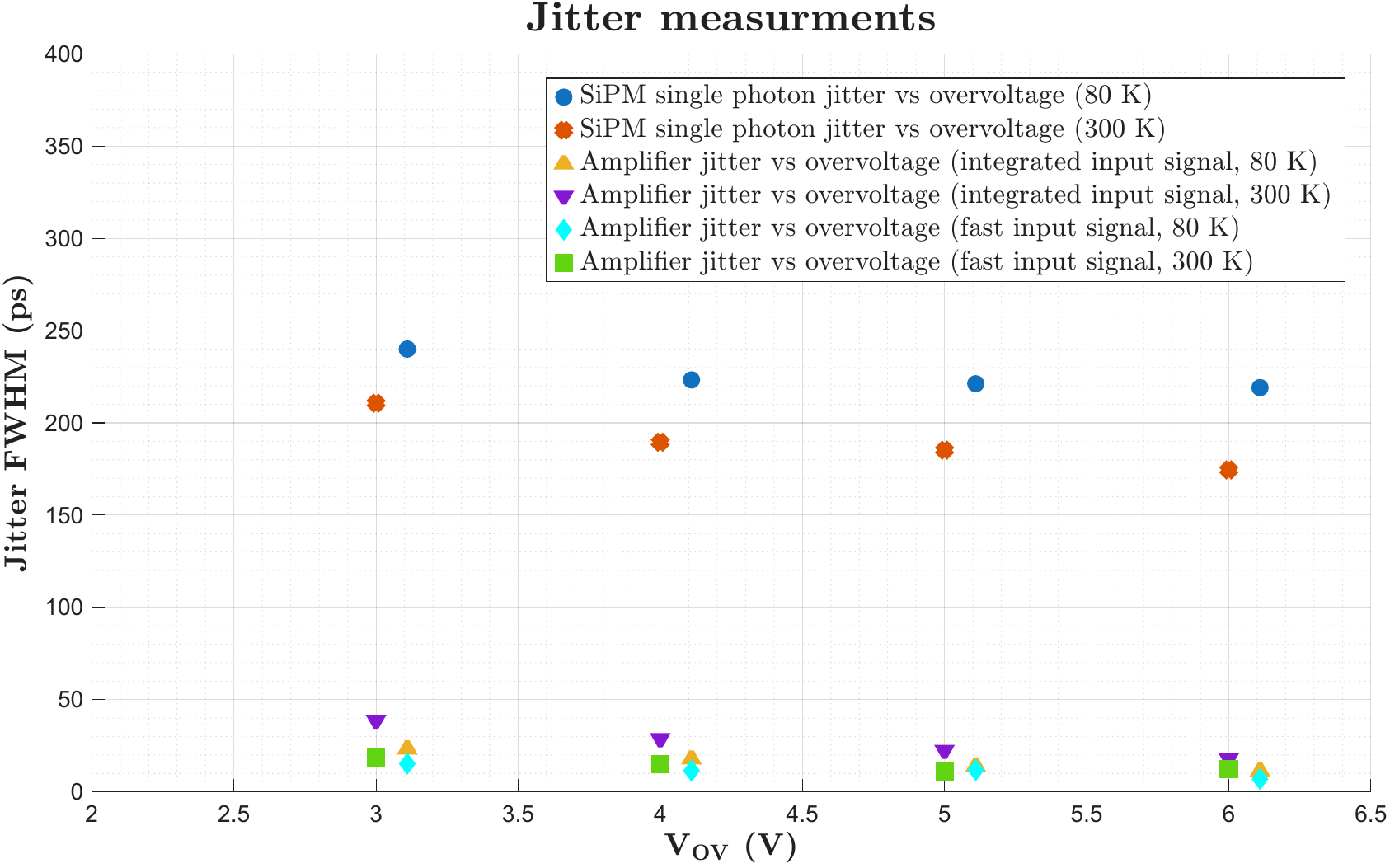}
    \caption{Summary of all the jitter contribution measurements, both at ambient temperature (300 K) and in liquid nitrogen (80 K). In this plot are the SiPM single photon jitter measurements and the contribution to the jitter of the amplifier alone. The amplifier jitter was measured either with a very fast (95 ps rise time) signal or with an integrated signal, with the latter better simulating the SiPM rise time.}
    \label{fig:summary_jitter}
\end{figure}

In figure \ref{fig:summary_jitter} are summarized all jitter measurements. 
In these measurements, we removed the systematic jitter contributions. 
The amplifier's performance was isolated by accounting for the $4.5\ ps_{RMS}$ intrinsic jitter of the signal generator and oscilloscope measurement chain.
Similarly, for the SiPM measurements, the 30 ps FWHM contribution from the laser pulse generator was removed from the total measured jitter.
Table \ref{tab:jitter_contributions} summarizes all systematic jitter sources.
We also measured the amplifier jitter contribution with no integration of the input signal whatsoever, so to evaluate the full performance of the amplifier. 

\begin{table}[tbp]
    \centering
    
    \caption{Systematic jitter contributions.}
    \vspace{2mm}
    
    \begin{tabular}{c|c|c}
        \toprule
         \textbf{Jitter contributions} & \textbf{$\mathbf{\sigma}$ ($\mathbf{ps})$} & \textbf{FWHM ($\mathbf{ps}$)} \\
         \midrule
         Signal generator+oscilloscope & $4.5\ ps_{RMS}$ &  $10.6\ ps$\\
         Laser pulse generator & $29.7\ ps_{RMS}$ & $70\ ps$\\
        
        \bottomrule
    \end{tabular}
    
    \label{tab:jitter_contributions}
\end{table}

As can be seen in figure \ref{fig:summary_jitter}, at all over-voltage values considered --- down to 3 V --- the amplifier jitter contributions are fairly negligible if compared to the SiPM signal jitter measurements. The jitter measurements at 80 K are slightly worse than at ambient temperature because of that tail in the Gaussian distribution that was previously mentioned in this section. 

The measured amplifier jitter aligns with results reported in literature using alternative front-end electronics for SiPM signal amplification \cite{article_comparison_jitter}. 
While a direct comparison is limited -- as jitter performance is sensitive to the specific rise time and characteristics of the SiPM used -- the contribution of approximately 20 ps FWHM observed for our 1.7 $mm^2$ device is consistent with or below state-of-the-art findings for sensors of this scale.

The impact that the amplifier has on jitter measurements can be evaluated with a percentage value $W$ as follows:

\begin{equation}
    W(\%)=\left(\frac{\sqrt{FWHM_{t_{AMP}}^2+FWHM_{t_{SiPM}}^2}}{FWHM_{t_{SiPM}}}-1\right)\cdot100
\end{equation}
\noindent
The amplifier impact on jitter diminishes at higher over-voltages (higher signal amplitudes). At $V_{OV}>3\ V$, $W<1.7\%\ @\ 300\ K$ and $W<0.5\%\ @\ 80\ K$. All these values are computed using the amplifier jitter measured with the integrated input signal. 

\section{Conclusions}
We have designed a fast, low-noise amplifier capable of operating between 80 K and 300 K, which can be used to characterize SiPMs over a broad temperature range.
This amplifier consists of a two-stage amplifier; the first stage is a Heterojunction Bipolar Transistor (HBT), whose output is fed into the second stage, a Current-Feedback Operational Amplifier (CFOA).

We simulated the complete network and validated it experimentally with a $\delta(t)$-like current test signal, achieving satisfactory performance at both ambient and cryogenic temperatures. 
Depending on the temperature operation, the voltage output signal exhibits a rise time of approximately $ 380-540 \ ps$ and fall time of about $ 1.5-2.5\ ns$. 
For single-photon signals modeled as pure exponentials and producing a $1.5\ {Me^-}$ charge, typical of Hamamatsu S13360 SiPMs with large cell sizes $50\ \mu m$–$75\ \mu m$ operated at $3\ V_{OV}$, the amplifier’s equivalent noise charge is $<200\ {ke^-}$, corresponding to an equivalent noise photon level of $<0.13$ photons. 

Other designs, such as the one presented in \cite{article_comparison_jitter}, offer a SiPM readout with similar timing performance at ambient temperature, confirming that the achieved performance is aligned with the state-of-the-art for SiPM readouts. 
However, our setup operates in a closed-loop configuration, so the gain is fixed at the transimpedance value, and it also functions reliably at cryogenic temperatures down to 80 K. 

This performance was achieved by characterizing the CFOA over a wide temperature range and by using an HBT with a carefully adjusted bias current (via an external voltage $V_{POS}$) to maintain a stable bandwidth with minimal ringing.  
In addition, appropriate passive components and a purpose-designed dielectric were selected to ensure robust operation.  
As a result, the amplifier provides reliable and reproducible performance from ambient down to cryogenic temperatures. 
This allows accurate SiPM characterization under the cooling conditions required to suppress dark counts.

\appendix

\section{Current Feedback Operational Amplifier In-Depth Analysis}
\label{app:A}
 
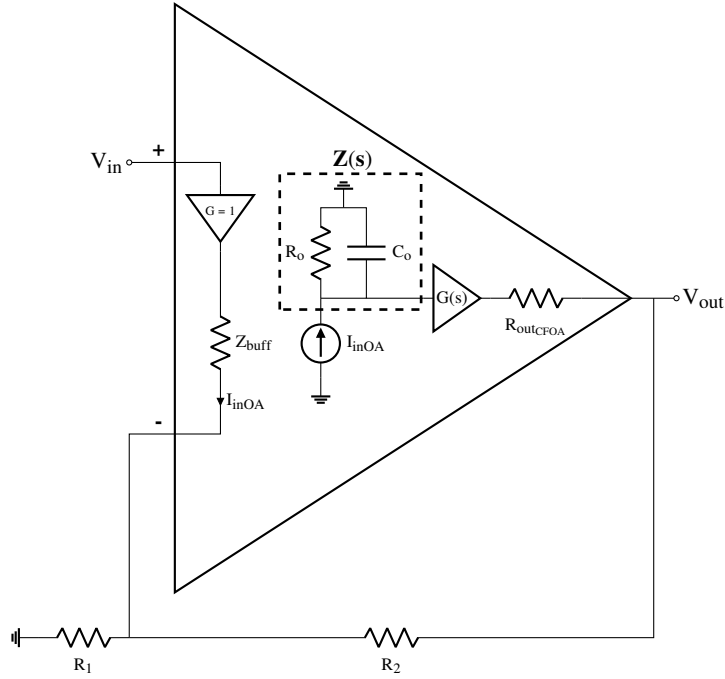
\begin{figure}[hbtp]

    \centering
    \ctikzset{amplifiers/thickness=2, transistors/thickness=3, amplifiers/scale=1.5}
    \begin{circuitikz}[scale=0.6 ,transform shape, american] 

        \coordinate (Tip) at (10, 1.5);       
        \coordinate (TopLeft) at (0, 8);    
        \coordinate (BotLeft) at (0, -5);    
        \coordinate (InvInput) at (0, -1.5);  
        \coordinate (NonInvInput) at (0, 4.5); 
    
        \draw [thick] (TopLeft) -- (BotLeft) -- (Tip) -- cycle;
    
        \node at (NonInvInput) [right, xshift=-10pt] [above] {\Large \textbf{+}};
        \node at (InvInput) [right, xshift=-10pt] [above] {\Large \textbf{-}};
    
        
        \draw (NonInvInput) -- ++(-1,0) node[ocirc]{} node[left]{\Large $V_{in}$};
        \draw (NonInvInput) -- ++(1, 0) -- ++(0,-0.5) node[rotate=270][buffer, anchor=in, scale=0.7](buf1){};
        \node at (buf1.center) [yshift=5pt] {\scriptsize $G=1$};
        
        \draw (buf1.out) to[R, l=$Z_{buff}$, i=$I_{inOA}$] (InvInput -| buf1.out) -- (InvInput);
    
        \coordinate (InternalNode) at (3.2, 1.5);
        
        \draw (InternalNode) -- ++(0,-0.25) to[I, l=$I_{inOA}$, invert] ++(0,-1.5) node[ground]{};
        \draw (InternalNode) to[R, l=$R_o$] ++(0,2) -- ++(0.5,0);
        \draw (InternalNode) -- ++(1,0) coordinate(C_o) to[C, l_=$C_o$] ++(0,2) -- ++(-0.5,0) node[rotate=180][ground]{};
        \draw (C_o) -- ++(1.25,0) coordinate(out_voltage);
        
        \draw (out_voltage) node[buffer, anchor=in, scale=0.7](buf2){};
        \node at (buf2.center) [xshift=-3pt] {$G(s)$};
        \draw (buf2.out) to[R, l_=$R_{out_{CFOA}}$] ++(2,0) coordinate(outR);
        
        \draw (outR) -- (Tip);
    
        \draw (Tip) -- ++(0.5, 0) coordinate(feedback) -- ++(0.5,0) node[ocirc]{} node[right]{\Large $V_{out}$};
    
        \draw (InvInput) -- ++(-1, 0) -- ++(0,-4.5) coordinate (ExtNode);
        \draw (ExtNode) to[R, l^=$R_1$] ++(-2,0) node[rotate=270][ground]{};
    
        \draw (ExtNode) to[R, l_=$R_2$] (ExtNode -| feedback) -- (feedback);

        \coordinate (lowerLeftRect) at ($(InternalNode)+(-0.9,-0.25)$);
        \coordinate (upperRightRect) at ($(lowerLeftRect)+(3.1,3)$);
        \coordinate (ZsWriting) at ($(upperRightRect)+(-1.5, 0.3)$);
        \draw[line width=1pt, dashed] (lowerLeftRect) rectangle (upperRightRect);
        \node at (ZsWriting) {\Large $\mathbf{Z(s)}$};
        
    \end{circuitikz}

    \caption{Detailed schematic of the internal structure of a CFOA, used in a non-inverting feedback configuration.}
    \label{fig:CFOA}
\end{figure}

Current Feedback Operational Amplifiers (CFOAs) provide high slew-rates and very high bandwidth, making these devices ideal for high-speed applications.
A simplified circuit model of a CFOA, shown in figure \ref{fig:CFOA}, contemplates the presence of a buffer between the two inputs, with a high input impedance on the non-inverting node, while the inverting node has a low impedance that can be called \enquote{$Z_{buff}$}. 

When the CFOA is operated without feedback, the device collects small input current signals $I_{inOA}$ that flow inside the buffer and are amplified by the transimpedance $Z(s)$, which represents the gain of the CFOA.
The transimpedance gain can be expressed as $Z(s)=R_0  \parallel  \frac{1}{s C_0} = \frac{R_0}{1 + s C_0 R_0}$, while the  buffer transfer function is $G(s)=\frac{1}{1+s\tau_{HF}}$, a buffer with a very high bandwidth. 
The input current $I_{inOA}$, flowing through this network, produces an output voltage equal to ($R_{out_{CFOA}}$ is neglected): 
\[
V_{out} = I_{inOA} \cdot Z(s) \cdot G(s)= I_{inOA} \cdot \frac{R_0 }{1 + s C_0 R_0} \cdot \frac{1}{1 + s\tau_{HF}}
\]

In a feedback configuration, the CFOA loop gain $T_{CFOA}$ is largely determined by just one external parameter, its feedback resistor ($R_2$ in figure \ref{fig:CFOA}). 
Contrary to a conventional voltage-feedback OA, the closed-loop gain plays a minor role in determining the loop gain, which can be expressed as ($R_{out_{CFOA}}$ is neglected) \cite{cfoa_book}:

\begin{equation}
    T_{CFOA}(s)=-\frac{Z(s)}{R_2\cdot \left(1+\frac{Z_{buff}}{R_2 \parallel  R_1}\right)} \frac{1}{1 + s\tau_{HF}}\simeq - \frac{R_0}{R_2}\frac{1}{1+sC_oR_o}\frac{1}{1 + s\tau_{HF}} \simeq - \frac{1}{sC_oR_2}\frac{1}{1 + s\tau_{HF}}
    \label{eq:T_CFOA_new}
\end{equation}

\noindent
where $Z_{buff}\sim 10\ \Omega$ is a small impedance and it can usually be neglected, except when second-order effects are considered, and $sC_oR_o\gg 1$.
When operating a CFOA, one fixes the bandwidth with only the feedback resistor and the closed-loop gain can be whatever without greatly affecting the device bandwidth.

When used in a non-inverting configuration, neglecting the direct transmission term and assuming $C_oR_2\gg \tau_{HF}$, the CFOA gain is:
\begin{equation}
   \frac{V_{out}(s)}{V_{in}(s)}\simeq \left(1+\frac{R_2}{R_1}\right) \cdot\frac{-T_{CFOA}(s)}{1-T_{CFOA}(s)}\simeq\left(1+\frac{R_2}{R_1}\right)\frac{1}{1+sC_oR_2}\frac{1}{1+s\tau_{HF}}
   \label{eq:closed_loop_CFOA}
\end{equation}

\section{LMH6702 bandwidth in-depth study when operated with high feedback resistor}
\label{app:B}
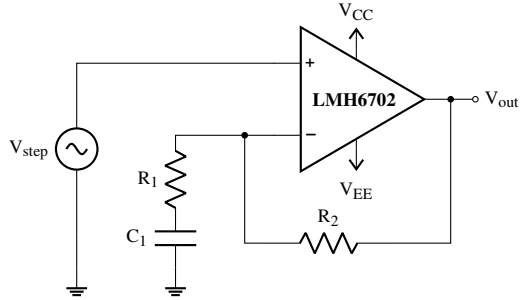
\begin{figure}[hbtp]

    \centering
    \ctikzset{amplifiers/thickness=2, transistors/thickness=3, amplifiers/scale=1.5}
    \begin{circuitikz}[scale=0.65 ,transform shape, american]       
          \draw node[op amp, anchor=+,noinv input up](opamp){$\textbf{LMH6702}$};
          \draw (opamp.-) to[short] ++(-2,0) -- ++(0,-0.3) to[R,l_=$R_1$] ++(0,-1.2) to[C, l_=$C_1$] ++ (0,-1.2) node[ground]{} coordinate(gnd2); 
          \draw (opamp.-) ++(-0.6,0) node[circ]{} coordinate(R2_node) to[short] ++(0,-2.2) to[R=$R_2$]  ++(3.4,0) -| (opamp.out) node[circ]{};
          \draw (opamp.up) -- ++(0,0.2) node[vcc]{$V_{CC}$};
          \draw (opamp.down) -- ++(0,-0.2) node[vee]{$V_{EE}$};
          \draw (opamp.out) -- ++(0.5, 0) node[ocirc,label=right:$V_{out}$]{} ; 
          \draw (opamp.+) -- ++(-4,0) to[vsourcesin, l_=$V_{step}$] ++(0,-3.5) coordinate(gnd); 
          \draw (gnd) -- (gnd2 -| gnd) node[ground]{};
    \end{circuitikz}

    \caption{Schematic of the test bench for the LMH6702 bandwidth in-depth study.}
    \label{fig:LMH6702_highfeedback_evaluation}
\end{figure}

In the ODP configuration, the local CFOA (LMH6702) bandwidth must be evaluated independently using a dedicated test bench (shown in figure \ref{fig:LMH6702_highfeedback_evaluation}). 
This is necessary because high feedback resistors ($R_2$) are not conventionally used, and the CFOA datasheet provides no information on its behavior under these conditions.
The test bench consists of the CFOA alone operated in a non-inverting configuration. 
A step voltage signal is injected at the CFOA non-inverting input, while a resistor $R_2=9.1\ k\Omega$ is used as the CFOA feedback and $R_1=20\ \Omega$ is attached between the CFOA inverting input and $C_1=150\ nF$, which is finally connected to ground.  

\begin{figure}[hbtp]
    \centering
    \includegraphics[width=0.7\linewidth]{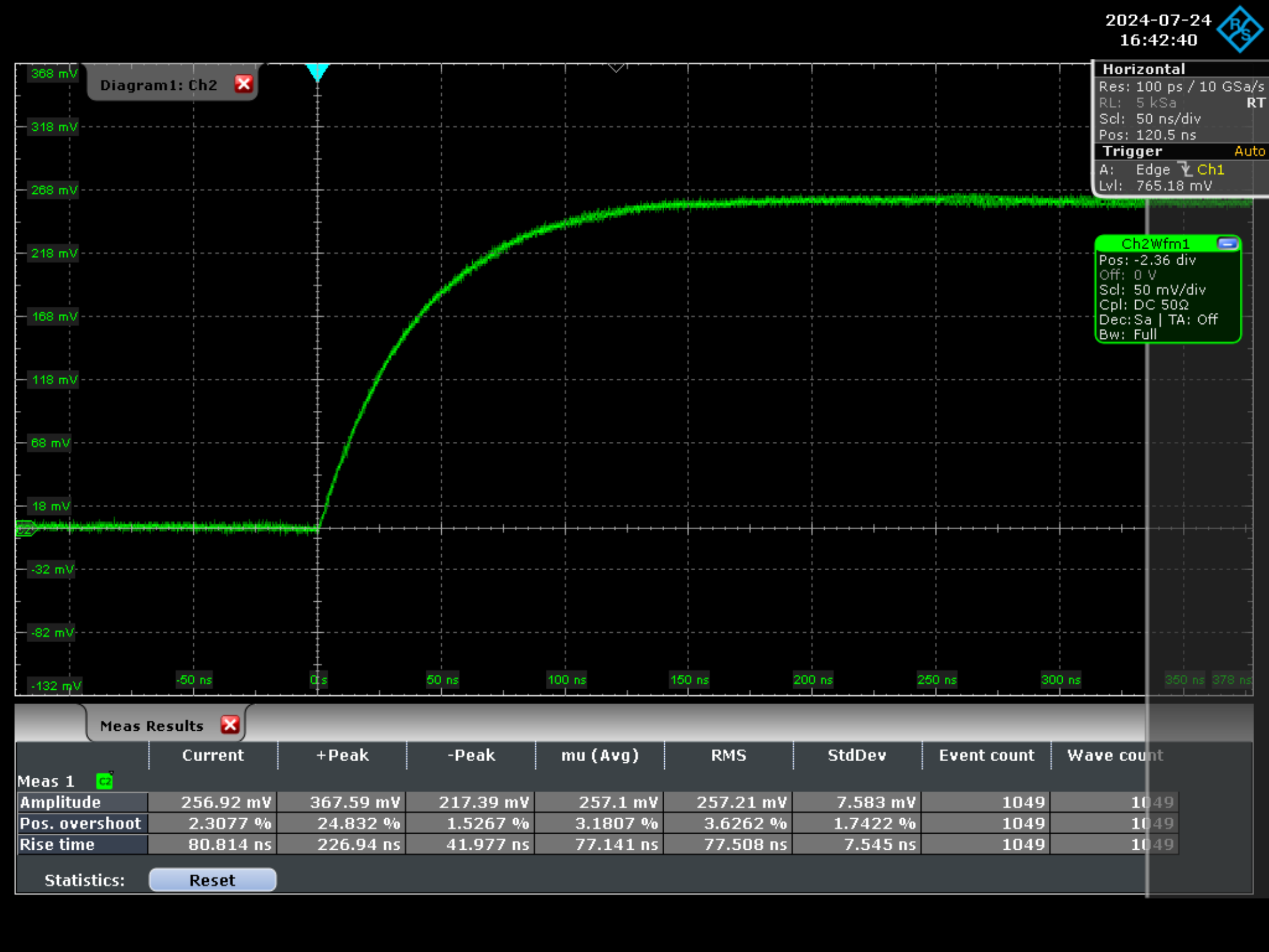}
    \caption{LMH6702 characterization with a high resistor in feedback ($9.1\ k\Omega$). This is the response of the CFOA to a step voltage. Rise time of $\sim 77\ ns$ and amplitude of $\sim 514\ mV$.}
    \label{fig:LMH6702_characterization}
\end{figure}

Both gain and bandwidth of the LMH6702 output signal were studied. From the gain measurement, it could be assessed that there is some deterioration in the ideal gain value. Knowing that the input signal is $V_{in}=1.25 \  mV$, $V_{out}$ should ideally be $V_{out}=\left(1+R_2/R_1\right)\cdot V_{in}=570 \  mV$ ($G_{CFOA_{ideal}}=1+R_2/R_1=456\  V/V$). 

Real measurements, shown in figure \ref{fig:LMH6702_characterization}, indicate an output signal of $\sim 514\ mV$, corresponding to a measured value of $257\ mV$ that must be multiplied by two due to the oscilloscope's $50\ \Omega$ termination, which halves the actual amplitude.
This means that the real gain is approximately 10\% lower than the ideal gain ($G_{CFOA_{real}}\simeq411\  V/V$).  
This phenomenon can be explained by the CFOA loop gain, which can be calculated using equation (\ref{eq:T_CFOA_new}), with $R_o\simeq 110\ dB\Omega$ — as specified in the manual — and assuming $Z_{buff}\simeq 10\ \Omega$. The resulting CFOA loop gain is $T_{CFOA}\simeq-23\ V/V$, hence the output gain is:
\begin{equation}
    G_{CFOA_{corrected}}=\frac{-T_{CFOA}}{1-T_{CFOA}}\left(1+\frac{R_2}{R_1}\right)\simeq 0.958\cdot \left(1+\frac{9.1\ k\Omega}{20\ \Omega}\right)\simeq 437\ V/V
\end{equation}
$G_{CFOA_{corrected}}$ is still $\sim6\%$ higher than $G_{CFOA_{real}}$. 

Another factor to be considered is the CFOA output impedance. 
The manufacturer indicates that this is $R_{out_{CFOA}}=30\ m\Omega$ with $R_{2}=R_1=237\ \Omega$.
By evaluating the loop gain for both the manufacturer-recommended resistor configuration and the one adopted in our design (using equation (\ref{eq:T_CFOA_new}), with $R_o\simeq110\ dB$ and assuming $Z_{buff}\approx 10 \ \Omega$), we can estimate an appropriate value for the output resistor.
With our configuration of resistors, $R_{out_{CFOA}}\simeq 1.7\ \Omega$.
Hence, we can see the CFOA as a voltage generator $V=V_{in}\cdot G_{CFOA_{corrected}}$ with in series $R_{out_{tot}}=R_{out_{CFOA}}+50\ \Omega \simeq 51.7\ \Omega$. 
The output voltage is then partitioned between $R_{out_{CFOA}}$ and the scope input resistor $R_{scope}=50\ \Omega$. 
In this way we obtain that the computed gain is:

\begin{equation}
    G_{CFOA_{corrected\ 2}}=G_{CFOA_{corrected}}\cdot \frac{R_{scope}}{R_{scope}+R_{{out}_{tot}}} (\cdot2) \simeq 430\ V/V
\end{equation}

\noindent
$G_{CFOA_{corrected\ 2}}$ still is not the same value as $G_{CFOA_{real}}$, with the former being $\sim 5\ \%$ higher. 

In addition, the input buffer impedance of the CFOA, $Z_{buff}$, also may play a significant role in modifying the CFOA output gain.
Using equation (\ref{eq:T_CFOA_new}) and considering $Z_{buff}$, the CFOA gain should be written as: 

\begin{equation}
    G_{CFOA_{withZbuff}}=\frac{R_o\left(1+R_2/R_1\right)}{R_2\left(1+\frac{Z_{buff}}{R_2 \parallel R_1}\right)}\cdot\frac{1}{1+\frac{R_o}{R_2\left(1+\frac{Z_{buff}}{R_2 \parallel R_1}\right)}} 
\end{equation}

Unfortunately, since the manufacturer does not provide a value for $Z_{buff}$, it is difficult to estimate the effect of this impedance on the gain.

The measured gain, regardless of its underlying cause, can be modeled considering a slightly lower closed-loop gain, with $R_1\simeq22\ \Omega$ instead of $20\ \Omega$. 

Setting aside the amplitude, from the analysis of the oscilloscope sampling in figure \ref{fig:LMH6702_characterization}, the CFOA closed-loop bandwidth can be simply extrapolated from from the signal rise time measurement and it is $\sim 4.5\ MHz$, which means that $C_oR_2\simeq 35\  ns$. 
This measured value agrees with the approximate value that can be estimated from the CFOA manual.
Together with the corrected value of $R_1$, we have all that is necessary to compute the amplifier loop gain in equation (\ref{eq:loop_gain_open_loop_configuration}). 

\section{Measurement of the parasitic capacitance \texorpdfstring{$\mathbf{C_p}$}{Cp}}
\label{app:C}

\begin{figure}[hbtp]

    \centering
    \ctikzset{amplifiers/thickness=2, transistors/thickness=3, amplifiers/scale=1.2}
    \begin{circuitikz}[scale=0.6 ,transform shape, american]       
          \draw (-1,0) coordinate(in) to[short, i =$I_{in}$] ++(2,0) coordinate(V_B); 
          \draw (in) to[C, l_=$C_t$, a^=\SI{1}{pF}, l2 halign=c] ++(-2,0) node[ocirc,label=left:Test signal]{};
          \draw (V_B) node[npn,anchor=base](bjt){$\textbf{HBT}$};  
          \draw (bjt.emitter) to[short] ++(0,-0.5) coordinate(gnd);
          \draw (bjt.collector) to[short,-*] ++(0,0.5) node[left]{$V_c$} coordinate(coll) -- ++(0,0.3) to[R=$R_c$] ++(0,2.5)  to[battery2,name=myB,l=$0.5\ V$] ++(0,1) node[ground,rotate=180]{} coordinate(Vcc);
          \draw (coll) -- ++(5,0) node[op amp, anchor=+,noinv input up](opamp){$\textbf{CFOA}$};
          \draw (coll) ++(2,0) node[circ]{} to[C=$C_x$] ++(0,3.8) coordinate(parasitc_gnd);
          \draw (coll) ++(4,0) node[circ]{} to[C=$C_p$] ++(0,3.8) coordinate(parasitc_gnd2);
          \draw (parasitc_gnd) -- (Vcc -| parasitc_gnd) node[ground,rotate=180]{};
          \draw (parasitc_gnd2) -- (Vcc -| parasitc_gnd2) node[ground,rotate=180]{};
          \draw (opamp.-) to[short] ++(-2,0) -- ++(0,-0.3) to[R,l_=$R_1$] ++(0,-1.2) to[C, l_=$C_1$] ++ (0,-1.2) node[ground]{} coordinate(gnd2); 
          \draw (opamp.-) ++(-0.6,0) node[circ]{} coordinate(R2_node) to[short] ++(0,-2.2) to[R=$R_2$]  ++(3.4,0) -| (opamp.out) node[circ]{};
          \draw (opamp.up) -- ++(0,0.2) node[vcc]{$V_{CC}$};
          \draw (opamp.down) -- ++(0,-0.2) node[vee]{$V_{EE}$};
          \draw (gnd) -- (gnd2 -| gnd) node[ground]{};
          \draw (opamp.out) -- ++(0.5, 0) node[circ]{} -- ++(0,-4.5) coordinate(FB);
          \draw (opamp.out) ++(0.5,0) -- ++(0.5,0) node[ocirc,label=right:$V_{out}$]{} ; 
          \draw (FB) to[R=$R_f$] (FB -| bjt.base) coordinate(FB2) -- (bjt.base) node[circ]{} node[above]{$V_b$};
          \draw (FB2) node[circ]{} -- ++(-0.4,0) to[R=$R_b$] ++(-1.5,0) to[C=$C_b$] ++(-1.5,0) node[ground,rotate=270]{};
    \end{circuitikz}

    \caption{Schematic of the test-bench for the $V_c$ node parasitic capacitance evaluation.}
    \label{fig:parasitic_cp_evaluation}
\end{figure}
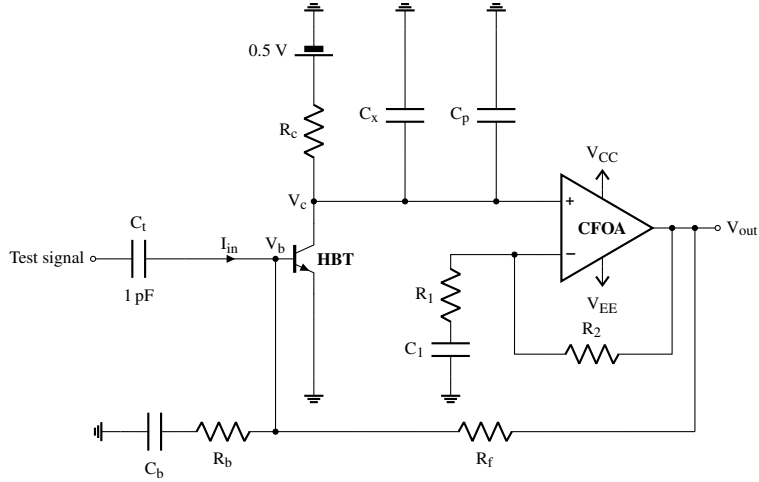

The capacitance $C_p$ is the sum of the parasitic capacitance introduced by the HBT collector, $R_c$, the CFOA non-inverting input terminal and the PCB. We minimized the latter by removing the ground plane just below the collector traces. We measured $C_p$ using the bench depicted in figure \ref{fig:parasitic_cp_evaluation}, with the employed passive and active elements listed in table \ref{tab:parasitic_eval}.

\begin{table}[tbp]
    \centering
    
    \caption{List of all active/passive elements used in the test bench for the $C_p$ capacitance evaluation.}
    \vspace{2mm}
    
    \begin{tabular}{c|c}
        \toprule
         \textbf{Network elements} & \textbf{Values}  \\
        \midrule
        \textbf{$HBT$} & $BFP640$ \\
        \textbf{$R_c$} & $100\ k\Omega$  \\
        \textbf{$R_d$} & $\infty$ \\
        \textbf{$C_d$} & 0 \\
        \textbf{$C_x$} & $0.4\ pF/0.9\ pF/1.5\ pF$ \\
        \textbf{$CFOA$} & $LMH6702/LMH6703$ \\
        \textbf{$R_1$} & $91\ \Omega$ \\
        \textbf{$C_1$} & $1140\  nF$ \\
        \textbf{$R_2$} & $390\ \Omega$  \\
        \textbf{$R_f$} & $10\ k\Omega$  \\
        \textbf{$R_b$} & $10\ \Omega$  \\
        \textbf{$C_b$} & $940\ nF$  \\
        \bottomrule
    \end{tabular}
    
    \label{tab:parasitic_eval}
\end{table}

The strategy is to maximize the CFOA bandwidth by operating it at relatively low gain ($\sim 5\ \mathrm{V/V}$), while an external capacitance at the collector node ($C_x$) sets the dominant pole of the open-loop gain at $(C_x + C_p)R_c$.
The rise time is approximately $0.5\ ns$, whereas the targeted fall times are on the order of $50\ ns$ or longer.
Hence the output signal is an impulse with an ideally infinitely fast rise time and slow fall time which depends on $R_c$ and $C_{c}=C_x+C_p$.
Using equations (\ref{eq:open_loop_gain_ol_unified}), (\ref{eq:generic_closed_loop}), and (\ref{eq:poles_closed_loop}), and substituting $G_\infty=-R_f$, $\tau_{1_{LG}} = C_c R_c$, $\tau_{2_{LG}}\rightarrow 0$ (neglecting $C_0 R_2$ and $\tau_{HF}$), and $\beta \simeq R_b / R_f$, the output signal can be expressed as:

\begin{equation}
\begin{aligned}
&v_{out}(s)=-R_f\frac{1}{1+s\tau_{1_{CL}}}\cdot i_{in}(s) \
&v_{out}(t)=-R_f\ e^{-t/\tau_{1_{CL}}}\cdot i_{in}(t)
\end{aligned}
\label{eq:vout_measure_cp}
\end{equation}

\noindent
where $\tau_{1_{CL}} \simeq \left| \frac{A_{{AMP}0}\beta}{\tau{1_{LG}}}\right| = \frac{g_m}{C_c},(1+R_2/R_1),R_b/R_f$, and $i(s)$ represents a $\delta(t)$-like current signal.
From equation (\ref{eq:vout_measure_cp}), the dominant pole of the system can be identified as $P = \frac{1}{2\pi \tau_{1_{CL}}}$.

By measuring the signal fall time with and without $C_x$ connected to the collector, the value of $C_p$ can be extracted.
The different measured fall times are called $t_{f1}$ and $t_{f2}$ and depend on the loop gain dominant pole $P$ as $t_f\simeq ln(9)/(2\pi \cdot P)$.
The pole without $C_x$ is labeled as $P_1$, while the second pole $P_2$ has $C_x$ attached to the collector node.
Then, the ratio $R$ of the two fall times can be linked to the parasitic capacitance as:
\begin{equation}
    R=\frac{t_{f2}}{t_{f1}}=\frac{P_1}{P_2}=\frac{C_p}{C_p+C_x}\longrightarrow C_p=\frac{C_x}{R-1}
\end{equation}

This procedure was performed three times with three different $C_x$ capacitors, as indicated in table \ref{tab:parasitic_eval}. 
The estimated $C_p$ was the average of the computed values with the three different $C_x$ capacitors. 
The $C_p$ capacitance was calculated with either the LMH6702 or the LMH6703 attached to the collector node.
With rise times $<1\ ns$ and fall times $>100\ ns$, the estimated $C_p$ are $C_{p_{\, LMH6702}}=1.95\ pF$ and $C_{ p_ {\, LMH6703}}=1.69\ pF$. 
The temperature has a tiny effect on the parasitic capacitance, with a small increase of $0.2\ pF$ between ambient temperature and $80\ K$.

\bibliographystyle{elsarticle-num-names} 
\bibliography{biblio}

\end{document}